\documentclass[a4paper,11pt]{article}

\usepackage{jheppub} 
\usepackage[style=numeric-comp,
            sorting=none, 
            url = false,
            hyperref=false,
            backref=false
]{biblatex}
\usepackage{atlasbiblatex_updated}
\usepackage[autostyle]{csquotes}

\usepackage{microtype}
\usepackage{amsmath}
\usepackage{amssymb}
\usepackage{mathtools}
\usepackage{multirow}

\usepackage{xfrac}

\usepackage[
  locale=US,
  separate-uncertainty=true,
  exponent-product=\cdot
]{siunitx}
\DeclareSIUnit{\electronvolt}{\text{e\hspace{-0.1em}V}}
\DeclareSIUnit{\eV}{\text{e\hspace{-0.1em}V}}

\usepackage{booktabs}       

\usepackage{graphicx}

\usepackage{scrhack}
\usepackage{float}
\usepackage{wrapfig}
\usepackage{mathtools}
\usepackage{caption}
\usepackage{subcaption}

\usepackage[section, below]{placeins}

\usepackage{tikz}
\usepackage[hyphenbreaks]{breakurl}

\usepackage{enumitem}
\usepackage{xspace}

\newcommand{\sinthetaW}{\ensuremath{\sin^2(\theta_\text{W}^{\text{eff}})}\xspace}
\newcommand{\AFBbeauty}{\ensuremath{A_{\mathrm{FB}}^b}\xspace}
\newcommand{\AFBbeautyzero}{\ensuremath{A_{\mathrm{FB}}^{b,0}}\xspace}
\newcommand{\AFBmuon}{\ensuremath{A_\text{FB}^\mu}\xspace}
\newcommand{\Rb}{\ensuremath{R_b}\xspace}
\newcommand{\dCb}{\ensuremath{\Delta C_b}\xspace}
\newcommand*\circled[1]{\tikz[baseline=(char.base)]{
            \node[shape=circle,draw,inner sep=1.5pt] (char) {#1};}}

\title{Measuring \boldmath{$A_\mathrm{FB}^b$} and \boldmath{$R_b$} with exclusive \boldmath{$b$}-hadron decays at the FCC-ee}

\author[a,b]{Lars Röhrig}
\author[a]{\!\!, Kevin Kröninger}
\author[b]{\!\!, Romain Madar}
\author[b]{\!\!, Stéphane Monteil}
\author[c]{\!\!, Fabrizio Palla}
\author[d]{\!\!, Emmanuel Perez}
\affiliation[a]{Department of Physics, TU Dortmund University, Dortmund, Germany}
\affiliation[b]{Université Clermont-Auvergne, CNRS, LPCA, 63000 Clermont-Ferrand, France}
\affiliation[c]{INFN Pisa, Pisa, Italy}
\affiliation[d]{CERN, EP Department, Geneva, Switzerland}

\emailAdd{lars.roehrig@tu-dortmund.de}

\abstract{
    This paper presents a novel tagging technique to measure the beauty-quark partial decay-width ratio $R_b$ and its forward-backward asymmetry \AFBbeauty at the FCC-ee, using $\mathcal{O}(10^{12})$ $Z$-boson decays. The method is based on the exclusive reconstruction of a selected list of $b$-hadron decay modes in $Z\to b\bar{b}$ events at the $Z$ pole, which 
    can provide the flavour and possibly the charge of the hemisphere. This approach effectively eliminates the contamination from light-quark physics events and reduces the leading systematic uncertainties arising from background contamination, tagging-efficiency correlations, and gluon-radiation corrections by exploiting the geometric and kinematic properties of beauty hadrons. This results in a total relative uncertainty of the order of \SI{0.01}{\percent} for both observables. Furthermore, this precision allows to obtain a commensurate 
    precision on the weak mixing angle \sinthetaW compared to the muon forward-backward asymmetry on the order of \SI{0.002}{\percent}.
}

\begin{document}
\maketitle
\flushbottom

\section{Introduction}

The discovery of the Higgs boson at CERN's Large Hadron Collider (LHC) in 2012 marked an important step towards the completion of the Standard Model of Particle Physics (SM)~\cite{Higgs_discovery_ATLAS, Higgs_discovery_CMS}. 
Despite its indisputable success to describe short-scale fundamental interactions, it does neither account for the existence of dark matter nor the observed magnitude of the asymmetry of matter and antimatter in the universe, among others. 
In absence of a definite energy scale for Beyond Standard Model (BSM) processes, it is therefore mandatory to push the limits of precision physics further by studying the properties of electroweak (EW) particles with even greater accuracy. 
However, even though the LHC experiments recently provided breakthroughs in precision~\cite{CMS_W_boson_mass, ATLAS_top_quark_mass, LHCb_sin2thetaW, CMS_sin2thetaW}, the latest benchmarks in this field are mostly set by measurements from the Large Electron Positron (LEP) collider that was in service until 2000. It allowed to study the electroweak interaction from centre-of-mass energies ranging from $\sqrt{s} \approx m_Z$ up to $\sqrt{s} = \SI{209}{\giga\eV}$. From about \num{2e7} $Z$-boson decays, fundamental parameters of the SM like the weak mixing angle \sinthetaW could be determined with a precision of \SI{0.1}{\percent}. Furthermore, the data revealed a $2.9\,\sigma$ tension with the SM prediction of the beauty-quark forward-backward asymmetry \AFBbeauty, an observable that is sensitive to radiative corrections to the $Z$-boson propagator and $Zb\bar{b}$ vertex-corrections. 
The latter are better captured by the partial decay-width ratio \Rb. The measurements of these two observables at the Future Circular Collider (FCC) as an electron-positron collider (FCC-ee) are discussed in this article.  

The FCC-ee features operations at energy stages from $\sqrt{s} = m_Z$ up to the $t\bar{t}$ energy threshold~\cite{FCC_ee_the_lepton_collider}. The sheer amount of $Z$-boson decays would allow to study beauty-quark EW precision observables (EWPO) with exceptionally high statistical precision. However, methods to overcome the systematic limitations of the measurement of \Rb and \AFBbeauty are required to improve the overall uncertainty of the measurement. One of these methods is presented in this paper, which is structured as follow: Sec.~\ref{sec:motivation} motivates the measurement of EWPOs at FCC-ee; the measurement principle and quark-identification techniques from LEP-times are discussed in Sec.~\ref{sec:measurement_principle}; the proof of concept of the exclusive hemisphere-tagger is provided in Sec.~\ref{sec:exclusive_reconstruction} with a performance evaluation in Sec.~\ref{sec:performance}; the application for the measurement of \Rb and \AFBbeauty is presented in Sec.~\ref{sec:application_to_Rb} and Sec.~\ref{sec:application_to_AFBb}, respectively;  the paper is concluded in Sec.~\ref{sec:Conclusions}.
\section{Motivation for the \boldmath{$Z$} pole at FCC-ee}\label{sec:motivation}

Beyond its role in precision EW measurements, the immense dataset provided by the FCC-ee at the $Z$ pole offers the potential for discovery, including the identification of new long-lived~\cite{long_lived_particles_reference} or axion-like~\cite{axion_like_particles_future_collider} particles, as well as the detection of subtle deviations from SM predictions in EWPOs. The couplings of the $Z$ boson to quarks and leptons are therefore useful to directly probe BSM physics, but the abundant production of heavy flavours at the $Z$ pole makes FCC-ee a multiple heavy-flavour factory as well: it is simultaneously a tau, a charm, and a beauty factory. Measurements of rare $c$- and $b$-hadron decays, as well as $\tau$ decays, will complement the EWPO constraints, offering a coherent picture of New Physics.

Fundamental SM parameters will be measured with exquisite statistical precision, such as \sinthetaW, which can be inferred from the measurement of the forward-backward asymmetry $A_\text{FB}^f$ of a fermion $f$, arising from the vectorial-axial parity-violating coupling of the $Z$ boson to fermions. 
Although the standard way to measure \sinthetaW is through \AFBmuon, it would require a precise validation in the case of a potential deviation from the SM prediction. This can be brought about by \AFBbeauty, which, among any other forward-backward asymmetry, has the highest sensitivity to \sinthetaW. Additionally, the combined measurement of LEP~\cite{ALEPH_AFB_measurement, OPAL_AFB_measurement, L3_AFB_measurement, DELPHI_AFB_measurement} still has the highest tension of $2.9\,\sigma$ with the SM prediction out of all EWPOs~\cite{Precision_electroweak_measurement_on_the_Z_resonance}. However, the precision of \AFBbeauty at the $Z$ pole is strictly limited by the systematic uncertainty and requires a fundamental revision to reach competitive statistical and systematic uncertainties.

Furthermore, the partial $b$-quark decay width with respect to all hadronic $Z$ decays, \Rb, provides direct access to vertex corrections at the $Zb\bar{b}$ vertex from top-quark and $W$-boson loops, since higher-order radiative corrections to the $Z$ propagator cancel out in
\begin{equation}
    \Rb = \frac{\Gamma_{b\bar{b}}}{\Gamma_{Z\to\text{had.}}}\,.
    \label{eqn:Zbb:Rb}
\end{equation}
This allows for unique tests of modifications to the $Wtb$ coupling (as well as the indirect $Zt\bar{t}$ coupling), potentially offering a higher precision than any direct top-quark measurement.
Although both observables will be measured with outstanding statistical precision, the limitations are given by the control over systematic uncertainties. In the following chapters, a new measurement philosophy will be presented and discussed in detail. If not stated otherwise, the index $Z\to\text{had.}$ as in Eq.~\eqref{eqn:Zbb:Rb} is replaced by simply stating $Z$ as the index, which accounts for the hadronic fraction of $Z$-boson decays.
\section[Measurement principle and lessons learnt from LEP studies]{Measurement principle and lessons learnt from history}\label{sec:measurement_principle}

Both $b$-quark observables, \Rb and \AFBbeauty, share the need for an unambiguous identification of the quark flavour, also called \textit{tag}. 
Although this is sufficient for \Rb, for \AFBbeauty it is necessary to identify the charge of the quark and its direction.
Since the most precise measurements have been made at LEP and SLD, it is worth looking back at the tagging techniques used at the time. Principally, two main methods of flavour identification have been used in decays of the $Z$ boson, whose event topology is briefly highlighted before going into further detail of the tagging techniques used by the time.

\subsection{Event topology and equations}\label{subsec:Zb:event_topology}

The initial quarks produced by the $Z$-boson decay can radiate high-energetic gluons before they hadronise and form at least two sprays of particles; the hadronic decay products therefore emerge back-to-back from the interaction point. 
The plane perpendicular to the direction of the thrust defines the two hemispheres of the event. The thrust axis in the first approximation models the direction of the initial quark. It has often been used at LEP in analyses of \AFBbeauty~\cite{ALEPH_AFB_measurement, DELPHI_AFB_measurement, OPAL_AFB_measurement, L3_AFB_measurement} or the strong coupling constant $\alpha_\text{S}$~\cite{ALEPH_alphaS}.

For the measurement of \Rb, the $b$-flavour identification of the hemispheres is required and is based on a \textit{double-tag method}. This allows for the simultaneous determination of \Rb as well as the $b$-tagging efficiency $\varepsilon_b$ directly from the data. The number of single- and double-tagged events $N_b$ and $N_{b\bar{b}}$ is given by
\begin{align}
    \begin{split}
        N_b &= 2N_Z\cdot (R_b\varepsilon_{b_{1,2}} + R_c\varepsilon_{c_{1,2}} + (1-R_b-R_c)\varepsilon_{{uds}_{1,2}})\,,\\
        N_{b\bar{b}} &= \hphantom{2}N_Z\cdot (R_b\varepsilon_{b_1}\varepsilon_{b_2}C_b + R_c\varepsilon_{c_1}\varepsilon_{c_2}C_c + (1-R_b-R_c)\varepsilon_{{uds}_1}\varepsilon_{{uds}_2}\,C_{uds})\,.
    \end{split}
    \label{eqalg:Zbb:single_double_tag_equation}
\end{align}
In Eq.~\eqref{eqalg:Zbb:single_double_tag_equation}, $\varepsilon_{i_{1,2}}$ and $\varepsilon_{i_1}\varepsilon_{i_2}$ are the single- and double-tagging efficiencies to identify the flavour of the quark $i$ and $C_i$ is the \textit{hemisphere efficiency correlation} (further simply referred to as \textit{hemisphere correlation}). The correlation term accounts for a biased tagging efficiency of the other hemisphere, if the first hemisphere has been identified to originate from a quark of flavour $i$. Its mathematical expression can be derived from Eq.~\eqref{eqalg:Zbb:single_double_tag_equation}
\begin{equation}
    C_i = \frac{\varepsilon_{i_1}\varepsilon_{i_2}}{\varepsilon_{i_{1,2}}^2}\,.
\end{equation}
The efficiencies $\varepsilon_{c_j}$ and $\varepsilon_{{uds}_j}$ account for the mis-identification (ID) of a $c$- or light quark as $b$ quark. 
Their size depends on the technique for tagging $b$ quarks, where state-of-the-art methods are presented below. 
However, the impact of actual $b$ quarks in the hemisphere from gluon radiations is discussed in Sec.~\ref{sec:application_to_Rb} in the case of the novel approach introduced below.

While $\varepsilon_{b_j}$ and \Rb are determined from data, $\varepsilon_{c_j}$ and $\varepsilon_{{uds}_j}$ must be estimated from Monte-Carlo (MC) simulations. The same is true for $C_b$. Due to the small values of $\varepsilon_{c_j}$ and $\varepsilon_{{uds}_j}$, $C_c$ and $C_{uds}$ have been assumed to be unity in the former measurements. 

In addition to the simple knowledge of the flavour of the hemisphere, the charge information as well as the direction of the inital $b$ quark have to be known with high precision for a measurement of the $b$-quark forward-backward asymmetry \AFBbeauty. 

For both observables, an effective reduction of systematic uncertainties to the scale of the statistical one for \Rb and \AFBbeauty requires a more accurate $b$-hemisphere tagging. Therefore, the latest $b$-flavour tagging techniques that have been used in measurements at LEP are discussed in the following paragraphs, and their limitations in an application at the Tera-$Z$ programme at FCC-ee will be described afterwards. However, it must be stated that the main goal and challenge of tagging the flavour and charge of the hemisphere is to use flavour-specific properties of the hemispheres, such as longer lifetimes of $b$ hadrons or higher masses, to reduce misidentification from $udsc$ physics as much as possible. Due to the similar physics properties of $c$ quarks compared to $b$ quarks (lifetime, semileptonic decays, etc.), the main challenge is therefore the suppression of the contribution of $c$ quarks compared to $b$ quarks.

\paragraph{Lifetime-mass tag (hemisphere-flavour tag)}
The lifetime-mass tag combines two tags into a single one. The sole lifetime tag is based on the large displacement of the $b$ hadron from the primary vertex (PV) due to its comparatively long lifetime of about \SI{1.6}{\pico\second} and the boost at the $Z$ pole. 
However, due to the similar decay lengths of $b$ and $c$ hadrons, additional information is required to purify the event selection. This information is taken from the invariant mass of the particles that form secondary vertices, since $b$ hadrons have a significantly larger mass than $c$ hadrons. The highest $b$-tagging purity following this flavour technique has been achieved by the OPAL Collaboration with \SI{98.6}{\percent} with an efficiency of \SI{29.6}{\percent}~\cite{OPAL_Rb_measurement}.

\paragraph{Lepton tag (hemisphere-flavour and charge tag)}
The decay signatures of heavy $b$- and $c$-hadrons can provide (additional) identification and separation power. The identification of high-momentum leptons produced in semileptonic decays is an example of such a tagging property.  
Although both quark flavours produce high-momentum leptons, the transverse momentum $p_\text{T}$ is larger for $b$ hadrons, since it is kinematically limited to $\large\sfrac{m_\text{hadr.}}{2}$.
Nevertheless, a flavour identification using only the lepton tag on its own is not competitive with the lifetime-mass tag, but is used to identify the quark charge. Actually, the charge of the lepton corresponds to the flavour of the decaying $b$ hadron. It must be noticed that the correspondence to the initial quark charge is diluted in the presence of neutral $b$-meson $B^0\!-\!\bar{B}^0$ mixing or secondary semileptonic $b \to c \to \ell^+$ cascades.

\paragraph{Jet charge (hemisphere-charge tag)}
From the average charge of particles in a jet or hemisphere, the initial quark charge can be inferred. In combination with other taggers, such as the vertex charge, very high purities have been reached for the measurement of \AFBbeauty~\cite{ALEPH_AFB_measurement}.
\par\bigskip
Although the tags using techniques such as the lifetime, mass, high-energetic leptons, or the vertex charge have made the most of the statistics available at the former lepton collider generation, their application at a Tera-$Z$ programme becomes challenging: the estimation of the quantities $\varepsilon_c$ and $\varepsilon_{uds}$ would simultaneously require enormous amounts of simulated events and much more accurate control of the physics details of the simulation to achieve a precision comparable to that obtainable with data on $\varepsilon_b$.

\subsection{Limitations}
In the following, a new hemisphere-flavour tagger is motivated in the context of the Tera-$Z$ programme at FCC-ee with approximately $10^{12}$ $Z\to b\bar{b}$ events. Although with this amount of data at hand, statistical precision is no longer a limiting factor, efficient control over the (sources of) systematic uncertainties becomes inevitable to improve the measurement uncertainty for \Rb and \AFBbeauty to actually reach $\mathcal{O}(\sigma_\text{syst.}) = \mathcal{O}(\sigma_\text{stat.})$.

The breakdown of systematic uncertainties from the ALEPH measurement of \Rb points to the region in the measurement, which can bring the largest improvement to shrink the respective source of systematic uncertainty. The three main sources of systematic uncertainties are briefly summarised below, indicating their percentage weight in parentheses~\cite{ALEPH_Rb_measurement_lifetime_mass}.

\paragraph{Monte-Carlo statistics (16\,\%)} The finite number of MC events leads to a small uncertainty in determining $\varepsilon_{{c}_j}$ and $\varepsilon_{{uds}_j}$. Studying systematic effects from $c$ and light-quark physics modelling would require unfeasible amounts of simulated events. This is directly linked to the next point.

\paragraph{\boldmath{$udsc$} physics (62\,\%)} Systematic uncertainties on $\varepsilon_{{c}_j}$ and $\varepsilon_{{uds}_j}$ arise from two main sources: in the simulation of tracking and in the physics modelling of charm- and light-quark events. Momentum and angular dependencies on the impact-parameter resolution affect the tagging efficiency, and have been treated as systematic uncertainty. Furthermore, uncertainties on the physics inputs to model $udsc$ events have been propagated to estimate the impact on \Rb. 
The modelling of hadronisation fractions, which control the production of different charm states, is particularly important due to the hierarchy in lifetimes and, therefore, influence $\varepsilon_{{c}_j}$. The main uncertainty in $\varepsilon_{{uds}_j}$ originates from the modelling of gluon-splitting events, where $\varepsilon_{{c}_j}$ and $\varepsilon_{{uds}_j}$ depend on the $g\to b\bar{b}$ rate.

\paragraph{Hemisphere correlation (22\,\%)} The departure of the hemisphere correlation value $C_b$ from unity is a source of systematic uncertainty, which contributes to the total uncertainty budget. A detailed study of $C_b$ discussing its sources and how to overcome its implications for \Rb is given in Sec.~\ref{sec:application_to_Rb}.

For \Rb, it can be concluded that approximately \SI{80}{\percent} of the systematic uncertainty arises from the contamination of $udsc$-physics events and the estimation of their respective tagging uncertainties in $b$-quark events. This leads to the two main requirements for \Rb at FCC-ee:
\begin{enumerate}
    \item $b$-quark events need to be identified with a purity of \SI{100}{\percent}, which in turn results in $\varepsilon_{{c}_j} = \varepsilon_{{uds}_j} = \SI{0}{\percent}$.
    \item The hemisphere correlation $C_b$ must be controlled to the per-mille level around one.
\end{enumerate}
For \AFBbeauty, the systematic uncertainty budget consists of about \SI{50}{\percent} referred to corrections that have to be applied to account for gluon radiations from the $b$ quark (Quantum Chromodynamics (QCD) corrections). Further uncertainties arise from the knowledge on hadronisation and modelling parameters, contamination from $udsc$-physics events, and detector-related uncertainties.
Similarly to \Rb, two main conclusions can be drawn for the measurement of \AFBbeauty at FCC-ee:
\begin{enumerate}
    \item The charge and the flavour of the $b$-quark events need to be identified with a purity of \SI{100}{\percent}.
    \item The QCD corrections need an effective reduction up to a level such that their impact on the systematic uncertainty does not inflate the overall measurement uncertainty.
\end{enumerate}
All of the aforementioned lessons from the LEP measurements are addressed with a new hemisphere-flavour tagger, which is based on the exclusive reconstruction of $b$-hadrons in the hemispheres. This leads to a background-free, up to a charge-unambiguous tag when using only non-mixing $b$ hadrons, which reduces the systematic uncertainty budget for both measurements by about \SI{70}{\percent}. 
In the following section, the exclusive reconstruction and its implications for \Rb and \AFBbeauty are detailed.
\clearpage\section{Exclusive \boldmath{$b$}-hadron reconstruction}\label{sec:exclusive_reconstruction}

The following section describes the fundamental principle of tagging hemispheres with exclusively reconstructed $b$-hadrons. For this, the statistical precision is evaluated first for the tagger before the simulated events used for the reconstruction and further studies are detailed.

\subsection{Statistical precision}
The basic principle lies in the reconstruction of a list of $b$-hadron decay modes that, if one of them has been reconstructed in an event, gives an unambiguous tag of the $b$-quark flavour in $Z\to q\bar{q}$ events with light-quark contribution only from $g\to b\bar{b}$ gluon-splitting. Furthermore, the charge ambiguity in the application for \AFBbeauty can be removed by considering only charged $b$-mesons and -baryons as a flavour and charge tagger, namely $B^\pm$ and $\Lambda_b^0$. 

In turn, this means for \Rb, that $\varepsilon_{{c}_j} = \varepsilon_{{uds}_j} = 0$ and Eq.~\eqref{eqalg:Zbb:single_double_tag_equation} simply reduces to\footnote{Here, gluon radiations and splitting into a $b\bar{b}$ pair has been neglected in this first, simplified approach}
\begin{align}
    N_b &= 2N_ZR_b\varepsilon_{b_{1,2}}\\
    N_{b\bar{b}} &= \hphantom{2}N_ZR_b\varepsilon_{b_1}\varepsilon_{b_2}C_b\,.
    \label{eqalg:Zbb:updated_Rb_formula}
\end{align}
With these updated equations at hand, the statistical uncertainty can be calculated to serve as a benchmark for the systematic uncertainty.

\paragraph{Statistical uncertainty of \boldmath{$R_b$}}
Since charge information is not required for \Rb, the list of $b$ hadrons to be used can be extended to neutral $b$ mesons so that it covers the decays of $B^0$, $B_s^0$, $B^\pm$ and $\Lambda_b^0$. Due to the limited branching ratio (Br), only decay modes with sufficiently large probabilities (typically greater than \num{e-3}) are considered. In addition, a maximum number of two neutral pions in the final state and no leptonic modes have been selected. The complete list of the decay modes included is presented in App.~\ref{sec:app:decay_modes}. 
In conclusion, an overall tagging efficiency of $\varepsilon_{b_{1,2}} = \SI{1}{\percent}$ is within reach. From this, the statistical precision is calculated from the known Gaussian uncertainty propagation
\begin{equation}
    \sigma_\text{stat.}(\Rb) = \sqrt{\sum_{i\in[Z, b, b\bar{b}]}^{} \left(\frac{\partial \Rb}{\partial N_i}\cdot \sqrt{N_i}\right)^2 + \sum_{i,j,i\neq j}^{} \kappa_{N_i,N_j}}\,,
    \label{eqn:Zbb:stat_precision}
\end{equation}
with the correlation expressions
\begin{equation}
    \kappa_{N_i,N_j}(\Rb) = 2\,\text{cov}(N_i, N_j) \frac{\partial \Rb}{\partial N_i}\frac{\partial \Rb}{\partial N_j}\,.
\end{equation}
Taking into account the correlations between $N_b$, $N_{b\bar{b}}$ and $N_{Z}$, $\sigma_\text{stat.}(\Rb)$ is derived numerically with the \texttt{ForwardDiff} package~\cite{ForwardDiff} and results to
\begin{equation}
    \sigma_\text{stat.}(\Rb) = \num{2.22e-5}\,,
\end{equation}
which is an improvement of a factor of 30 with respect to the most precise measurement~\cite{DELPHI_Rb_measurement}. 

\paragraph{Statistical uncertainty of \boldmath{$A_\text{FB}^b$}}
In case of \AFBbeauty the list of $b$-hadrons is reduced, which results in a lowered $b$-tagging efficiency of $\varepsilon_{b_{1,2}} \approx \SI{0.45}{\percent}$. However, for \AFBbeauty, only single-tagged \textit{forward} and \textit{backward} events $N_\text{F}$ and $N_\text{B}$ are needed, and $\varepsilon_{b_{1,2}}$ does not scale to the square as for \Rb. In this context, forward and backward refer to the angle between the incoming electron and the outgoing $b$-quark.
The statistical uncertainty follows from the definition of \AFBbeauty expressed in terms of $N_\text{F}$ and $N_\text{B}$
\begin{equation}
    \AFBbeauty = \frac{N_\text{F} - N_\text{B}}{N_\text{F} + N_\text{B}}\,.
    \label{eqn:Zbb:AFB_from_counting}
\end{equation}
Again, the statistical uncertainty is derived numerically and gives 
\begin{equation}
    \sigma_\text{stat.}(\AFBbeauty) = \num{1.56e-5}\,.
    \label{eqn:Zbb:AFB_stat}
\end{equation}
This bare statistical precision translates into an improvement of about a factor \num{60} compared to the statistically most precise measurement~\cite{ALEPH_AFB_measurement}.

In the following sections, the new tagging method is applied to simulated events in order to test the purity assumption of \SI{100}{\percent} in an FCC-ee environment. 

\subsection{Event samples}
Simulated events have been used to perform different stages of the analysis. A summary is given is Tab.~\ref{tab:event_samples}.
\begin{table}[h]
    \centering
    \caption{The tabular summary of all samples used throughout this paper.}
    \label{tab:event_samples}
    \small
    \begin{tabular}{l|ccc}
        \toprule
        Dataset and analysis & Simulation type & Exclusive/Inclusive & Sample size \\
        \midrule
        \circled{1} Hemisphere-tagger performance & Fast (IDEA) & Inclusive   & \num{4e7} \\
        \circled{2} Application for \Rb                   & Full (CLD)  & Exclusive   & \num{e6} \\
        \circled{3} Application for \AFBbeauty            & Fast (IDEA) & Exclusive   & \num{5e7} \\
        \bottomrule
    \end{tabular}
\end{table}
The hard scattering as well as the hadronisation of inclusive $Z\to q\bar{q}$ events have been centrally simulated using \texttt{PYTHIA8}~\cite{Pythia8_reference} with a parameterised IDEA detector~\cite{IDEA_detector} response (dataset \circled{1}). Observable-specific samples have been exclusively simulated forcing the decay in both hemispheres to be
\begin{itemize}
    \item Hemisphere 1: $b \to B^+ \to [K^+\pi^-]_{\bar{D}^0}\,\pi^+$
    \item Hemisphere 2: $\bar{b} \to B^- \to [K^-\pi^+]_{D^0}\,\pi^-$
\end{itemize}
For this, \texttt{EvtGen}~\cite{EvtGen_reference} has been used together with either a fully simulated CLD detector~\cite{CLD_detector} response using \texttt{GEANT4}~\cite{Geant4_reference} (dataset \circled{2}) or the IDEA detector card using \texttt{DELPHES}.
In the following, statistical uncertainties for \Rb and \AFBbeauty are derived and the principle of the exclusive $b$-hadron reconstruction is detailed.

\subsection{Representative decays: one of the six}\label{subsec:Zbb:exclusive_reconstruction_mode}
Out of the $\mathcal{O}(\num{200})$ decay modes to be considered, a comprehensive selection of six has been made to serve as representative modes for the rest. These six modes are characterised by the number of tracks and the number of neutral pions in their respective final state and have been exclusively reconstructed from the simulated dataset \circled{1} as indicated in Tab.~\ref{tab:event_samples}. They are grouped into three classes.
\begin{description}
    \item[Including one \boldmath{$c$} meson] $B^+ \to \bar{D}^0\pi^+$ with $\text{Br} = \num{4.61e-3}$, considering different $\bar{D}^0$ decays:
    \begin{description}
        \item[\normalfont\textit{Fully charged:}] $\bar{D}^0 \to K^+\pi^-$, with $\text{Br} = \SI{3.947}{\percent}$
        \item[\normalfont\textit{One neutral pion:}] $\bar{D}^0 \to K^+\pi^-\pi^0$, with $\text{Br} = \SI{14.4}{\percent}$
        \item[\normalfont\textit{Two neutral pions:}] $\bar{D}^0 \to K^+\pi^-\pi^0\pi^0$, with $\text{Br} = \SI{8.86}{\percent}$
        \item[\normalfont\textit{Four charged tracks at the decay vertex:}] $\bar{D}^0 \to K^+\pi^-\pi^-\pi^+$, with $\text{Br} = \SI{8.22}{\percent}$
    \end{description}
    \item[Including two \boldmath{$c$} mesons] $B^+ \to \bar{D}^0D_s^+$ with $\text{Br} = \num{9e-3}$ and the fully charged decay of $\bar{D}^0 \to K^+\pi^-$ and $D_s^+ \to K^+K^-\pi^+$, which has a $\text{Br}$ of \SI{5.37}{\percent}
    \item[Including a \boldmath{$c\bar{c}$} meson] $B^+\to J/\psi\,K^+$ with $\text{Br} = \num{1.02e-3}$ and fully leptonic decay of $J/\psi \to \ell^+\ell^-$ with $\text{Br} = \SI{11.932}{\percent}$ for $\ell\in[e, \mu]$
\end{description}
In the following, the mode $B^+ \to \bar{D}^0\pi^+ \to [K^+\pi^-]_{\bar{D}^0}\pi^+$ has been chosen to exemplary present the reconstruction process and to quantify its tagging performance.
If not stated otherwise, the charge-conjugated decay is considered likewise. The results of the remaining five decay modes are presented in App.~\ref{app:subsec:Zbb:rest_of_the_modes}.

\paragraph{\boldmath{$B^+$} reconstruction}
A pair of oppositely charged kaon and pion is required to have a common vertex with less than \SI{50}{\micro\meter} disagreement. At this stage, no vertex-fitting tools have been applied, since neutral pseudotracks have not yet been made available in the \texttt{DELPHES} tool~\cite{Delphes_reference}. Hence, the sole four-vectors of the reconstructed particles have been used. The invariant mass of the $\bar{D}^0$ candidates has been modelled with the sum of three Gaussian distributions from which the mass window of $(1790 \leq m(\bar{D}^0) < 1940)\,\si{\mega\electronvolt}$ has been chosen to accept the candidates for the $B^+$ reconstruction.

\begin{figure}[t]
    \centering
    \begin{subfigure}[t]{0.48\textwidth}
        \centering
        \includegraphics[width = 1\textwidth]{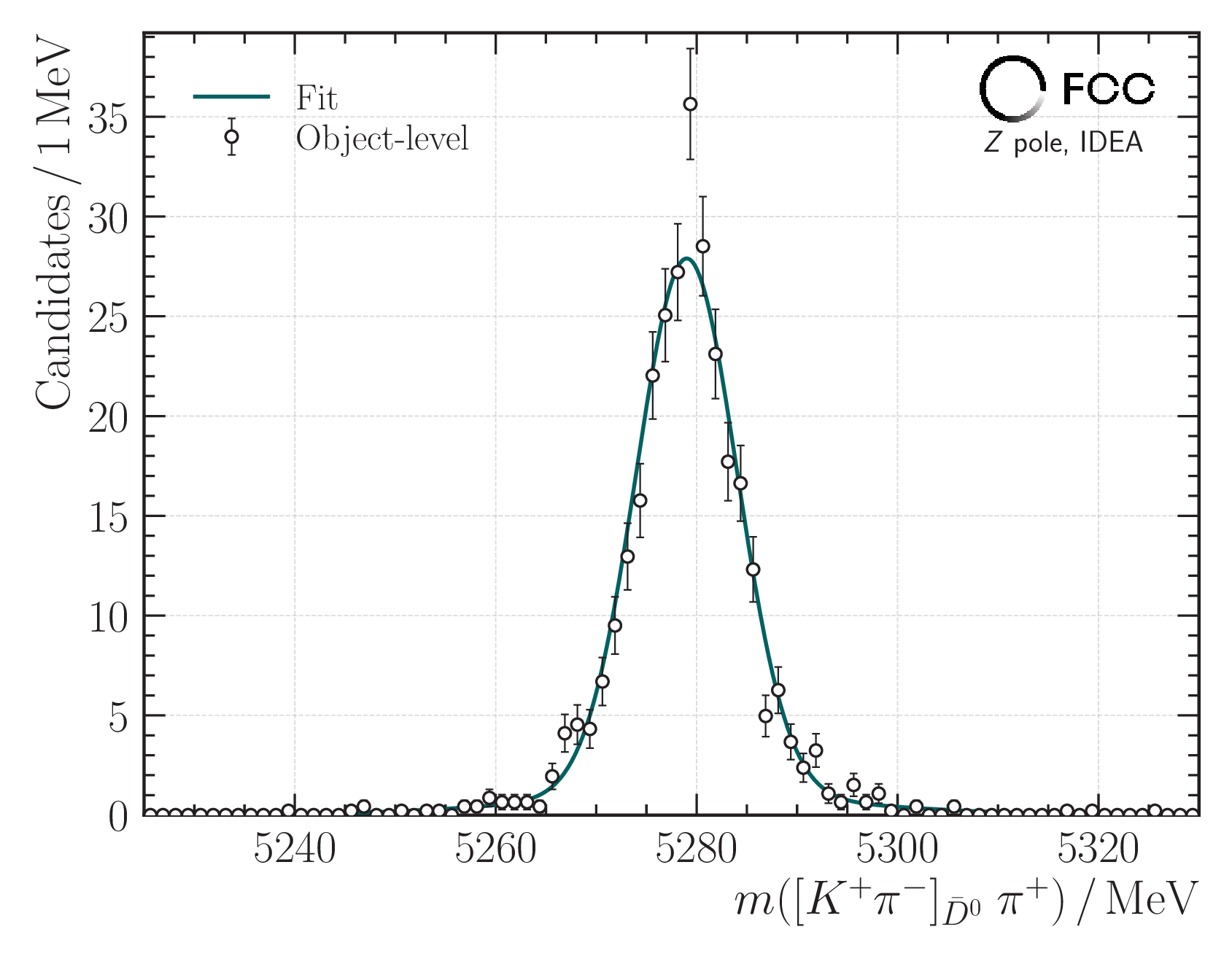}
        \caption{Truth-matched $B^+$ candidates with an unbinned maximum likelihood fit, showing the distributions at the particle- and object level in orange and black dots, respectively.}
        \label{subfig:Zbb:B_mass_peak_fit}
    \end{subfigure}\hfill
    \begin{subfigure}[t]{0.48\textwidth}
        \centering
        \includegraphics[width = 1\textwidth]{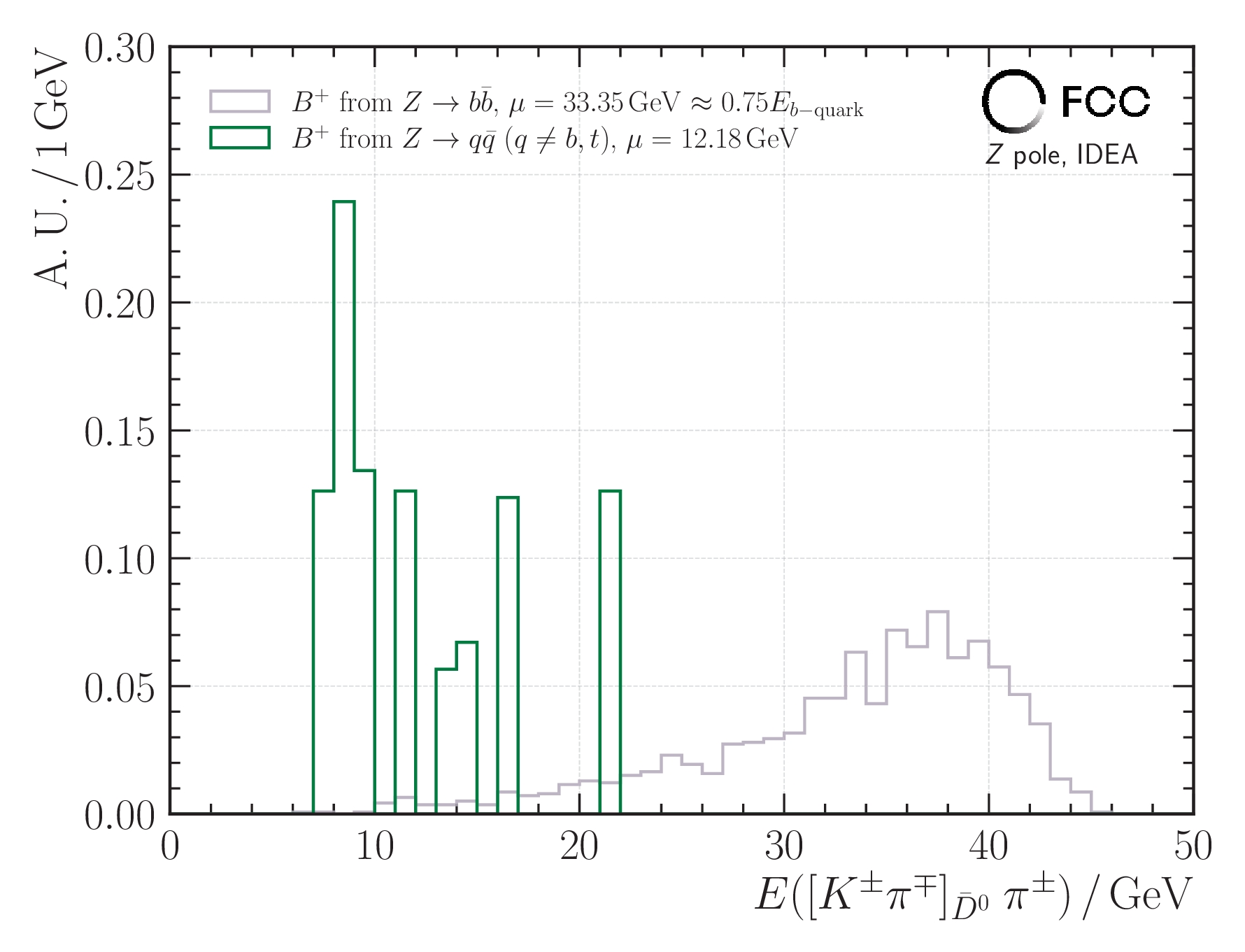}
        \caption{The energy distribution of the $B^+$ candidates, for the signal (grey) and background (green) candidates. The background events originate only from gluon-splitting events and have much less energy.}
        \label{subfig:Zbb:B_energy_comparison}
    \end{subfigure}
    \caption{Fit to the invariant-mass distribution around the signal peak in Fig.~\subref{subfig:Zbb:B_mass_peak_fit} and the $B^+$-energy distribution in Fig.~\subref{subfig:Zbb:B_energy_comparison}. An opportunistic cut on the energy is set to $E_{B^+} \geq \SI{20}{\giga\electronvolt}$.}
    \label{fig:Zbb:B_mass_and_energy}
\end{figure}

In the second step, an additional charge-matching pion has been added to the $\bar{D}^0$ to form $B^+$ candidates. Similarly to the $\bar{D}^0$ reconstruction, a vertex resolution of \SI{50}{\micro\meter} has been emulated. To further remove the contribution from background events and to make use of the boost at the $Z$ pole, a cut on the $B^+$ flight distance of \SI{300}{\micro\meter} with respect to the PV has been applied. 
In Fig.~\ref{fig:Zbb:B_mass_and_energy}, the left panel shows the invariant-mass distribution at the particle and object level of the truth-matched $B^+$ signal candidates, including an unbinned maximum likelihood fit to the object-level distribution. 
Further exclusion of contamination from background events in the signal region, which enter only from gluon splitting through $q \to q + [b\bar{b}]_g$ with $q\in[u,d,s,c]$, has been achieved by inspecting the energy spectrum of the background candidates, which is expected to be softer than from the signal, as shown in the right panel of Fig.~\ref{fig:Zbb:B_mass_and_energy}. The limited amount of data at hand does not allow for a proper statistical evaluation of a suitable energy cut; therefore, an opportunistic one has been set at $E_{B^+} > \SI{20}{\giga\electronvolt}$, which removes most of the background $B^+$ candidates. 

In the following section, the performance of the tagger in terms of purity and reconstruction efficiency is evaluated from the invariant-mass spectrum of the reconstructed $B^+$ candidates.
\section{Performance of the tagger}\label{sec:performance}

\begin{figure}[t]
    \centering
    \includegraphics[width = 0.7\textwidth]{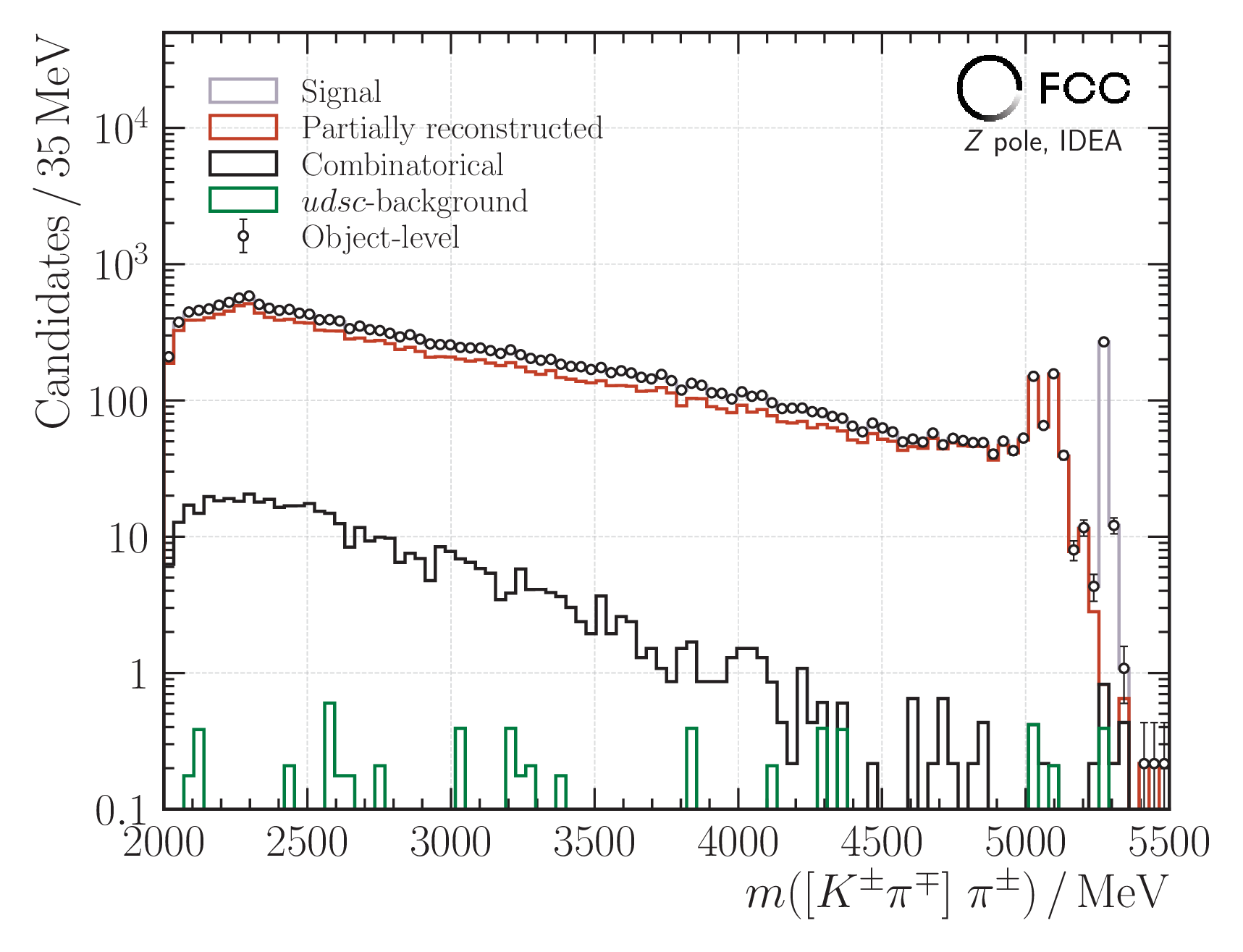}
    \caption{Invariant mass-spectrum for the $B^+\to [K^+\pi^-]_{\bar{D}^0}\pi^+$ decay mode. The different contributions from the signal, partially reconstructed, and combinatorial background, as well as from the $udsc$-physics background events are shown in grey, red, black, and green, respectively. The purity in the mass region of $(5150\leq m_{B^+} \leq 5400)\,\si{\mega\electronvolt}$ has been evaluated to be \SI{99.89(9)}{\percent} with $E_{B^+} \geq \SI{20}{\giga\electronvolt}$. Lowering the mass constraint to also include partially reconstructed events greatly increases the efficiency of the exclusive tagger.}
    \label{fig:Zbb:B00_full_mass}
\end{figure}

So far, neither the direction of the hemisphere has been considered nor has the question of whether events have one or two tagged hemispheres been considered. However, the bare reconstruction and tagger performance can be evaluated from the invariant $B^+$-mass spectrum, which serves as observable to quantify the purity of the hemisphere-flavour tagger. The spectrum is presented in the range from $(2000 \leq m([K^+\pi^-\pi^0]\pi^+) \leq 5500)\,\si{\mega\electronvolt}$ in Fig.~\ref{fig:Zbb:B00_full_mass}, distinguishing between different contributions: the grey peak shows the candidates from the signal $B^+$ mesons, while the partially reconstructed\footnote{Partially reconstructed particles refer to the (intermediate) particles where not all decay products are fully reconstructed. In case of $B^+ \to \bar{D}^0 \pi^+$, this can include $B^+ \to \bar{D}^0\pi^+\pi^+\pi^-$ with a Br of \num{5.5e-3} and two missing charged pions.} and combinatorial background candidates from $Z\to b\bar{b}$ events are coloured red and black, respectively. The overall background contribution from $Z\to q\bar{q}$ events with $q\in[u,d,s,c]$ is shown in green, while each contribution is weighted with their respective fraction $R_q$.

For the purpose of flavour (and charge) tagging the event, candidates within the mass-peak region of $(5100 \leq m_{B^+} \leq 5500)\,\si{\mega\electronvolt}$ have been selected, also to first assess systematic uncertainties in this region. The purity $P$, where every contribution except the $udsc$-physics background is taken as signal $N_\text{S}$, results in
\begin{equation}
    P = \frac{N_\text{S}}{N_\text{S} + N_\text{B}} = \SI{99.81(7)}{\percent}\,,
    \label{eqn:Zbb:purity}
\end{equation}
where the uncertainty given is statistical and $N_\text{B}$ refers to the number of background events. Here, it can already be concluded that the exclusive reconstruction as tagger achieves ultra-high purities, which are only contaminated from physical, non-reducible background arising from gluon splitting. Their impact on the systematic uncertainty, also in comparison to the hemisphere correlation, is studied in further detail in Sec.~\ref{sec:application_to_Rb}.

The performance of the other representative decay modes is summarised in Tab.~\ref{tab:Zbb:purity_efficiency}. As can be seen, for all decay modes a purity above $\SI{99.7}{\percent}$ has been reached, where the uncertainty stated in the table refers to the finite statistical precision of the dataset. For the decay $B^+\to D_s^+\bar{D}^0$, no energy cut has been applied, since no background events have been found, also due to the smallest Br among all modes. However, the energy cut will probably be required with the full event statistics in place.

The reconstruction efficiencies $\varepsilon_\text{reco}$ are calculated as the ratio of reconstructed candidates with respect to the generated ones. Therefore, the efficiency includes all cut efficiencies, namely the cut on the flight distance, cuts on invariant masses of intermediate particles, and the final $B^+$-meson energy.
\begin{table}[t]
    \centering
    \caption{Reconstruction efficiencies and purities for the six representative decay modes in the mass-peak region. In total, purities above \SI{99.7}{\percent} are in reach for all considered $B^+$ decay modes.}
    \label{tab:Zbb:purity_efficiency}
    \begin{tabular}{l|S[table-format=2.2(2.2)] S[table-format=2.2(2)]}
        \toprule
        $B^+$ decay mode & {$\varepsilon_\text{reco}$\,/\,\%} & {Purity\,/\,\%} \\
        \midrule
        $\bar{D}^0\pi^+ \to [K^+\pi^-]_{\bar{D}^0}\pi^+$           & 77.17(299) & 99.93(11) \\
        $\bar{D}^0\pi^+ \to [K^+\pi^-\pi^0]_{\bar{D}^0}\pi^+$      & 64.89(141) & 99.89(09) \\
        $\bar{D}^0\pi^+ \to [K^+\pi^-\pi^0\pi^0]_{\bar{D}^0}\pi^+$ & 49.95(268)& 99.81(07) \\
        $\bar{D}^0\pi^+ \to [K^+\pi^-\pi^-\pi^+]_{\bar{D}^0}\pi^+$ & 72.63(690)& 99.73(27) \\
        $D_s^+\bar{D}^0 \to [K^+K^-\pi^+]_{D_s^+}[K^+\pi^-]_{\bar{D}^0}$ & 78.57(2239) & 100.00(0) \\
        $J/\psi\,K^+ \to [\ell^+\ell^-]_{J/\psi}K^+$                & 85.87(413) & 99.90(24) \\
        \bottomrule
    \end{tabular}
\end{table}
This section closes the motivation, description and evaluation of a new $b$-hemisphere tagger for the application at the Tera-$Z$ programme at FCC-ee. Its validity and feasibility have been shown and the principle has been demonstrated with the exclusive reconstruction of the $B^+\to[K^+\pi^-\pi^0]_{\bar{D}^0}\pi^+$ decay. The next section presents the application of the reconstructed $b$-hadrons for the measurement of \Rb and \AFBbeauty.
\section{Application to the measurement of \boldmath{$R_b$}}\label{sec:application_to_Rb}

For \Rb, only the flavour tag of the hemisphere is of interest; therefore, information about the direction or about the charge of the hemisphere is not necessary. This specificity allows to significantly increase the tagger efficiency $\varepsilon_{{b}_{1,2}}$ from the targeted \SI{1}{\percent} by including also partially reconstructed candidates as hemisphere taggers. This approach is mainly driven by the absence of $udsc$ contributions outside the signal-peak region. The possibility of releasing the mass-peak constraint and its impact on the purity and tagging efficiency are discussed below.

\begin{figure}[t]
    \centering
    \begin{subfigure}[t]{0.48\textwidth}
        \includegraphics[width = 1\textwidth]{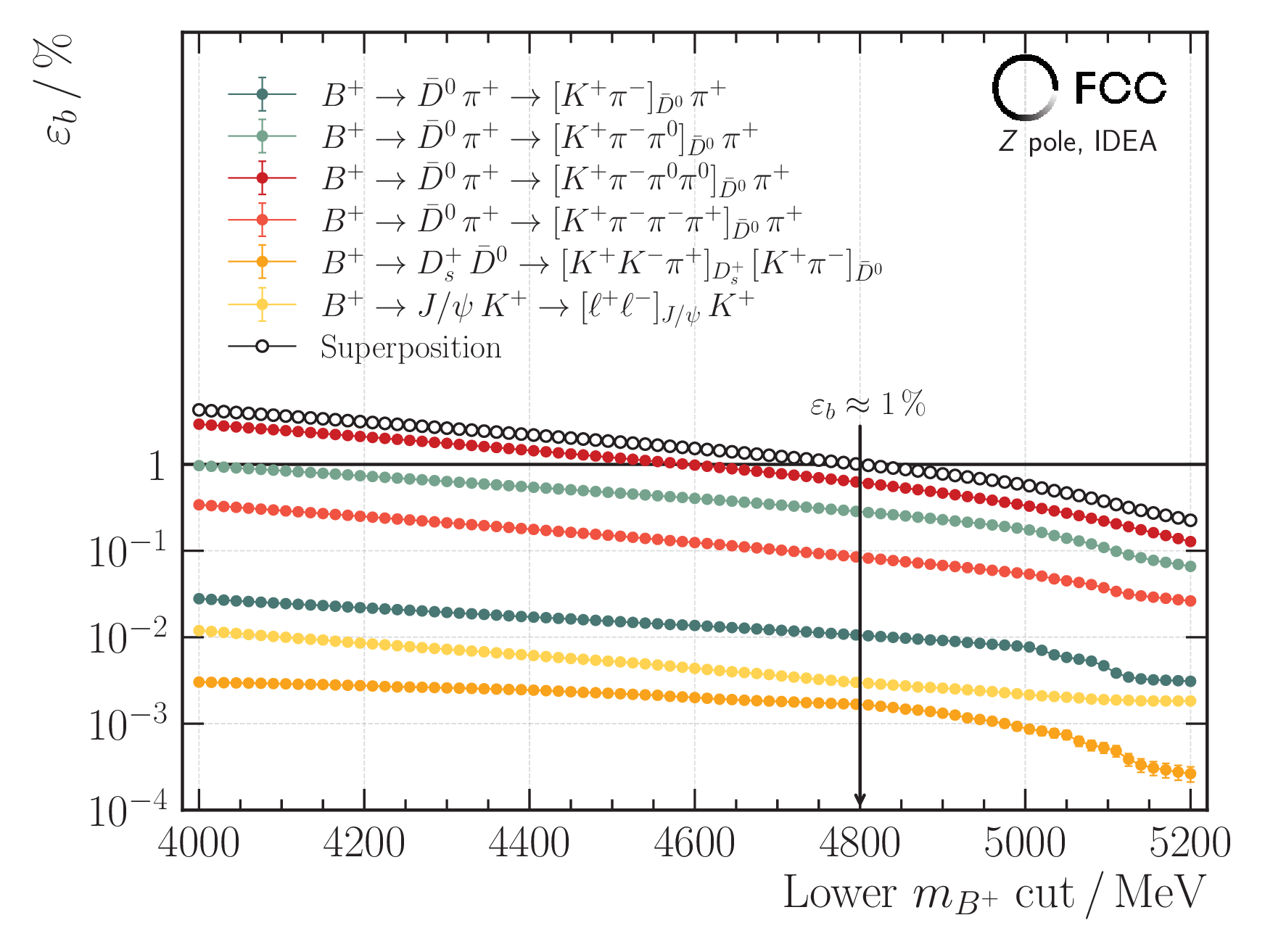}
        \caption{Lower invariant-mass cuts highly increase the efficiency of the tagger. The sum of the different modes is shown in black dots and is labelled \textit{Superposition}.}
        \label{fig:enter-label}
    \end{subfigure}\hfill
    \begin{subfigure}[t]{0.48\textwidth}
        \includegraphics[width = 1\textwidth]{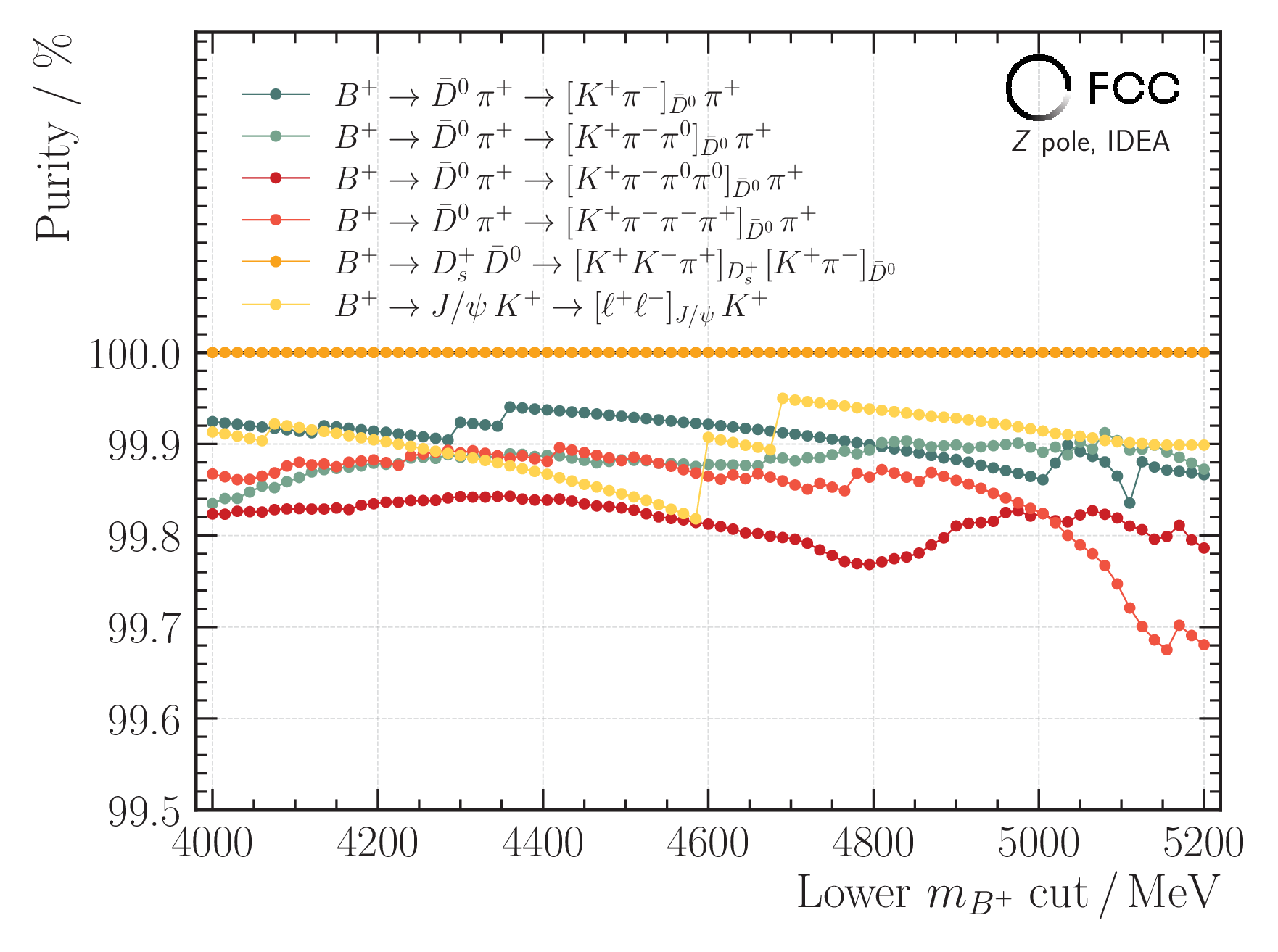}
        \caption{The purity as function of the lower invariant mass cut. All of the values converge to purities above \SI{99.8}{\percent}. The more neutral pions are included in the final state, the lower the purity of the mode.}
        \label{fig:enter-label}
    \end{subfigure}
    \caption{Already with the six representative decay modes and a cut at $m_{B^+} \geq \SI{4800}{\mega\electronvolt}$, the targeted \SI{1}{\percent} tagging efficiency is reached without loss in purity.}
    \label{fig:Zbb:B_mass_cut_efficiency_purity}
\end{figure}

The left panel of Fig.~\ref{fig:Zbb:B_mass_cut_efficiency_purity} illustrates $\varepsilon_{{b}_{1,2}}$ as a function of the invariant $B^+$-mass cut, where $\varepsilon_{{b}_{1,2}}$ is determined by
\begin{equation}
    \varepsilon_b = \frac{N^{\text{all}}_\text{S}}{N_\text{gen}}\,.
    \label{eqn:Zbb:epsilon_b}
\end{equation}
In Eq.~\eqref{eqn:Zbb:epsilon_b}, $N_\text{gen} = \num{9.5e6}$ represents the number of generated events and $N^{\text{all}}_\text{S}$ is the number of all candidates originating from the $Z\to b\bar{b}$ decay, respectively. 
As anticipated, the efficiency for all decay modes increases significantly with the mass threshold, even reaching the \SI{1}{\percent} threshold for the modes with neutral pions within the mass window studied. In addition, the combined efficiencies of the decay modes studied are highlighted in black and are referred to as \textit{Superposition}. The plot indicates that the six representative decay modes with a lower invariant-mass cut of $m_{B^+} > \SI{4800}{\mega\electronvolt}$ are sufficient to achieve a tagging efficiency of $\varepsilon_b \approx \SI{1}{\percent}$. 

Consequently, the purity at the same invariant-mass cuts, as calculated in Eq.~\eqref{eqn:Zbb:purity}, is displayed in the right panel of Fig.~\ref{fig:Zbb:B_mass_cut_efficiency_purity}. A convergence towards purities exceeding $99.8\,\%$ can be observed for all decay modes. Even tighter cuts on the $B^+$ energy can be applied when the mass window constraint is released, which would further reduce the impact from gluon radiation and would lead to even higher purities.
However, the release of the invariant-mass constraint would require a dedicated investigation into the impact on the systematic uncertainties for \Rb, which is beyond the scope of this paper. The impact of systematic uncertainties has only been examined assuming the candidates in the signal-peak region. Their sources and handling are discussed in the following sections.

\paragraph{Sources and impact of systematic uncertainty}
Systematic uncertainties arise from two main origins: 
\begin{enumerate}
    \item Gluon radiation from a light quark and subsequent splitting into a $b\bar{b}$ pair through $q \to q[b\bar{b}]_g$ for $q\in[u,d,s,c]$. These enter the signal region when the $b$ quark hadronises and decays into the channel under study. The probability of gluon splitting is given by $g_{b\bar{b}}$ and its precision and impact on $\sigma_{\text{syst.}}(\Rb)$ are examined.
    \item Correlation of single- and double-tagging efficiencies between the two hemispheres, $C_b$. The sources of this correlation and the appropriate methods for addressing it in the measurement of \Rb are discussed.
\end{enumerate}
The impact of both sources of systematic uncertainties on $\sigma_{\text{syst.}}(\Rb)$ is first worked out, assuming that each contribution adds in quadrature to the total systematic uncertainty
\begin{equation}
    \sigma_\text{syst.}(\Rb) = \sqrt{\left(\sigma^{\text{from\,}g_{b\bar{b}}}_\text{syst.}(\Rb)\right)^2 + \left(\sigma^{\text{from\,}C_b}_\text{syst.}(\Rb)\right)^2}
\end{equation}
For this analysis, the most precise measurements for $g_{b\bar{b}}$ (where the average value has been calculated from all LEP and SLD measurements~\cite{ALEPH_gluon_splitting, DELPHI_gluon_splitting, DELPHI_gluon_splitting_2, OPAL_gluon_splitting, SLD_gluon_splitting} as described in Ref.~\cite{DESY_gluon_splitting_summary}) and the hemisphere-correlation coefficient $C_b$ obtained by the ALEPH~Collaboration~\cite{ALEPH_Rb_measurement_lifetime_mass} are used as reference values
\begin{align}
    g_{b\bar{b}} &= \num{0.00247(56)}\,, \label{eqalg:Zbb:gbb_DESY}\\
    \Delta C_b^\text{ALEPH} &= 0.0376\,\pm\,0.0025(\text{stat.})\,\pm\,0.0027(\text{syst.}) = \num{0.0376(37)}\,.\label{eqalg:Zbb:dCb_ALEPH}
\end{align}
Because in most measurements of \Rb, the difference to unity, $\dCb = 1 - C_b$, is stated, it will also be used in the following. A value of $\dCb = 0$ would refer to no bias in the tagging of the hemispheres. 

\begin{figure}[t]
    \begin{subfigure}[t]{0.48\textwidth}
        \centering
        \includegraphics[width = 1\textwidth]{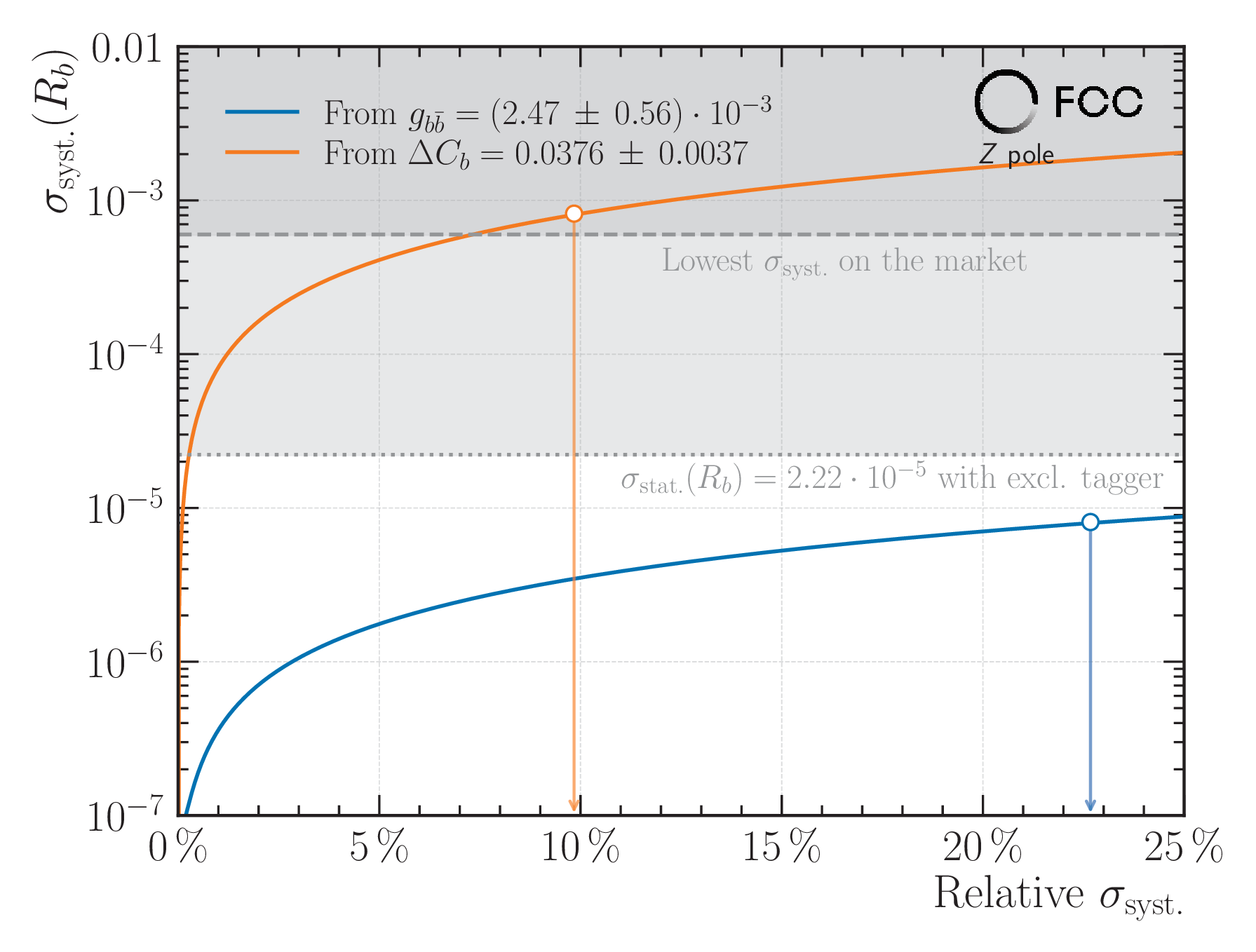}
        \caption{Impact of relative systematic uncertainties (expressed as a percentage) on the precision of \Rb values derived from two sources: the hemisphere correlation \dCb in orange and the gluon-splitting rate $g_{b\bar{b}}$ in blue. The dominating uncertainty in $\sigma_{\text{syst.}}(\Rb)$ originates from the hemisphere correlation.}
        \label{subfig:Zbb:gbb_and_Cb}
    \end{subfigure}\hfill
    \begin{subfigure}[t]{0.48\textwidth}
        \centering
        \includegraphics[width = 1\textwidth]{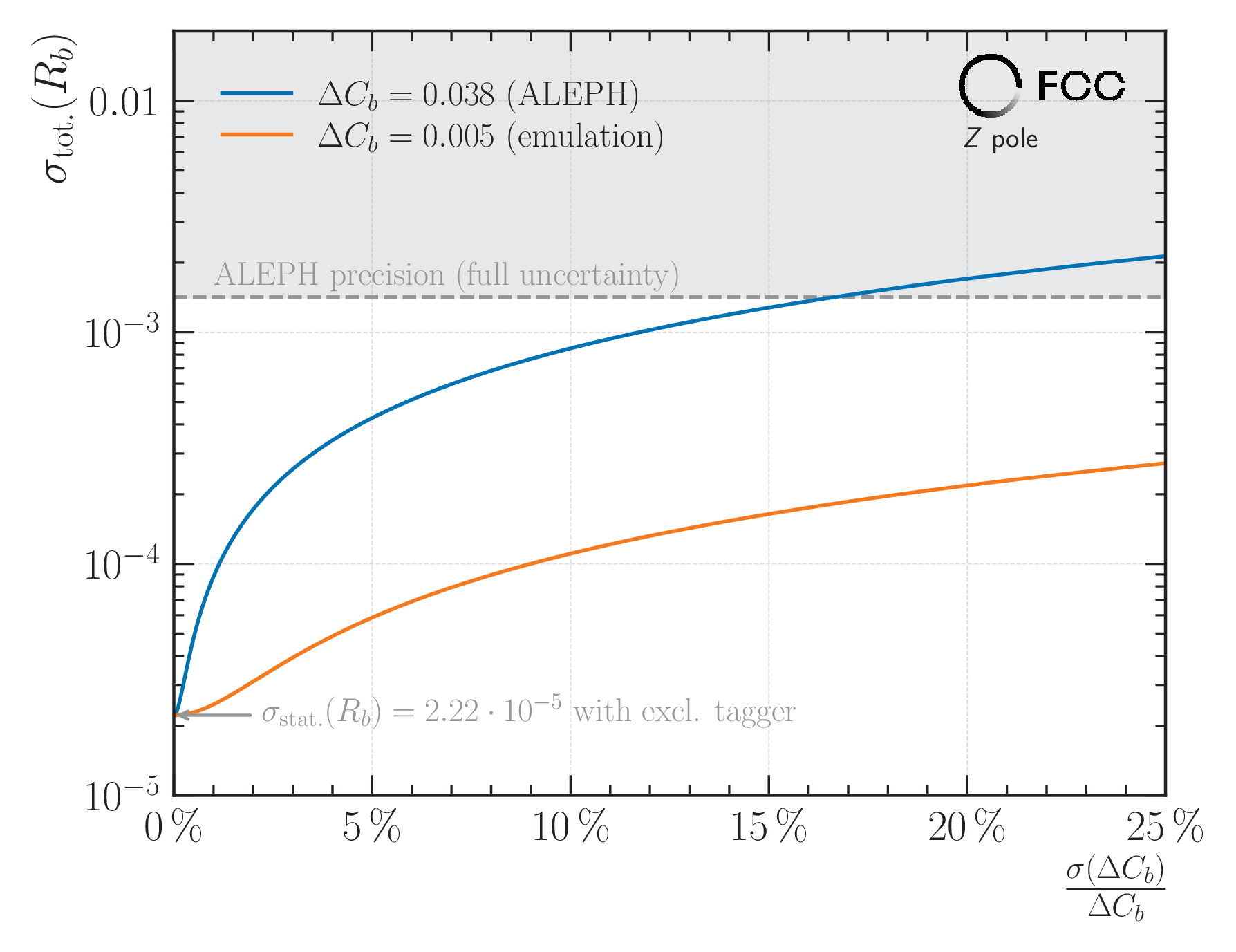}
        \caption{The total precision on \Rb as function of the relative uncertainty on \dCb for two cases: the value from the ALEPH measurement~\cite{ALEPH_Rb_measurement_lifetime_mass} and emulated \dCb value of \num{0.005}. The nominal value of \dCb has a strong impact on the uncertainty of \Rb, as well as the precision in the determination of \dCb.}
        \label{subfig:Zbb:Cb_comparison}
    \end{subfigure}
    \caption{The importance of the hemisphere correlation on the measurement precision of \Rb. Other sources such as the gluon splitting rate (contamination of the background in the signal region) become negligible.}
    \label{fig:Zbb:gbb_and_Cb}
\end{figure}

Both values have been used to compute the systematic uncertainty on \Rb. To individually study the effect of gluon splitting on \Rb, Eq.~\eqref{eqalg:Zbb:single_double_tag_equation} is adjusted as follows
\begin{equation}
    N_b = 2N_Z\cdot (\Rb\varepsilon_b^{Z\to b\bar{b}}\varepsilon^{Z\to b\bar{b}}_{E} + (1 - \Rb)g_{b\bar{b}}\varepsilon_b^{g\to b\bar{b}}\varepsilon^{g\to b\bar{b}}_{E})\,,
    \label{eqn:Zbb:N_b_from_gbb}
\end{equation}
with the efficiency of the energy cut on the $B$-meson candidates, individually for the signal and background events, $\varepsilon^{Z\to b\bar{b}}_{E} = \SI{88}{\percent}$ and $\varepsilon^{g\to b\bar{b}}_{E} = \SI{8}{\percent}$, respectively. Furthermore, $\varepsilon_b^{g\to b\bar{b}} \approx \varepsilon_b^{Z\to b\bar{b}} = \SI{1}{\percent}$ neglecting any kinematic difference for signal and background events. Rearranging Eq.~\eqref{eqn:Zbb:N_b_from_gbb} gives the following for \Rb
\begin{equation}
    R_b = \frac{N_b - 2N_Z \varepsilon_b^{g\to b\bar{b}}\varepsilon_E^{g\to b\bar{b}}}{2N_Z \left(\varepsilon_b^{Z\to b\bar{b}}\varepsilon_E^{Z\to b\bar{b}} - g_{b\bar{b}}\varepsilon_b^{g\to b\bar{b}}\varepsilon_E^{g\to b\bar{b}}\right)}\,.
    \label{eqn:Zbb:Rb_for_gbb}
\end{equation}
The left panel of Fig.~\ref{fig:Zbb:gbb_and_Cb} illustrates the systematic uncertainty of \Rb as a function of the relative uncertainties due to $g_{b\bar{b}}$ and \dCb, depicted in orange and blue, respectively, arising from the central values from Eqs.~\eqref{eqalg:Zbb:gbb_DESY} and~\eqref{eqalg:Zbb:dCb_ALEPH}. The figure indicates that the bias from the gluon-splitting uncertainty is suppressed by more than two orders of magnitude compared to the effect of the hemisphere correlation. Furthermore, the systematic uncertainty on \Rb is highly dependent on the precision of the respective inputs $g_{b\bar{b}}$ and \dCb. Although measured with a higher accuracy at FCC-ee, the current precision on $g_{b\bar{b}}$ is sufficient such that it does not limit the measurement of \Rb, where the statistical limit at $\sigma_\text{stat.}(R_b) = \num{2.22e-5}$ is indicated by the lighter grey colour. Therefore, the systematic uncertainty on \Rb simply reduces to 
\begin{equation}
    \frac{\sigma_\text{syst.}(\Rb)}{\Rb} = \frac{\sigma(\dCb)}{\dCb}\,.
\end{equation}
In conclusion, the primary influencing factor arises from the hemisphere correlation, whose effect on the measurement is described in the following discussion.

\subsection{Hemisphere correlation}\label{subsec:Zbb:hemisphere_correlation}

The hemisphere correlation \dCb measures the bias introduced in the probed hemisphere by the tagged one. The precision of \dCb has been shown to be a handle to reduce $\sigma_\text{syst.}(\Rb)$. Furthermore, its nominal value is another crucial factor in minimising its effect on the systematic uncertainty of \Rb. This is shown in the right graph of Fig.~\ref{fig:Zbb:gbb_and_Cb}, which shows the total uncertainty $\sigma_\text{tot.}(R_b) = \sqrt{\sigma^2_\text{syst.}(\Rb) + \sigma^2_\text{stat.}(\Rb)}$ as a function of the relative uncertainty on \dCb for two different cases: the first case represents the best current determination from the measurement of the ALEPH Collaboration~\cite{ALEPH_Rb_measurement_lifetime_mass}. The second case emulates a reduced hemisphere correlation by approximately a factor of ten, resulting in $\dCb = 0.005$. It can be seen that the reduction of the nominal value of \dCb directly impacts the measurement uncertainty of \Rb, reducing it by about a factor of ten.

Therefore, the primary causes of \dCb deviating from zero are examined in the next paragraph, starting with the findings from Ref.~\cite{ALEPH_Rb_measurement_lifetime_mass}. Four sources of hemisphere correlations have been identified:
\begin{description}
    \item[Detector-acceptance effects] Due to the back-to-back configuration of the two $b$ quarks initially, if one enters a region with lower detector acceptance, such as the very forward or backward region, the other hemisphere is likely to present a similar lower acceptance. 
    \item[Hard gluon radiation] In events where a high energetic gluon in the initial state has been radiated ($Z \to b \bar{b} g$), the momenta of the $b$ hadrons in each hemisphere will decrease, making the reconstruction of the other $b$ hadron less probable.
    \item[Shared PV] When both hemispheres share a single PV, increased measurement uncertainty of the PV affects the probability of tagging both $b$ hadrons; a bias of the PV measurement towards one hemisphere increases the likelihood of tagging the $b$ hadron in the opposite hemisphere.
    \item[Unequal flight distances] A longer flight distance of one $b$ hadron caused by a higher $b$-hadron momentum reduces the number of fragmentation tracks that form the PV, decreasing its measurement precision. As a consequence, the reconstruction of the second $b$ hadron is less likely.
\end{description}
The first two effects have been identified as less significant, whereas the third and fourth factors are particularly influential in determining the hemisphere correlations. The latter two share a similar origin, which can be attributed to the reconstruction of a common and shared PV in the event. Consequently, measurements at LEP have reconstructed two PVs (one in each hemisphere) to mitigate the bias caused by a single PV per event. Although this method has proven effective, a simpler alternative approach based on a different track-selection procedure has been pursued to address the limitations imposed by the measurement uncertainty of the PV. Nevertheless, all results are compared to the method of using a shared PV. To thoroughly investigate the impacts of detector imperfections, the following studies have been outlined using the fully-simulated dataset within the CLD detector, named dataset \circled{2} in Tab.~\ref{tab:event_samples}. The alternative track selection is described below.

\begin{figure}[t]
    \begin{subfigure}[t]{0.48\textwidth}
        \centering
        \includegraphics[width = 1\textwidth]{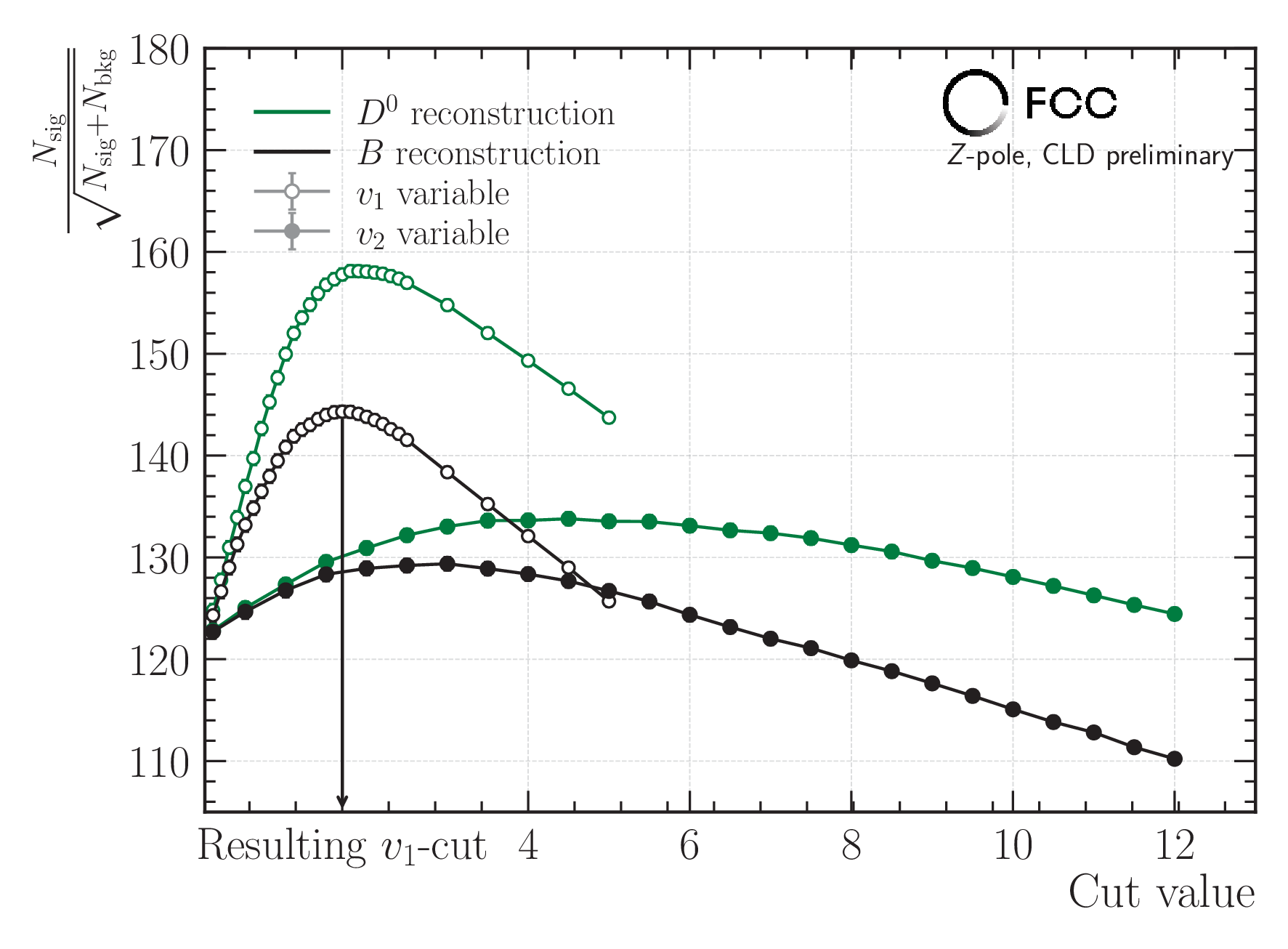}
        \caption{Signal significance $S$ as function of the $v_{1,2}$ variable thresholds. The resulting $v_1$ cut is extracted from the maximum of the $B^+$-reconstruction significance.}
        \label{subfig:Zbb:v1_v2_thresholds}
    \end{subfigure}\hfill
    \begin{subfigure}[t]{0.48\textwidth}
        \centering
        \includegraphics[width = 1\textwidth]{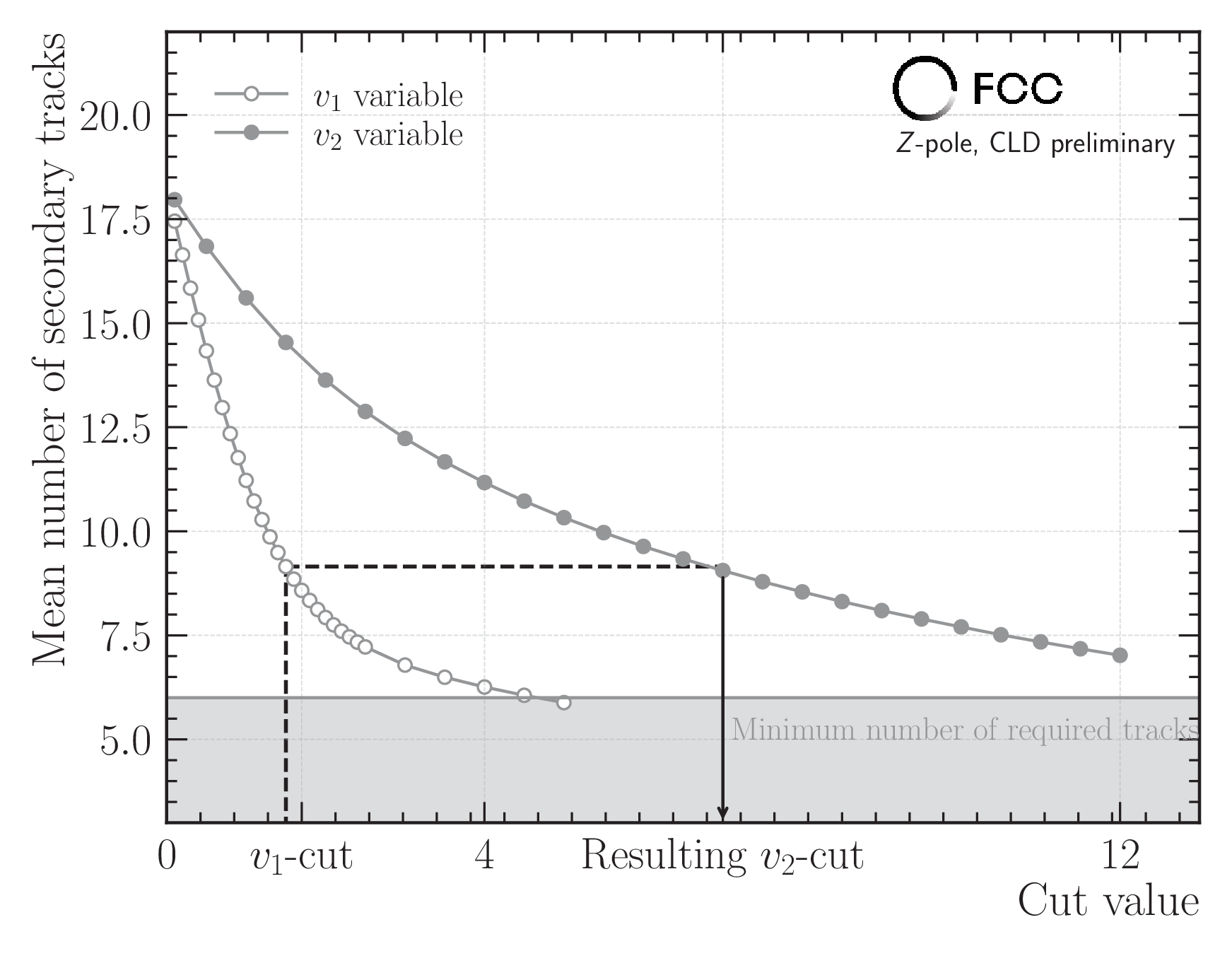}
        \caption{The cut on $v_2$ extracted from the mean number of secondary tracks slope where the same number is reached as for the given $v_1$ cut.}
        \label{subfig:Zbb:mean_number}
    \end{subfigure}
    \caption{Optimised parameters for the determination of the LR and to select tracks that are inconsistent with the LR. They have been further used to perform the reconstruction of $\bar{D}^0$ and $B^+$ mesons.}
    \label{fig:Zbb:significance_mean_number_of_secondary_tracks}
\end{figure}

\paragraph{Tracks outside the luminous region} The accurate knowledge of the beam-spot region (LR) as well as its smallness is used to select tracks issued from the primary and secondary vertices. The LR refers to the area where the two beams intersect, also referred to as the PV at the truth level. Given the finite sizes of the beams, their intersection point has specific dimensions in the $(x,y,z)^\top$ plane.
\begin{figure}[t]
    \centering
    \includegraphics[width=0.7\textwidth]{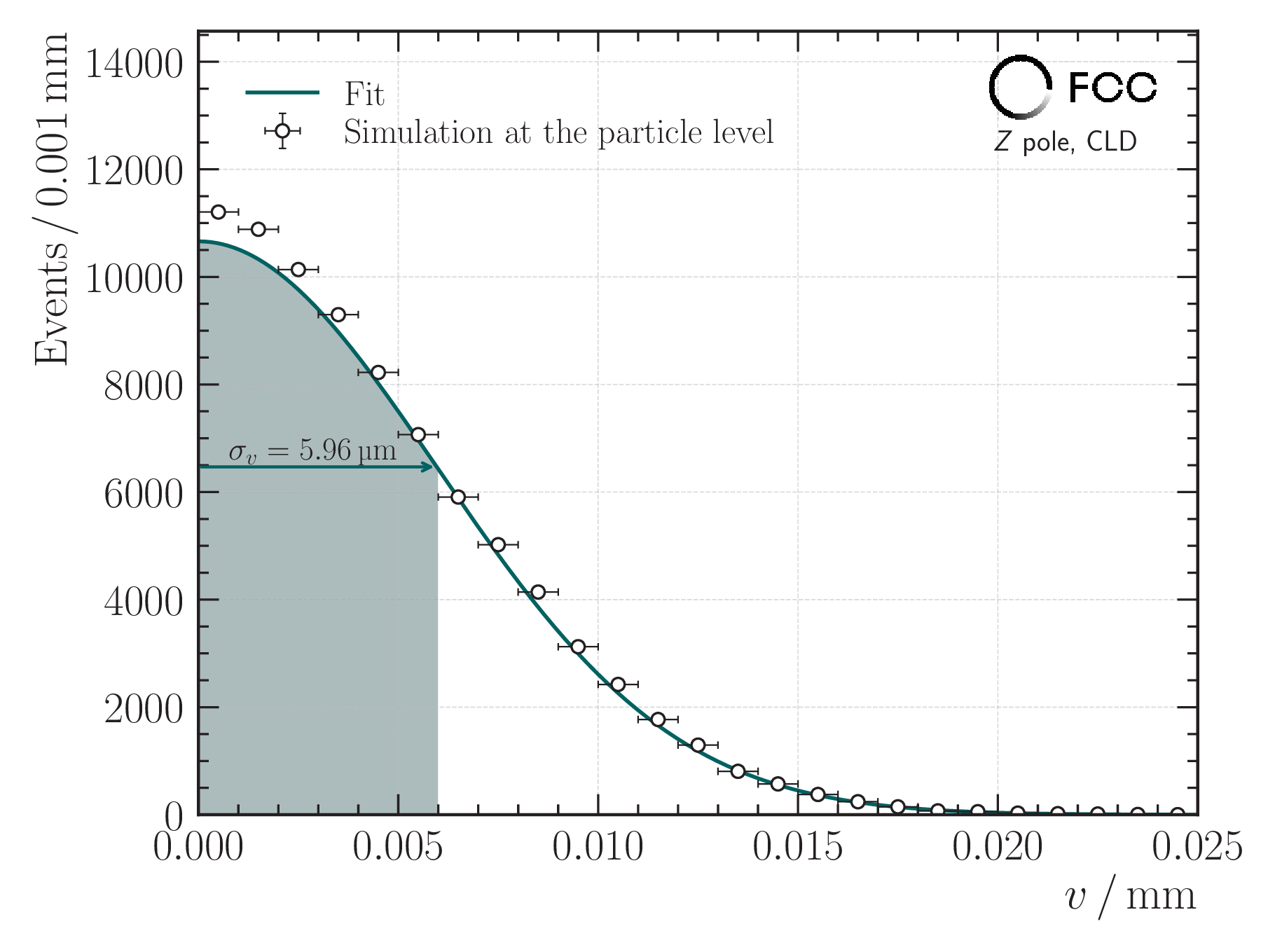}
    \caption{Distribution of $v$ as defined in Eq. \eqref{eqn:Zbb:v_distribution} with a fit of a truncated Gaussian function, from which the size of the LR of $\sigma_v = \SI{5.96}{\micro\meter}$ is extracted.}
    \label{fig:Zbb:v_distribution}
\end{figure}
In order to define variables that measure the agreement of track origins with the IP, the IP region in the $(x,y)^\top$ plane has been translated into a single transverse variable, $v$, defined as
\begin{equation}
    v = \sqrt{\left(\text{PV}_x^{\text{Particle-level}}\right)^2 + \left(\text{PV}_y^{\text{Particle-level}}\right)^2}\,,
    \label{eqn:Zbb:v_distribution}
\end{equation}
where the size of the transverse beam-spot region is given by the width of the distribution of $v$, $\sigma_v$, and is taken as the LR. The resulting distribution of $v$ is presented in Fig.~\ref{fig:Zbb:v_distribution} and has been derived from the exclusive dataset \circled{2}. The superscript \textit{Particle-level} indicates that the true collision point of the electron and positron beam has been used. In order to identify whether reconstructed tracks of an event are consistent with the beam-spot region, track-wise variables $v_{1,2}$ have been introduced
\begin{equation}
    v_1 = \frac{d_0}{\sqrt{\sigma_{d_0}^2 + \sigma_{v}^2}}\,,\quad v_2 = \frac{z_0}{\sqrt{\sigma_{z_0}^2 + \sigma_{v}^2}}\,,
\end{equation}
where $\sigma_{d_0}$ and $\sigma_{z_0}$ are the respective uncertainties of the impact parameters. Although $v_2$ uses the longitudinal impact parameter $z_0$ and quantifies its agreement with the transverse extension of the beam spot, at this stage of the analysis it has served its purpose to find thresholds up to which tracks are taken as consistent or inconsistent with the beam-spot region. These thresholds have been determined by independently varying $v_1$ and $v_2$ and maximising the significance $S$ of the $\bar{D}^0$- and $B^+$-meson reconstruction, along with the number of remaining tracks after selection. The significance is defined as
\begin{equation}
    S = \frac{N_\text{sig}}{\sqrt{N_\text{sig} + N_\text{bkg}}}\,,
\end{equation}
where $N_\text{sig}$ represents the number of truth-matched meson candidates and $N_\text{bkg}$ denotes the number of background events (where background refers to partially reconstructed and combinatorial background events, since only $Z\to b\bar{b}$ events have been used) in the region of interest. Here, in the considered mass region, the background rate is approximated to be uniform.
The dependence of $S$ as a function of $v_{1,2}$ thresholds and the mean number of tracks inconsistent with the LR are shown in Fig.~\ref{fig:Zbb:significance_mean_number_of_secondary_tracks}. The significance of the $B^+$ meson shows a maximum around \num{1.7} for $v_1$, which has been chosen as the optimal cut. In contrast, $v_2$ shows only a slight dependence on the significance, so its threshold has been determined based on the mean number of tracks that are inconsistent with the LR. This threshold, where a similar number of tracks is observed as with the given $v_1$ cut, has been set at eight. Given that the $v_1$ and $v_2$ distributions are symmetric around zero, the absolute value is used to decide whether tracks have been used in the $B^+$-meson reconstruction process.

\paragraph{Particle reconstruction}
The evaluation of \dCb requires a full reconstruction of the charged $B^+$ mesons in both hemispheres. In contrast to the exemplary reconstruction in Sec.~\ref{sec:exclusive_reconstruction}, the decay $B^+ \to [K^+\pi^-]_{\bar{D}^0}\,\pi^+$ has been simulated in the hemispheres (plus the charge-conjugate decay).
In contrast to the reconstruction method described in Sec.~\ref{sec:exclusive_reconstruction}, the vertexing features of the \texttt{DELPHES} package have been used~\cite{Vertexing_Franco}. 
During this phase of the analysis, the neutral vertexing capabilities have been made available for the reconstruction of neutral intermediate $\bar{D}^0$ tracks. As in the reconstruction method detailed in Sec.~\ref{sec:exclusive_reconstruction}, a pair of oppositely charged kaon and pion tracks has been combined to a common vertex, constraining their mass to the $\bar{D}^0$ pole-mass. The fit provides updated momenta for the tracks, which then have been used to form $\bar{D}^0$ candidates. Candidates with $(1800\leq m_{\bar{D}^0} \leq 1930)\,\si{\mega\electronvolt}$ have been further vertexed with another pion track, resulting in $B^+$ candidates that fall within $(5150 \leq m_{B^+} < 5400)\,\si{\mega\electronvolt}$ and have a vertex quality of $\chi^2_{B^+} < 25$.

Subsequently, the single- and double-tag efficiencies $\varepsilon_{{b}_{1,2}}$ and $\varepsilon_{b_1}\varepsilon_{b_2}$ have been calculated and then differentially assessed in distributions that are sensitive to deviations of \dCb from zero.

\subsection{Integrated and differential hemisphere-correlation} 
The integrated \dCb value has been determined for both the shared PV and the LR, taking into account detector-acceptance effects by excluding events where the absolute value of the thrust-axis polar angle $|\cos(\theta_\text{Thrust})|$ exceeds \num{0.9}. The inclusive values are determined to be
\begin{align}
    \begin{split}
        \Delta C_b^\text{shared PV} &= \hphantom{-}\num{0.035(3)}\,,\\
        \Delta C_b^\text{LR} &= \num{-0.001(3)}\,,
    \end{split}
    \label{eqalg:Zbb:dCb_LR}
\end{align}
where $\Delta C_b^\text{LR}$ is statistically consistent with zero, unlike the shared PV approach. Therefore, eliminating dependencies caused by intrinsic biases from the PV by choosing tracks independently of the PV already reduces the hemispheric correlation to the required level for an accurate measurement of \Rb. Nevertheless, the potential causes of a non-zero \dCb value are examined in the following, with an emphasis on detector-acceptance effects and displacement from the IP.

\paragraph{Dependence on detector acceptances} 
The influence of detector acceptance has been analysed in bins of the maximum allowed $|\cos(\theta_\text{Thrust})|$. A finer binning has been chosen for the extreme forward/backward region where $|\cos(\theta_\text{Thrust})| > 0.9$.
\begin{figure}[t]
    \centering
    \includegraphics[width = 0.7\textwidth]{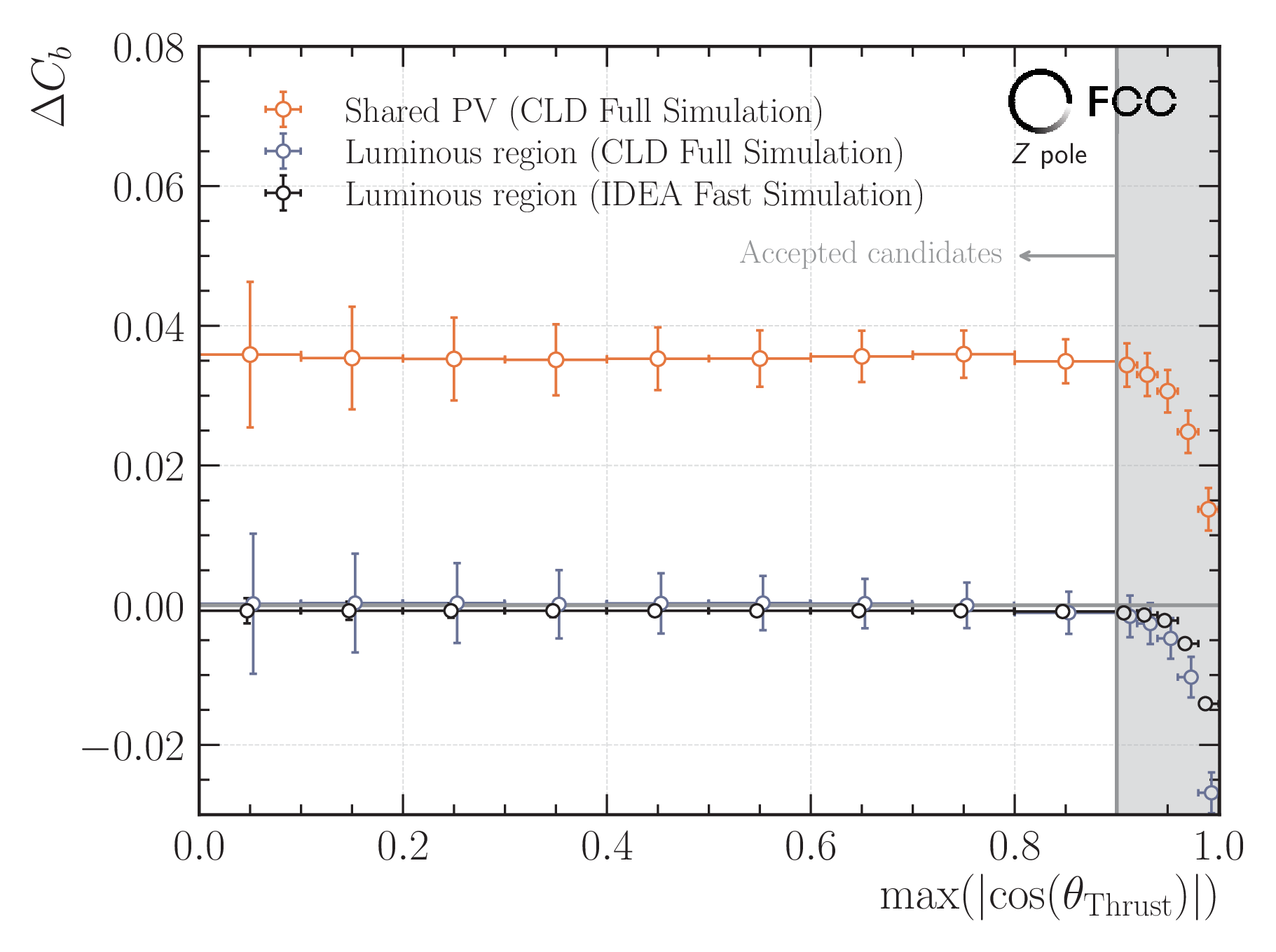}
    \caption{
        \dCb as function of the maximally allowed $|\cos(\theta_{\text{Thrust}})|$, comparing the shared PV and the LR approach. 
        Although both drop towards higher values in the very forward/backward region, \dCb becomes compatible with zero within the statistical precision when $|\cos(\theta_\text{Thrust})| < 0.9$ for the LR approach. The value of \dCb has been confirmed from the fast simulation dataset.
    }
    \label{fig:Zbb:dCb_max_costheta_thrust}
\end{figure}
The results are illustrated in Fig.~\ref{fig:Zbb:dCb_max_costheta_thrust}, indicating that for $|\cos(\theta_\text{Thrust})| > 0.9$, \dCb decreases for both methods and converges for $|\cos(\theta_\text{Thrust})| < 0.9$, which also sets the cut value. The dependence and inclusive value of \dCb have been validated using the high-statistics fast simulation dataset within the IDEA detector (dataset \circled{3}), also depicted in Fig.~\ref{fig:Zbb:dCb_max_costheta_thrust} with black dots. It shows a less pronounced drop in the extreme forward/backward region due to the larger acceptance area of the IDEA detector compared to CLD. Therefore, a precise measurement of \Rb also requires a detector with a wide acceptance range, which would increase the fraction of accepted events while reducing the hemisphere correlation at the same time.

\begin{figure}[t]
    \begin{subfigure}[t]{0.48\textwidth}
        \centering
        \includegraphics[width = 1\textwidth]{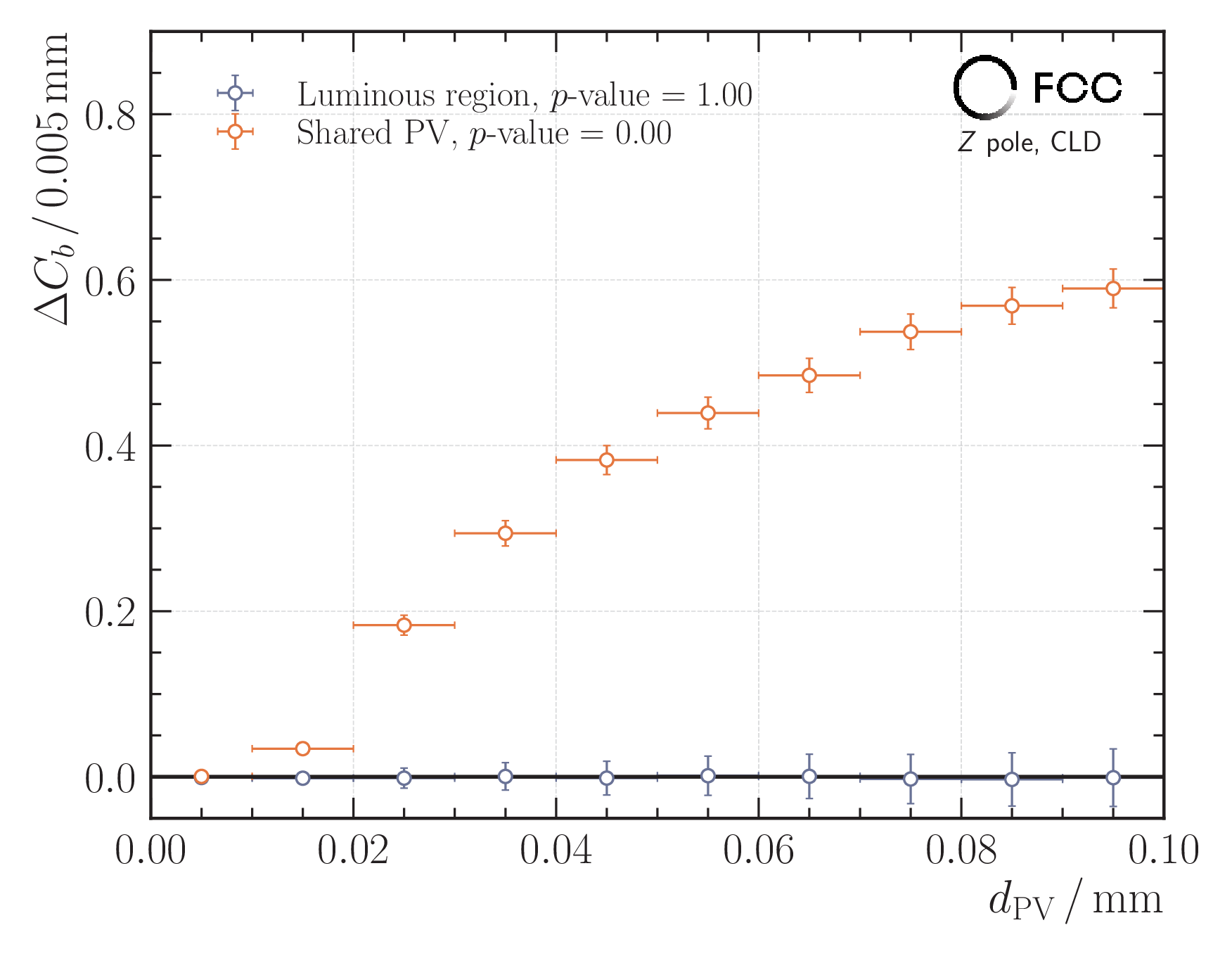}
        \caption{\dCb as function of the displacement of the reconstructed PV from the true collision point. When the displacement is zero (most precise determination), there is no correlation for both methods. For further displaced PVs, the correlation increases for the shared PV, while staying zero for the LR approach.}
        \label{subfig:Zbb:dCb_PV_displacement}
    \end{subfigure}\hfill
    \begin{subfigure}[t]{0.48\textwidth}
        \centering
        \includegraphics[width = 1\textwidth]{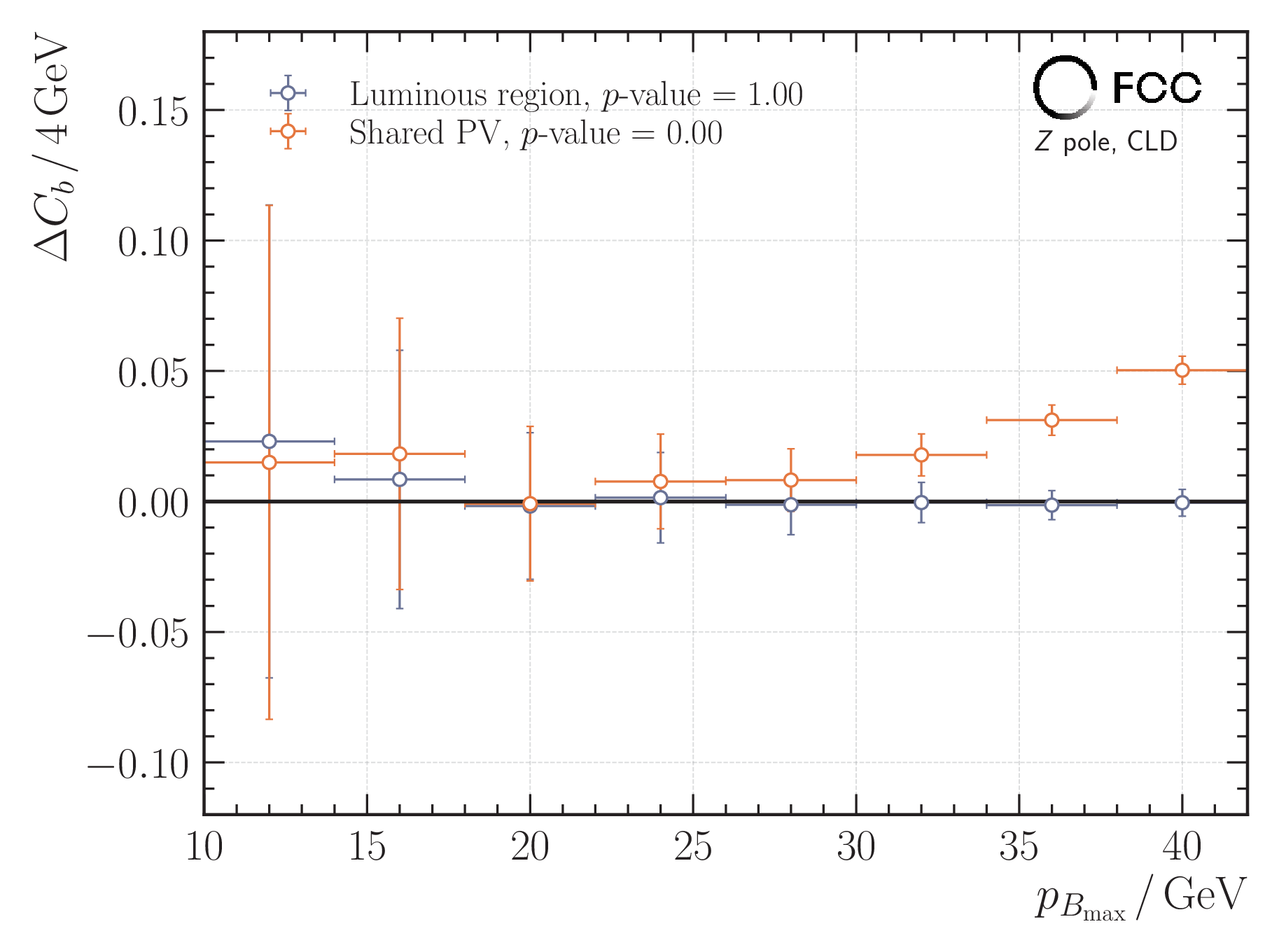}
        \caption{\dCb in bins of the $B$-meson momentum. Higher momenta mesons induce a correlation when a shared PV is used, due to the bias in the other hemisphere coming from the PV which is shifted towards the tagged hemisphere.}
        \label{subfig:Zbb:dCb_momentum}
    \end{subfigure}
    \caption{The displacement of the PV has strong impact on the hemisphere correlation. When the PV has been reconstructed preferring one hemisphere, the $b$-hadron reconstruction probability decreases for the other hemisphere. All dependencies have been removed when using tracks for the reconstruction which have been selected due to their inconsistency with the LR.}
    \label{}
\end{figure}

\paragraph{Displacement from the PV} 
The deviation of the PV at the object level from the true collision point (PV at the particle level), defined as
\begin{equation}
    d_\text{PV} = \sqrt{\sum_{i\in[x,y,z]}^{} \left(\text{PV}_i^{\text{Object-level}} - \text{PV}_i^{\text{Particle-level}}\right)^2}
\end{equation}
encapsulates two metrics simultaneously. Firstly, $d_\text{PV}$ introduces a bias in one hemisphere, increasing the likelihood of tagging the $b$-hadron in that hemisphere, while decreasing the reconstruction probability for the $b$-hadron in the opposite hemisphere. Secondly, $d_\text{PV}$ serves as an indicator of the PV reconstruction quality, which diminishes with increasing $d_\text{PV}$. Fig.~\ref{subfig:Zbb:dCb_PV_displacement} shows \dCb in bins of $d_\text{PV}$, separately for the shared PV and the LR approach. In addition, a $\chi^2$ test has been used to evaluate the agreement of the points with zero, with the result presented in the legend as $p$-value. Values near one indicate support for the hypothesis that there is no deviation from zero, whereas $p$-values less than \num{0.05} generally lead to the rejection of this hypothesis. 

It can be seen that there is a strong dependence of \dCb for the shared PV approach, already for PV displacements above \SI{0.01}{\milli\meter}, for which an ALEPH-like size of the hemisphere correlation is reached. As already discussed in Fig.~\ref{subfig:Zbb:Cb_comparison}, the overall uncertainty in \Rb would be significantly affected. In contrast, no dependence within the statistical precision can be observed for the LR ansatz.

\paragraph{Flight distance\,/\,Momentum} 
The flight distance is directly related to the momentum, which allows to use the momentum as a proxy for the flight distance. This resolves any ambiguities with respect to the reference point for the flight distance (either $(0,0,0)^\top$ or the PV) in the two methods. 
Concerning \dCb, the higher momentum $B$-meson introduces a bias, reducing the likelihood of reconstructing the oppositely-charged $B$ meson in the opposite hemisphere. Fig.~\ref{subfig:Zbb:dCb_momentum} displays \dCb in different $B^+$-meson momentum bins. Both methods show good agreement within the statistical uncertainty in the low-momentum range, but the correlation increases at higher momenta for the shared-PV method. This also leads to the rejection of the hypothesis that \dCb is consistent with zero for the shared PV.

\subsection[Conclusions for \Rb]{Conclusions for \boldmath{$R_b$}}

In summary, the use of exclusively reconstructed $b$-hadrons as $b$-hemisphere taggers enables unprecedented purity levels during the $Z$-pole run at FCC-ee. For six representative decay modes and a target efficiency of \SI{1}{\percent}, purities exceeding \SI{99.8}{\percent} are achievable, leaving \dCb as the sole unknown in the set of equations (refer to Eqs.~\eqref{eqalg:Zbb:updated_Rb_formula}) due to the minimal impact of gluon radiation on the systematic uncertainty. From fully-simulated events, \dCb has been found to be consistent with zero by eliminating dependencies from a shared PV and selecting tracks outside the LR. With the current dataset, \dCb has been found to be
\begin{equation}
    \dCb = -0.001\,\pm\,0.003(\text{stat.})\,.
    \label{eqn:Zbb:dCb_result}
\end{equation}
Given the nominal value of \dCb from Eq.~\eqref{eqn:Zbb:dCb_result} and assuming that the precision of \Rb only depends on the precision of \dCb, achieving a relative precision of \SI{10}{\percent} on \dCb is necessary to determine \Rb with exclusive $b$-hadron decays such that $\sigma_\text{stat.}(\Rb) \approx \sigma_\text{syst.}(\Rb)$. Consequently, \dCb must be derived from a simulation dataset of at least $N_Z \approx \num{e9}$ events, where in both hemispheres the $b$-hadron decays according to a list of approximately 200 decay modes. Based on this, \Rb results in
\begin{align*}
    \Rb &= \mu(R_b)\,\pm\,\num{2.22e-5}(\text{stat.})\,\pm\,\num{2.16e-5}(\text{syst.})\,,\\ 
        &= \mu(R_b)\,\pm\,\num{3.10e-5}(\text{tot.})\,.
\end{align*}
The simultaneous use of both, a novel ultra-pure tagger and a selection of the secondary tracks based on the inconsistency with the LR, which have been presented in this paper, enables the precision on \Rb to be improved by at least a factor of \num{20}~\cite{PDG} with respect to the state-of-the-art while keeping a measurement dominated by statistics of the sample.
\section{Application to the measurement of \boldmath{$A_\text{FB}^b$}}\label{sec:application_to_AFBb}

The forward-backward asymmetry of the $b$ quark is of particular interest for the hemisphere tagger based on the exclusive $b$-hadron reconstruction. To date, it still has the highest tension~\cite{Precision_electroweak_measurement_on_the_Z_resonance} among all EWPOs with the SM prediction of
\begin{equation}
    A_\text{FB}^{b,\text{SM}} = \num{0.1037(8)}\,,
\end{equation}
which is in $2.9\,\sigma$ tension with the average of the LEP measurements
\begin{equation}
    A_\text{FB}^{b} = \num{0.0992(16)}\,.\label{eqn:Zbb:AFB_LEP_world_average}
\end{equation}
Similarly to \Rb, the primary challenge of the measurement is the effective reduction of the systematic uncertainty given the raw statistical precision available at FCC-ee. However, in addition to the hemisphere flavour tag, which is sufficient to measure \Rb, an estimation of the charge and direction of the initial $b$-quark is necessary. This leads to two consequences:
\begin{enumerate}
    \item The criteria for the decay modes to be considered are more stringent. To mitigate one source of systematic uncertainty that comes from the charge confusion of neutral $B$-meson mixing, only the modes of the $B^+$ meson and the $\Lambda_b^0$ baryon can be used as taggers.
    \item Various estimators of the $b$-quark direction can be employed, such as jets, the thrust axis, or the flight direction of the $b$ hadron. In the latter, only fully reconstructed $b$-hadrons can be used, restricting the selection of candidates to those from the mass-peak region.
\end{enumerate}
Nevertheless, the exclusive $b$-hadron reconstruction provides all tools in order to overcome limitations induced by the main source of systematic uncertainties, which is the accounting for the correction of the direction estimation from high-energetic gluon radiations, the QCD corrections. Therefore, the application is described in this section. However, the principle of the \AFBbeauty measurement is discussed first, before going into detail about the remaining systematic uncertainty.

\subsection{Measurement principle}\label{subsec:Zbb:measurement_principle}
\begin{figure}[t]
    \centering
    \begin{subfigure}[t]{0.48\textwidth}
        \centering
        \includegraphics[width = 1\textwidth]{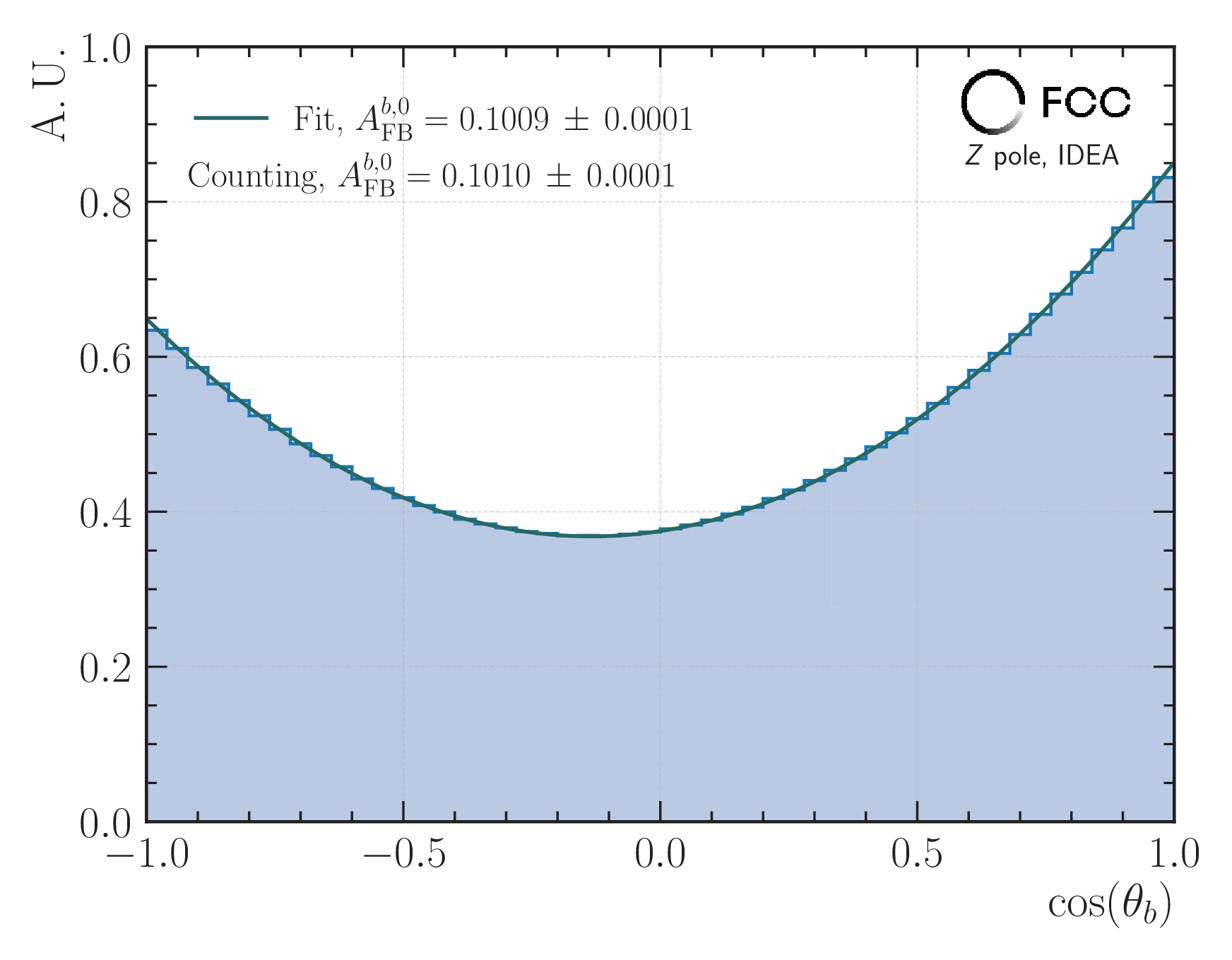}
        \caption{The polar-angle distribution of the $b$ quark before gluon radiation. The fit to Eq.~\eqref{eqn:Zbb:AFB_full_fit} results within the statistical uncertainty in the same value as counting the forward and backward hemispheres.}
        \label{subfig:Zbb:costheta_b_quark_fit_count}
    \end{subfigure}\hfill
    \begin{subfigure}[t]{0.48\textwidth}
        \centering
        \includegraphics[width = 1\textwidth]{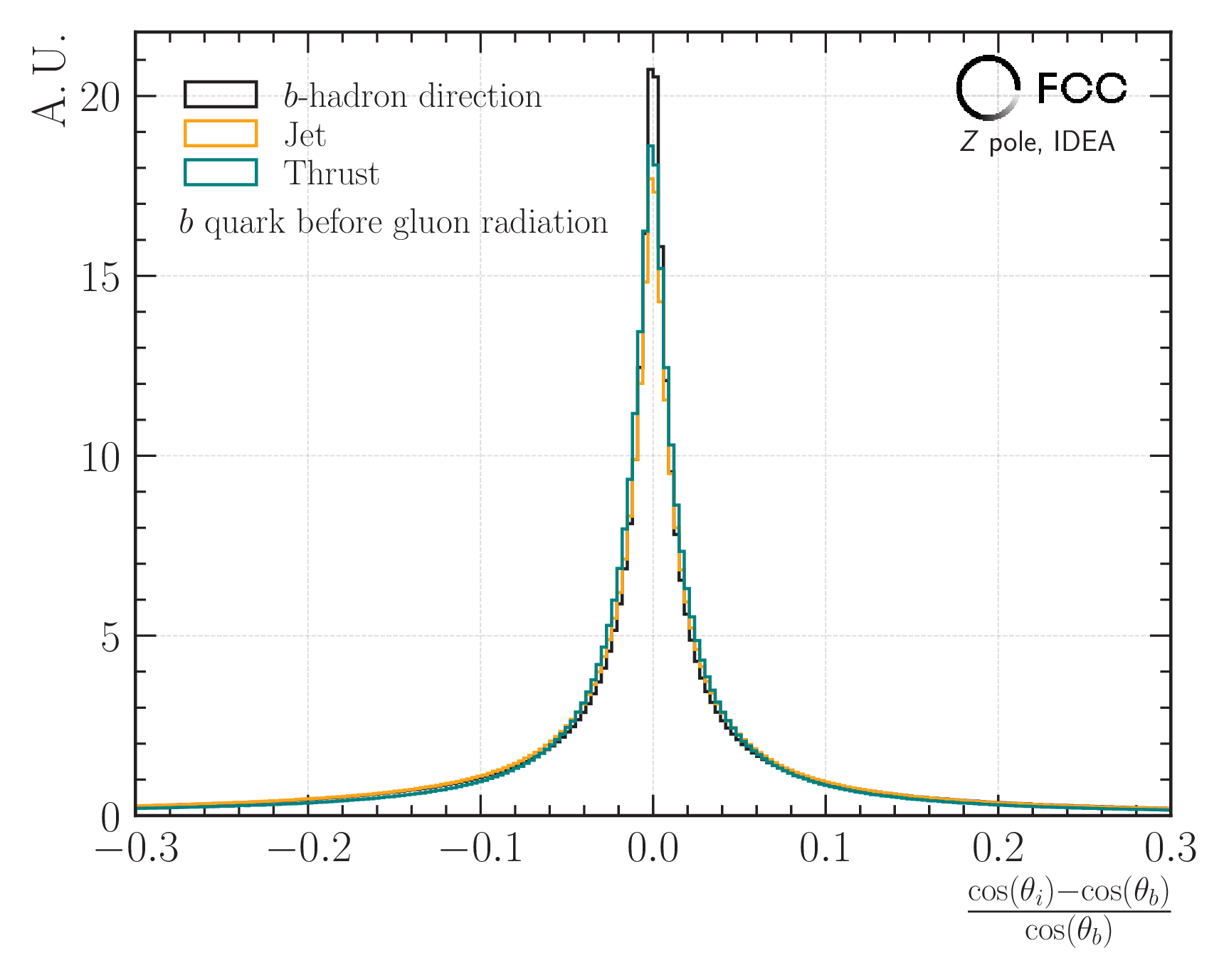}
        \caption{Relative difference of the polar angle of different estimators to the $b$-quark polar angle before gluon radiation, with the $b$-hadron direction being the precise approximation.}
        \label{subfig:Zbb:direction_estimators}
    \end{subfigure}
    \caption{Proof of concept, that extracting \AFBbeautyzero from the fit and counting leads to the same result in Fig.~\subref{subfig:Zbb:costheta_b_quark_fit_count}. In Fig.~\subref{subfig:Zbb:direction_estimators}, different approximations of the $b$-quark direction have been examined.}
    \label{fig:Zbb:AFB_and_jet_costheta}
\end{figure}

The forward-backward asymmetry can be calculated by counting the number of \textit{forward} and \textit{backward} events (see Eq.~\eqref{eqn:Zbb:AFB_from_counting}) or by extracting the value from a fit to the differential cross-section distribution
\begin{equation}
    \frac{\text{d}\sigma}{\text{d}\cos(\theta_b)} = \frac{1}{1 + f_\text{L}} \left(\frac{3}{8}(1 + \cos^2(\theta_b)) + \frac{3}{4} f_\text{L} (1 - \cos^2(\theta_b))\right) + \AFBbeauty \cos(\theta_b)\,.
    \label{eqn:Zbb:AFB_full_fit}
\end{equation}
The parameter $f_\text{L}$ represents the fraction of longitudinally polarised $Z$ bosons along the quark's flight path and has been set to zero in the fit\footnote{In a first approach, $f_\text{L}$ is expected to be zero without considering effects that might distort the $b$-quark's direction of flight from the radiation of gluons.}. The $b$-quark polar-angle distribution before gluon radiations, shown in the left panel of Fig.~\ref{fig:Zbb:AFB_and_jet_costheta}, has been taken to extract \AFBbeautyzero. For all the following studies, the exclusively simulated dataset \circled{3} has been used. 
Although at this stage the kinematic properties of the $b$ quark are not affected by radiation effects in the final state, ISR from the colliding beams can reduce the energy of the $b$ quarks. In order to account for effects from ISR, a minimal energy-cut on the $b$ quarks has been set to \SI{45}{\giga\eV}. In addition, the correction factor to account for $\gamma$ exchange and $Z/\gamma$ propagator interference even at $\sqrt{s} = m_Z$ has not been considered here, since it introduces a constant bias.\\
The result of the fit and the counting leads to
\begin{align}
    \begin{split}
        \text{Fit:}\quad \AFBbeautyzero &= \num{0.1009(1)}\,, \\
        \text{Counting:}\quad \AFBbeautyzero &= \num{0.1010(1)}\,,
    \end{split}
\end{align}
where the fit result is shown as green line in Fig.~\ref{subfig:Zbb:costheta_b_quark_fit_count}.
Both results are in precise agreement with each other. However, since the fitting procedure is insensitive to angular acceptance and/or efficiency effects and provides a generally smaller statistical uncertainty, it has been used as default method to compute \AFBbeauty from angular distributions. Yet, since the $b$-quark polar angle is experimentally inaccessible, different estimators have been used to approximate the initial quark direction $\cos(\theta_b)$ in Eq.~\eqref{eqn:Zbb:AFB_full_fit}. They are presented below.

\subsubsection{Experimental access to the quark direction}
In the following paragraphs, different quark-direction estimators are presented, before their accuracy to model the $b$-quark direction is examined.\\
First, the thrust axis is considered, which has been the conventional event-shape variable to assess the $b$-quark direction at LEP~\cite{ALEPH_AFB_measurement, OPAL_AFB_measurement, L3_AFB_measurement, DELPHI_AFB_measurement}.
Second, a revision of \AFBbeauty at FCC-ee~\cite{Revisiting_QCD_correction_really_a_problem} has explored the possibility of utilising the reconstructed $b$-tagged jet direction, also to reduce the impact of QCD corrections. A recapitulation of this approach is discussed in Sec.~\ref{subsec:Zbb:acollinearity}. 
Finally, the list is extended by incorporating the reconstructed $b$-hadron, which, when fully reconstructed, also approximates the original $b$-quark direction. All of the aforementioned direction estimators are discussed in the following.

\paragraph{Thrust}
The thrust has been calculated from all particles at the object level
\begin{equation}
    \tilde{T} = \max_{\vec{T}}\!\left(\frac{\sum_{i}^{}|\vec{p}_i\cdot \vec{T}|}{\sum_{i}|\vec{p}_i|}\right)\,,
\end{equation}
with $i$ running over all particles in the event with their momentum vector $\vec{p}$. The thrust direction corresponds to the thrust-axis polar angle, where the sign of the $z$ component is extracted from the hemisphere with the higher energy.

\paragraph{Jets}
Jets at both the particle and object levels have been reconstructed using the anti-$k_t$ jet-clustering algorithm with a cone-radius parameter of $R = 0.4$. To account for additional jets from radiated gluons, the jets have been clustered inclusively with a minimum transverse momentum of \SI{5}{\giga\electronvolt}. For this analysis, the flavour and charge of the jet have been determined from a matching procedure with the nearest $b$ hadron. However, dedicated algorithms must be installed for a more realistic analysis. Details of the matching are provided below. Furthermore, the jet polar-angle distribution at the object level has been corrected for acceptance effects in the very forward and backward region of the detector using the polar-angle distribution from the jets at the particle level.

\paragraph{\boldmath{$B^+$} meson}
Similar to the $b$-hadron reconstruction described in Sec.~\ref{subsec:Zbb:hemisphere_correlation}, $B^+$ candidates have been reconstructed by vertexing $\bar{D}^0$ candidates in a first stage from the fast-simulation dataset. They have been furthermore combined with a charged-pion track. The final $B^+$ candidates must have a reconstructed vertex-quality $\chi^2 < 25$ and must meet the mass criteria $(5150 \leq m_{B^+} \leq 5400)\,\si{\mega\electronvolt}$. In the following, it is generally referred to as the $b$ hadron.

An important aspect arises from the reconstructed $b$-hadron: usually, the $b$-jet flavour tagging achieves efficiencies of the order $(10-90)\,\si{\percent}$ depending on the background-rejection rate. This is at least an order of magnitude larger compared to the presence of a reconstructed $b$-hadron from the list of possible decay modes to consider with efficiencies of the order \SI{0.5}{\percent}. However, in the case of a reconstructed $b$-hadron in the event, valuable information can be obtained from it to identify the jet flavour and charge. This can be achieved by matching the $b$ hadron with the jet that is closest to it to minimise the effect from jet-charge confusion using traditional methods discussed in Sec.~\ref{subsec:Zb:event_topology} and the mis-ID of the jet flavour. In addition, the same procedure can be applied to tag the hemisphere charge and flavour when using the thrust axis as a direction estimator.
The matching criterion is based on a distance measure $\Delta R$, defined via
\begin{equation}
    \Delta R = \sqrt{(\eta_k - \eta_B)^2 + (\phi_k - \phi_B)^2}\,,\quad\text{where}\quad k\in[\text{Jet}_n, \text{Thrust}]\,,
    \label{eqn:Zbb:Delta_R}
\end{equation}
considering the pseudorapidity $\eta_i$ and the azimuth $\phi_i$. In case of at least two jets per event, the index $n$ considers all of them.
Given that the sample used is biased and has been produced so that each hemisphere contains a $b$ hadron, the jet or thrust-hadron pair with the smallest opening angle has been selected to act as the charge and flavour tagger for the hemisphere. 
This choice has been determined by finding the smallest angle $\omega$ between the $b$ hadron, represented as \mbox{$\vec{B} = (B_x, B_y, B_z)^\top$}, and either a jet or the thrust vector, $\vec{q} = (q_x, q_y, q_z)^\top$ using the known formula
\begin{equation}
    \omega = \sin^{-1}\left(\frac{|\vec{B}\times\vec{q}|}{|\vec{B}|\cdot|\vec{q}|}\right)
\end{equation}
This method ensures that the pairing reduces the impact of hemisphere confusion caused by high-energy gluon radiation that might change the $b$-hadron direction. Experimentally, with typically only one $b$ hadron per hemisphere, minimising $\Delta R$ from Eq.~\eqref{eqn:Zbb:Delta_R} is usually sufficient. The result, shown as the relative difference in the polar-angle distribution in the right panel of Fig.~\ref{fig:Zbb:AFB_and_jet_costheta}, compares the different estimators to the $b$-quark direction before gluon radiation, from which \AFBbeautyzero has been extracted. A narrow, zero-centred distribution is obtained for all estimators; however, the novel approach based on the $b$-hadron reconstruction method serves as the most precise approximation of the $b$-quark direction. The jet and thrust axes produce similar results in terms of precision.

Nevertheless, as pointed out earlier, the direction of flight can be distorted by the radiation of gluons, referred to as QCD corrections. An introduction is given in the following, with a focus on experimental handles to minimise their effects. 

\subsection{Angular distortions: QCD corrections}\label{sec:angular_distortions}

In the determination of \AFBbeauty at LEP~\cite{ALEPH_AFB_measurement, DELPHI_AFB_measurement, OPAL_AFB_measurement, L3_AFB_measurement} and their combined analysis~\cite{Precision_electroweak_measurement_on_the_Z_resonance}, QCD corrections have contributed to about \SI{50}{\percent} of the systematic uncertainty budget and are the leading uncertainty after excluding contamination from $udsc$-physics events. These QCD corrections are mainly due to the emission of high-energy gluons from the $b$ quark before it hadronises, causing the quark's initial direction to change, potentially even reversing it. The degree of this distortion depends on the energy of the emitted gluon(s) and the chosen method for estimating the $b$-quark's direction, since the quark direction cannot be directly measured in experiments. Consequently, a correction factor must be applied later to account for the distortion, which is further detailed below.

Typically, the uncorrected $b$-quark forward-backward asymmetry \AFBbeautyzero without gluon radiation is adjusted using a scaling factor $C_{\text{QCD}}(\mu)$ to derive the experimentally measurable \AFBbeauty
\begin{equation}
    \AFBbeauty = \left(1 - \frac{\alpha_\text{S}}{\pi}C_\text{QCD}(\mu)\right)\AFBbeautyzero\,,
    \label{eqn:Zbb:AFB_with_QCD}
\end{equation}
where $\large\mu = \sfrac{2m_q}{\sqrt{s}}$ is the quark-specific energy-scale parameter of a quark with mass $m_q$. In the following, $\mu$ is set to $0.107$ for $b$ quarks with $m_b = \SI{4.8}{\giga\eV}$ and has been neglected in notations. The main challenge is to reduce the impact of $C_\text{QCD}$, making sure that its uncertainty does not increase the total uncertainty of the measurement by using experimental methods that are sensitive to $C_\text{QCD}$.
\begin{figure}[t]
    \begin{subfigure}[t]{0.48\textwidth}
        \centering
        \includegraphics[width=\textwidth]{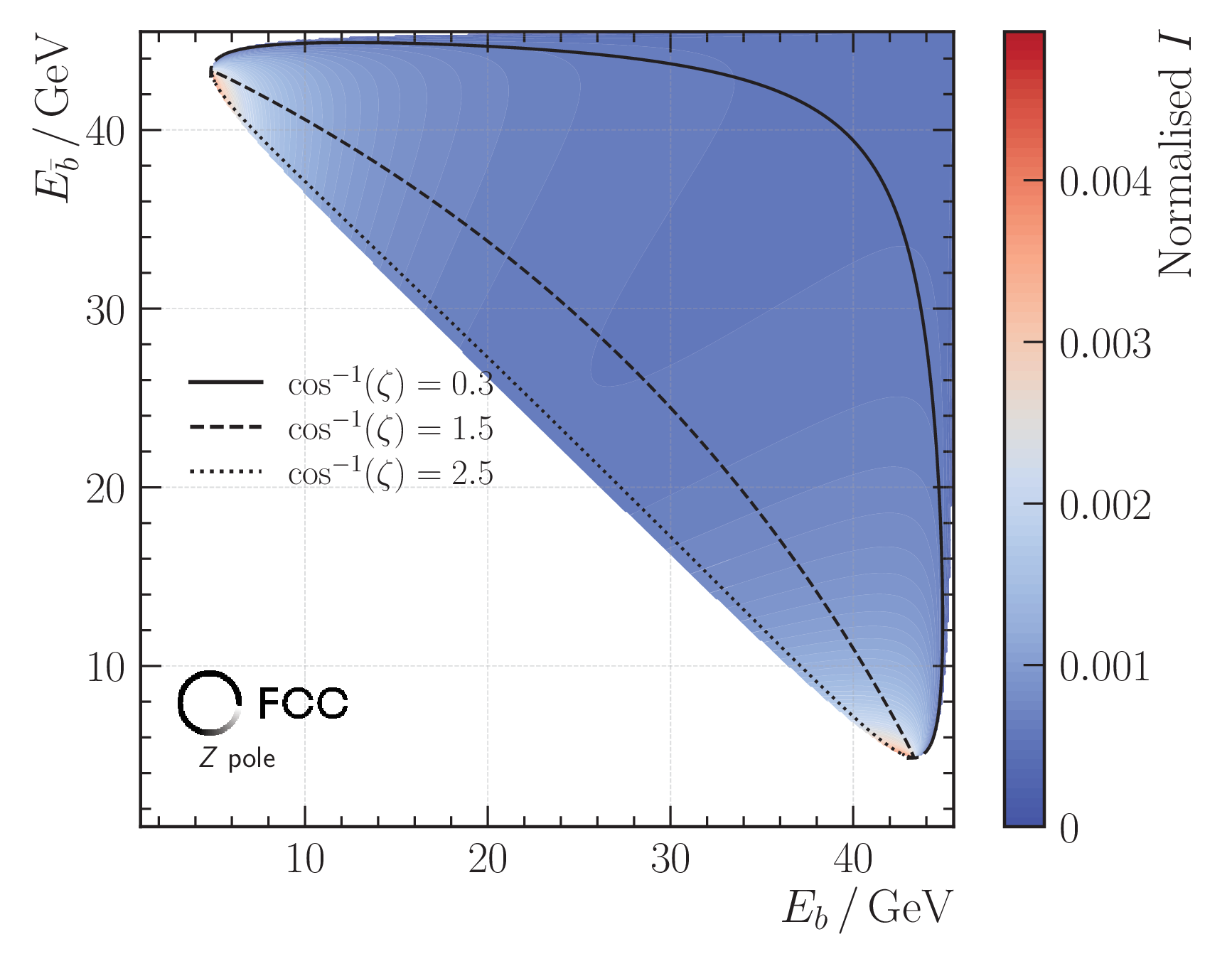}
        \caption{The normalised value of the integrand $I$ on the $z$ axis as function of the $b$- and $\bar{b}$-quark energy on the $x$- and $y$-axis, respectively. The lowest values are achieved for the highest-energetic quarks.}
        \label{fig:enter-label}
    \end{subfigure}\hfill
    \begin{subfigure}[t]{0.48\textwidth}
        \centering
        \includegraphics[width=\textwidth]{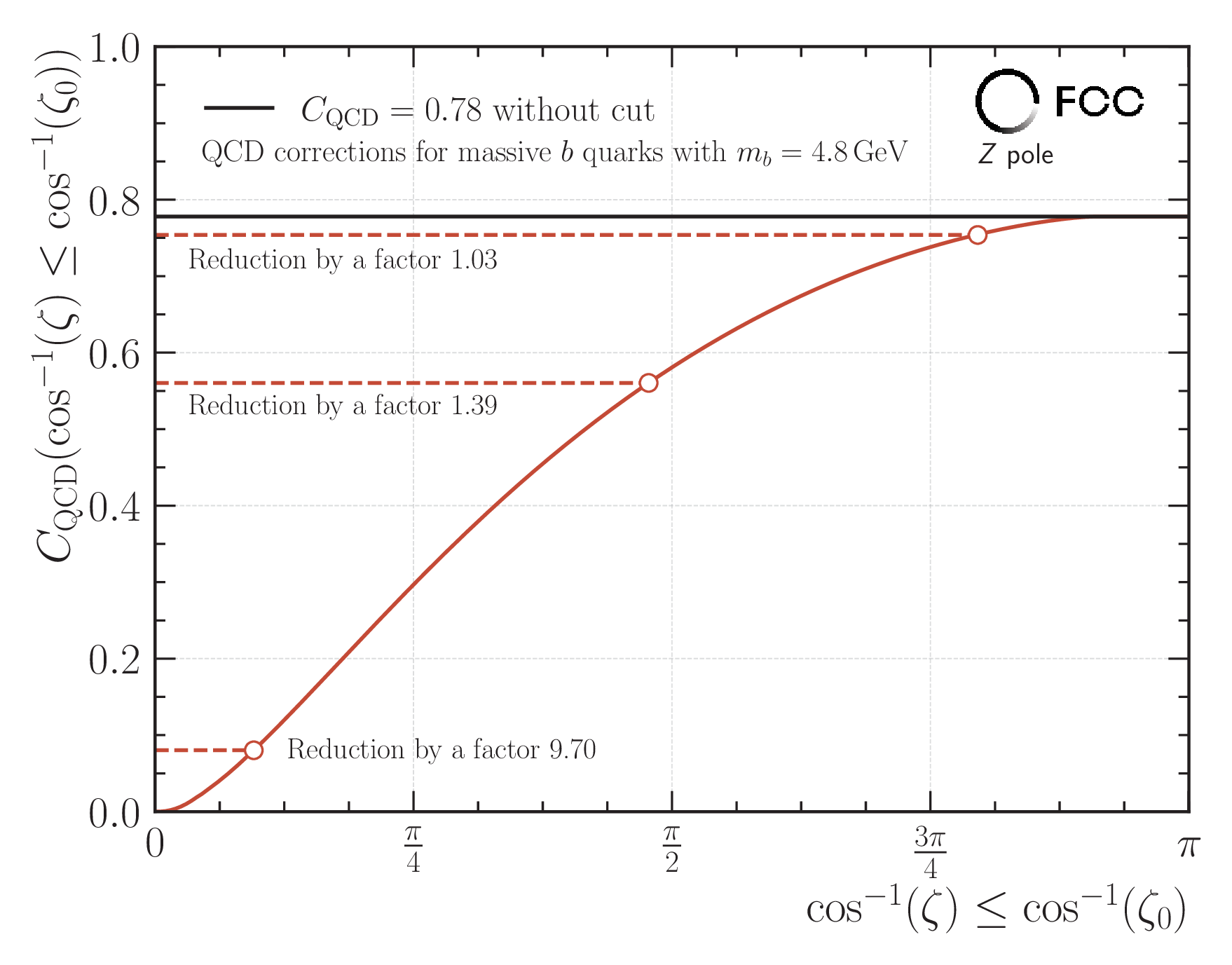}
        \caption{Cuts on the maximally allowed acollinearity angle effectively reduce the impact of the QCD corrections by about an order of magnitude for $\cos^{-1}(\zeta) \leq 0.3$.}
        \label{fig:enter-label}
    \end{subfigure}
    \caption{Since the QCD corrections $C_\text{QCD}$ affect the precision of \AFBbeauty, regions in the phase space are found, in which their impact is minimised.}
    \label{fig:Zbb:cos_zeta_and_C_QCD}
\end{figure}

A sensitive parameter identified is the acollinearity between the two $b$ quarks. This method has been explored in earlier studies (see, for example, Ref.~\cite{Note_on_QCD_corretions}), but has been first implemented in an experimental FCC-ee setting in Ref.~\cite{Revisiting_QCD_correction_really_a_problem}. Mathematically, the acollinearity angle $\cos(\zeta(x, \bar{x}))$, defined as 
\begin{equation}
    \cos(\zeta(x,\bar{x})) = \frac{x\bar{x} + \mu^2 + 2(1 - x - \bar{x})}{\sqrt{x^2 - \mu^2}\sqrt{\bar{x}^2 - \mu^2}}\,,
\end{equation}
is a measure for the angle between the two $b$ quarks when projected onto the plane orthogonal to the initial beam-direction. It solely depends on the energy of the $b$ and $\bar{b}$ quark, $x = \large\sfrac{2E_b}{\sqrt{s}}$ and $\bar{x} = \large\sfrac{2E_{\bar{b}}}{\sqrt{s}}$. Furthermore, the analytical expression of $C_\text{QCD}$ is written in App.~\ref{subsec:app:C_QCD_and_fL}. However, to illustrate the main conclusions of the formula, the differential correction $I(x, \bar{x})$ in the integral of $C_\text{QCD}$, expressed via 
\begin{equation}
    I(x, \bar{x}) = \frac{(x^2 + \bar{x}^2)\cdot(1 - \cos(\zeta(x, \bar{x})))}{3 (1 - x) (1 - \bar{x})}\,,
\end{equation}
is presented in the left panel of Fig.~\ref{fig:Zbb:cos_zeta_and_C_QCD} as a function of $E_b$ and $E_{\bar{b}}$ on the $x$- and $y$-axis, respectively. It represents a measure for the amount of QCD corrections, where higher values of $I$ correspond to larger values of $C_\text{QCD}$. The possibility of radiating gluons from either the $b$ and/or $\bar{b}$ quark is accounted for by symmetrising $I$ accordingly to the expression in Eq.~\eqref{eqn:app:C_mu_integral}. 

It can be seen that the QCD corrections are largest for the lowest energies possible for the $b$ quarks. The reason for this is the reduction of the $b$-quark energy when gluons have been radiated beforehand. In the figure, constant acollinearity angles at $\cos^{-1}(\zeta) = [0.3, 1.5, 2.5]$ are shown as black lines and indicate the cut in the phase space when the acollinearity of the $b$ quarks is required to have a certain value. \\
The effect on the actual QCD corrections $C_\text{QCD}$ as a function of the upper limit on the acollinearity angle is shown on the right side of Fig.~\ref{fig:Zbb:cos_zeta_and_C_QCD}. Without cuts, \mbox{$C_\text{QCD} = 0.78$}, indicated as a horizontal black line. For tighter cuts applied, $C_\text{QCD}$ reduces significantly by about an order of magnitude for $\cos^{-1}(\zeta)\leq 0.3$, therefore mitigating its impact on the systematic uncertainty of \AFBbeauty. However, a direct approach to reduce $I$ would be to select the highest energetic $b$-quarks, which requires, in addition to the direction estimation of the $b$ quark, an estimation of $E_{b}$ and/or $E_{\bar{b}}$.
The potential of the reconstructed $b$-hadron as an energy estimator is discussed in the second part of the following analysis.
Furthermore, $f_\text{L}$ also influences the measurement of \AFBbeauty, and becomes nonzero when including QCD corrections. The impact of $f_\text{L}$ has been examined in Refs.~\cite{Revisiting_QCD_correction_really_a_problem,rL_parameter_reference} and is expected to have a minimal effect on \AFBbeauty when kinematic cuts are applied with the aim of reducing the effect of QCD corrections. 
\clearpage
Two strategies have been followed, which are described in detail in the following:
\begin{enumerate}
    \item A recapitulation of the analysis in Ref.~\cite{Revisiting_QCD_correction_really_a_problem} is performed using acollinearity cuts on reconstructed jets as direction estimators. Since cuts on the jet acollinearity require the reconstruction of at least two objects in the event that approximate the $b$-quark direction, limits on the $b$-hadron acollinearity are not directly applicable because typically only one $b$ hadron per event is reconstructed. However, the $b$ hadron has been used to extend the study by serving as an unambiguous charge identifier for the hemisphere in the unlikely case of a reconstructed $b$-hadron in the event (details are given in Sec.~\ref{subsec:Zbb:measurement_principle}). 
    Furthermore, the impact of incorrect pairing of the $b$ hadron with a jet is greatly reduced when acollinearity cuts on the jets are applied, as these cuts are designed to lower the probability for scenarios with three or more jets originating from gluons.
    \item A more straightforward and novel approach is presented to experimentally reduce the effects of QCD corrections, which involves leveraging the kinematic properties of the reconstructed $b$-hadron. This approach is based on also estimating the $b(\bar{b})$-quark energy $E_{b}(E_{\bar{b}})$ from the energy of the $b$ hadron.
\end{enumerate}
For both, the proof of principle is demonstrated at the parton level first, before applying the concept at the object level.

\subsection{Acollinearity cuts}\label{subsec:Zbb:acollinearity}
In the following section, a recapitulation of the analysis in Ref.~\cite{Revisiting_QCD_correction_really_a_problem} is performed. As a further development of the presented method in Ref.~\cite{Revisiting_QCD_correction_really_a_problem}, the assumption of a perfect flavour tag in an inclusive $Z\to b\bar{b}$ sample can be adapted by using the exclusively reconstructed $b$-hadron of the event. However, since only one $b$ hadron is expected to be reconstructed for the measurement of \AFBbeauty, an inclusive tag of the other hemisphere jet would be needed to apply acollinearity cuts. In the following, the method is presented first at the parton level.

\paragraph{Parton level}
The impact of acollinearity cuts is first studied at the parton level, considering the $b$ quarks after gluon radiation. According to Ref.~\cite{Revisiting_QCD_correction_really_a_problem}, the quark acollinearity cuts have been chosen to
\begin{equation}
    \max\left(\cos^{-1}(\zeta)\right) = \{\pi, 1.5, 1.0, 0.5, 0.3, 0.2, 0.1\}\,.
\end{equation}
The size of QCD corrections has then been estimated from Eq.~\eqref{eqn:Zbb:AFB_with_QCD} to
\begin{equation}
    C_{\text{QCD}} = \frac{\pi}{\alpha_\text{S}}\frac{A_\text{FB}^{b,0} - A_\text{FB}^{b}}{A_\text{FB}^{b,0}}\,.
\end{equation}
\begin{figure}[t]
  \begin{minipage}[b]{.48\textwidth}
    \centering
    \includegraphics[width = \textwidth]{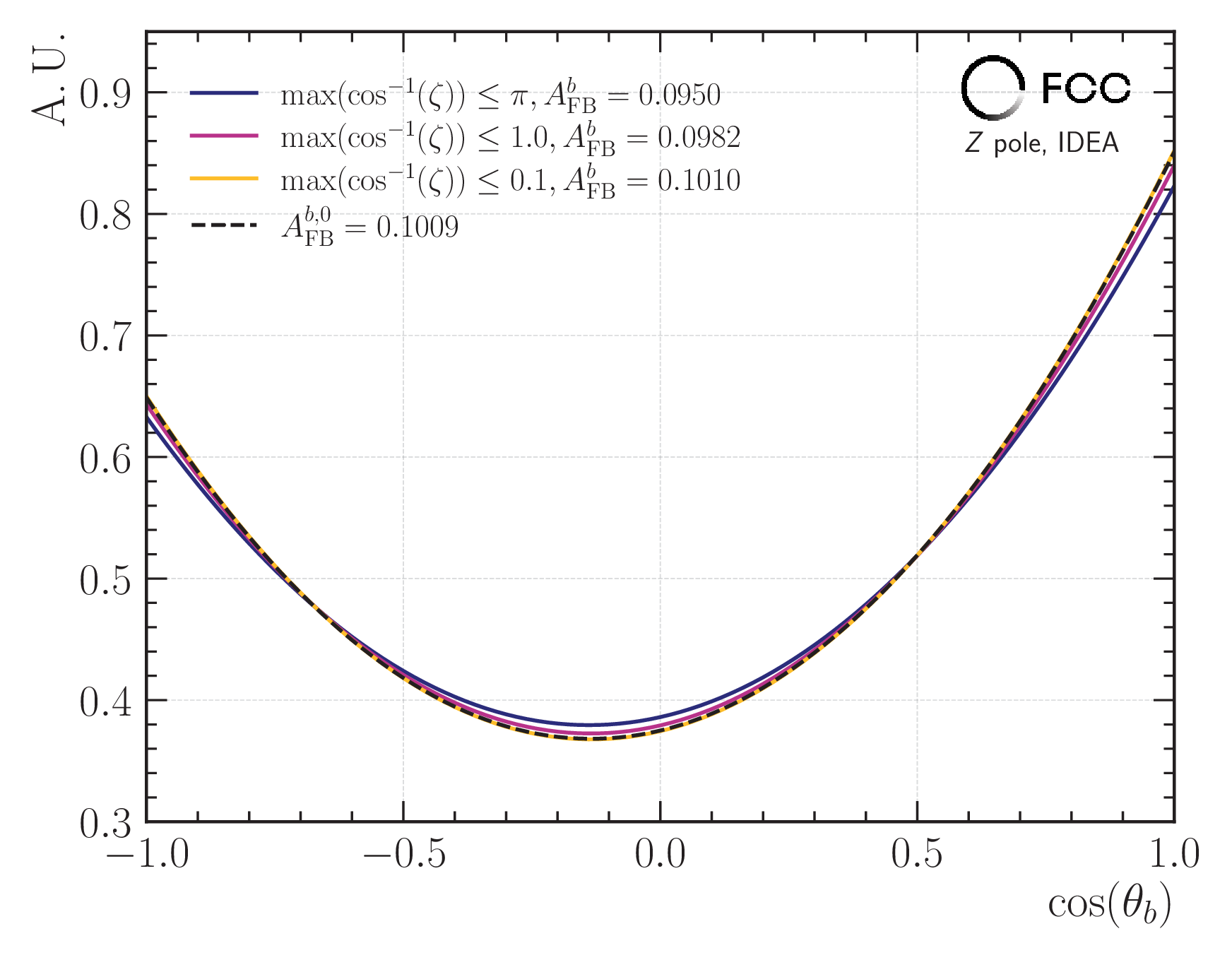}
    \captionsetup{width=0.9\textwidth}
    \caption{The fit result for different cuts on the $b$-quark acollinearity. The $b$-quark forward-backward asymmetry without gluon radiation is shown as a dashed line. For tighter cuts, the results converge.}
    \label{fig:Zbb:bquark_acollinearity}
  \end{minipage}\hfill
  \begin{minipage}[b]{.48\textwidth}
    \centering
    \begin{tabular}{l|S[table-format=1.4(4)] S[table-format=2.4]} 
        \toprule
        Cut & {$A_\text{FB}^b$} & {$\sfrac{\alpha_\text{S}}{\pi}\cdot C_\text{QCD}$} \\ \midrule 
        $\pi$ & 0.0950(1) & 0.0585  \\ 
        1.5           & 0.0975(1) & 0.0337  \\ 
        1.0           & 0.0982(1) & 0.0268  \\ 
        0.5           & 0.0990(2) & 0.0188  \\ 
        0.3           & 0.0997(2) & 0.0119  \\ 
        0.2           & 0.1003(2) & 0.0059  \\ 
        0.1           & 0.1010(4) & -0.0009 \\ 
        \bottomrule
    \end{tabular}
    \captionsetup{width=0.9\textwidth}
    \captionof{table}{\AFBbeauty and $C_\text{QCD}$ computed from the $b$ quarks after gluon radiation for different cuts on the acollinearity. A convergence of \AFBbeauty towards $\AFBbeautyzero = 0.1009$ can be observed due to a significant reduction of $C_\text{QCD}$.}
    \label{tab:Zbb:bquark_acollinearity}
  \end{minipage}
\end{figure}
Given that \AFBbeautyzero and \AFBbeauty have been derived from the same dataset, there is no statistical uncertainty associated with $C_{\text{QCD}}$. The numerical values for \AFBbeauty and $C_\text{QCD}$ are shown in Tab.~\ref{tab:Zbb:bquark_acollinearity}. The decrease in $C_\text{QCD}$ due to acollinearity cuts results in \AFBbeauty approaching \AFBbeautyzero. Consequently, this reduces the influence of $C_\text{QCD}$ on the overall uncertainty of \AFBbeautyzero, which is derived from the propagation of the uncertainty and is considered as the only systematic uncertainty
\begin{equation}
    \sigma_\text{syst.}\left(\AFBbeautyzero\right) = \frac{\pi\alpha_\text{S}\AFBbeauty\sigma(C_\text{QCD})}{(\pi - \alpha_\text{S}C_\text{QCD})^2}\,,
    \label{eqn:Zbb:AFB_systematic_uncertainty}
\end{equation}
with the uncertainty on $C_\text{QCD}$, $\sigma(C_\text{QCD})$. 
The impact of acollinearity cuts is visualised in Fig.~\ref{fig:Zbb:bquark_acollinearity}, which presents the interpolation result for cuts at $\max(\cos^{-1}(\zeta)) \leq [\pi, 1.0, 0.1]$ together with the result from \AFBbeautyzero in the dashed line.
It can be seen that towards tighter cuts, the effect of QCD radiations becomes negligible. 

\begin{figure}[t]
    \begin{subfigure}[t]{0.48\textwidth}
        \centering
        \includegraphics[width = \textwidth]{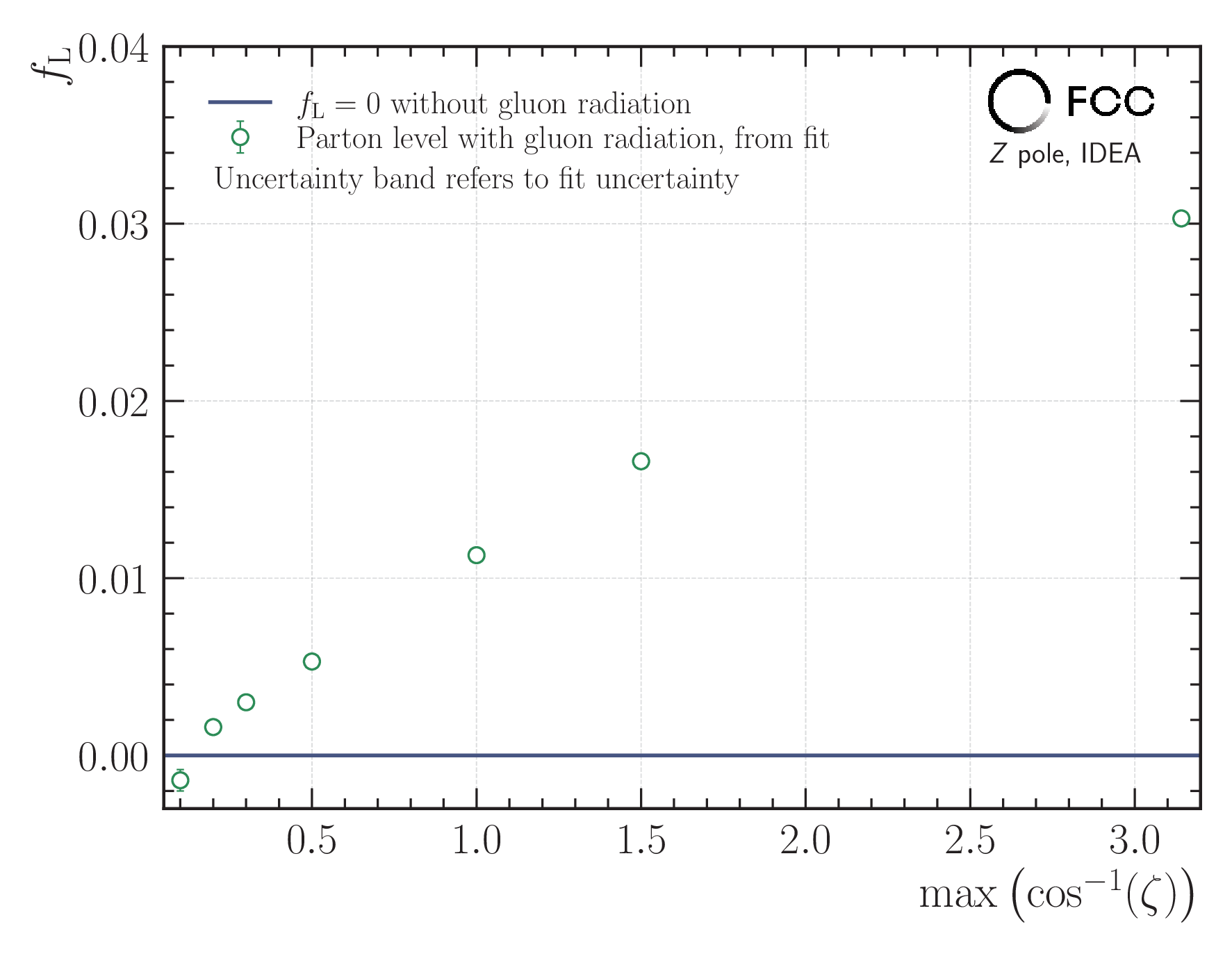}
        \caption{The $f_\text{L}$ parameter as function of the maximally allowed $b$-quark acollinearity.}
        \label{subfig:Zbb:fL_acollinearity}
    \end{subfigure}\hfill
    \begin{subfigure}[t]{0.48\textwidth}
        \centering
        \includegraphics[width = 1\textwidth]{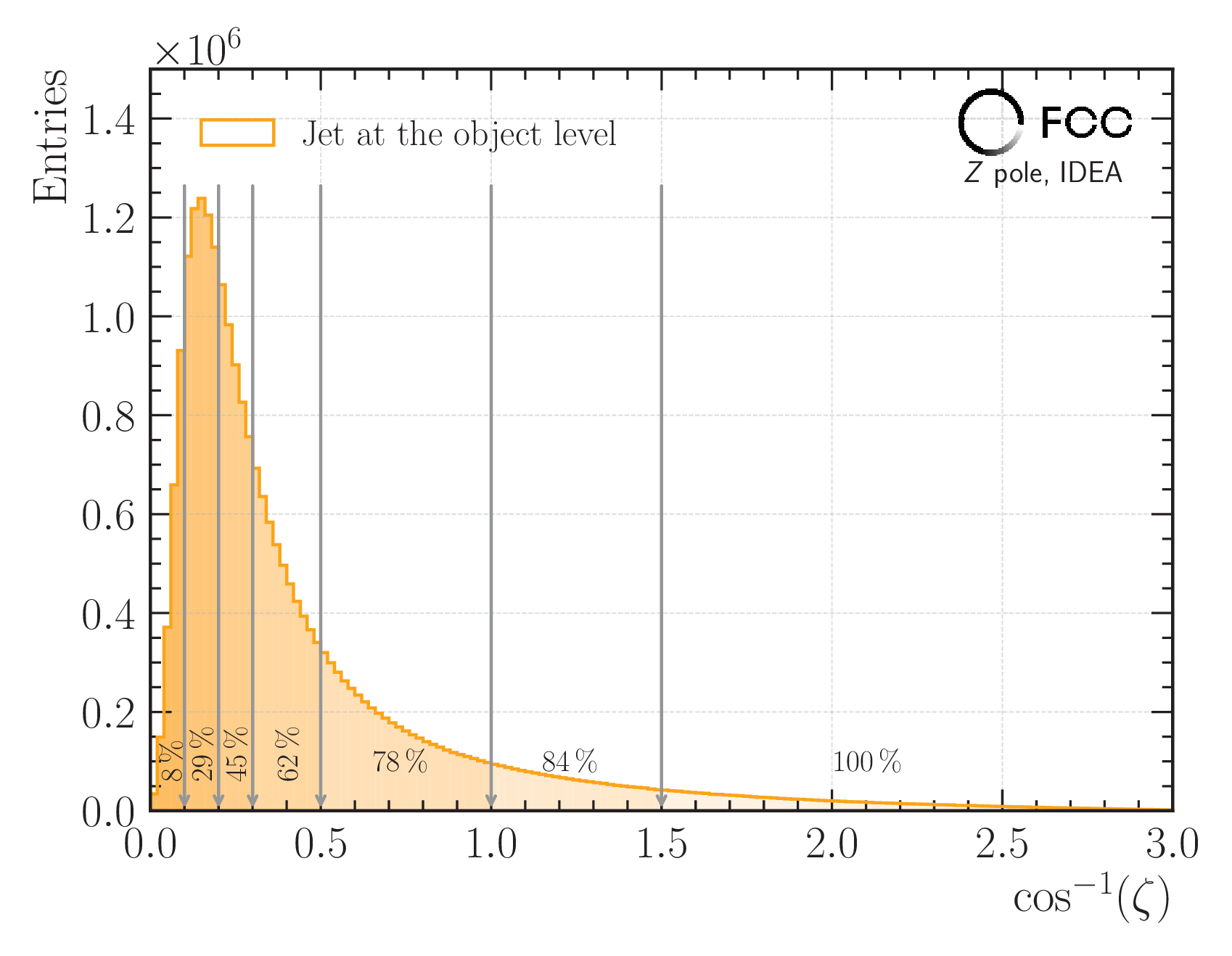}
        \caption{The jet-acollinearity distribution, highlighting also the fraction of events after the application of the cut. About \SI{30}{\percent} of the events remain for $\max(\cos^{-1}(\zeta)) \leq 0.2$, for which the effect of QCD corrections is reduced by about one order of magnitude.}
        \label{subfig:Zbb:jet_acollinearity_cuts}
    \end{subfigure}
    \caption{The longitudinal component $f_\text{L}$ as function of the acollinearity cuts at the parton level for the $b$ quarks after gluon radiation. Experimentally, cuts on the jet acollinearity reduce the number of events. The remaining fraction of events is presented in the one-dimensional distribution of the jet acollinearity in Fig.~\subref{subfig:Zbb:jet_acollinearity_cuts}.}
    \label{fig:Zbb:AFB_parton_level_result_and_jet_acollinearity}
\end{figure}

Consequently, cuts on the $b$-quark acollinearity impact the longitudinal fraction $f_\text{L}$ as shown in the left panel of Fig.~\ref{fig:Zbb:AFB_parton_level_result_and_jet_acollinearity}, while $f_\text{L}$ has been determined from the simultaneous fit at the parton level to Eq.~\eqref{eqn:Zbb:AFB_full_fit}. The errorbars represent the uncertainty of the fit. 
Again, a significant decrease towards $f_\text{L} = 0$ can be seen for tighter acollinearity cuts. With the result obtained at the parton level, the study has been extended to use the experimentally accessible jet acollinearity in the following.

\paragraph{Object level}

At the object level, acollinearity cuts have been implemented between the two jets identified as $b$ jets. In an analysis with actual data from an FCC-ee experiment, specialised tools are required to identify the flavour of both $b$ jets. Additionally, employing a jet-flavour tagging algorithm requires careful handling and consideration of factors such as detector-acceptance effects, since the performance of these algorithms typically depends on the polar angle of the jet. 

Compared to the acollinearity cuts applied at the quark level, the acollinearity between two reconstructed jets derived from the jet angles can be greater than one due to inaccuracies in the jet-energy estimation. Consequently, imposing $0 \leq \cos^{-1}(\zeta) \leq \pi$ inherently results in an event cut, which has been determined to be insignificant. The jet-acollinearity distribution for the $b$ jets at the object level is shown in the right panel of Fig.~\ref{fig:Zbb:AFB_parton_level_result_and_jet_acollinearity}. This distribution indicates that a substantial portion of events has jet acollinearities below \num{0.2}, thereby reducing $C_\text{QCD}$ by about an order of magnitude at this working point (refer to Tab.~\ref{tab:Zbb:bquark_acollinearity}).

In a manner similar to the parton-level studies, cuts on the jet acollinearity have been applied and \AFBbeauty has been determined from the fit. Due to the lack of experimental sensitivity to the longitudinal fraction, the values of $f_\text{L}$ have been fixed to those obtained at the parton level. The fit has been performed on the polar-angle distribution of the jet that has been assigned the negative charge in the event. 

To establish a threshold where the systematic uncertainty matches the statistical uncertainty, two scenarios have been considered for the systematic uncertainty due to $C_\text{QCD}$. For clarity, the QCD-corrected result from
\begin{equation}
    A_\text{FB}^{b,0} = \frac{1}{1 - \frac{\alpha_\text{S}}{\pi}C_\text{QCD}} A_\text{FB}^b\,,
    \label{eqn:Zbb:QCD_correction_application}
\end{equation}
has been evaluated below, and the uncertainty from Eq.~\eqref{eqn:Zbb:AFB_systematic_uncertainty} has been computed. The first scenario considers a relative uncertainty on $C_\text{QCD}$ from Tab.~\ref{tab:Zbb:bquark_acollinearity} to be \SI{1}{\percent}, while the second, more conservative scenario, assumes an uncertainty of \SI{5}{\percent}. The statistical precision is derived from the nominal statistical precision of $\sigma_\text{stat.}(\AFBbeauty) = \num{1.56e-5}$ and its reduction has been assumed to follow the remaining fraction of events from the jet-acollinearity distribution.
\begin{figure}[t]
    \begin{subfigure}[t]{0.48\textwidth}
        \centering
        \includegraphics[width = \textwidth]{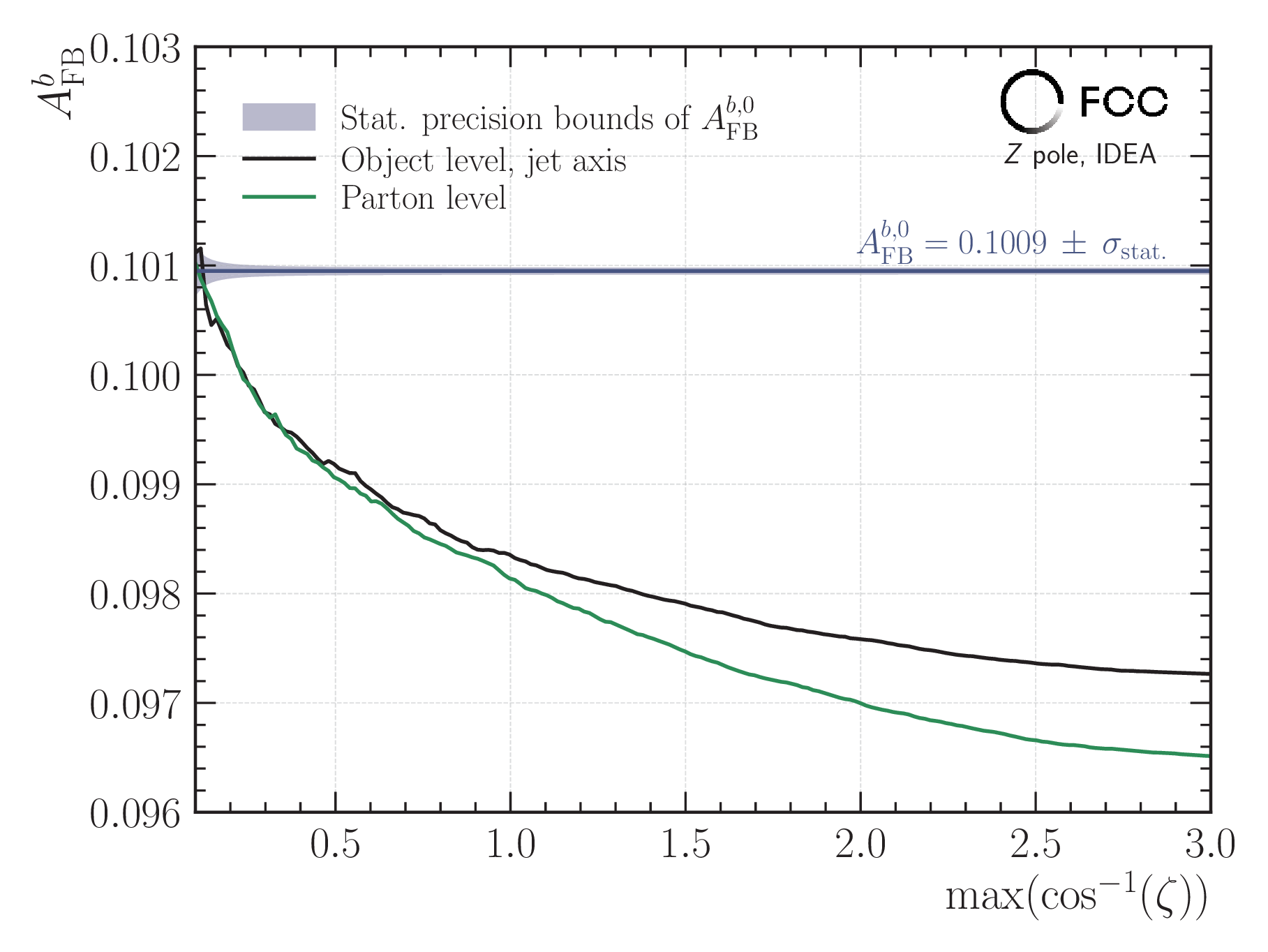}
        \caption{The non-QCD-corrected \AFBbeauty as function of the jet-acollinearity, comparing the jet direction at the object level and the $b$-quark direction at the parton level. Differences become negligible for jet acollinearities below \num{0.8}.}
        \label{subfig:Zbb:AFB_acollinearity_different_direction_estimators}
    \end{subfigure}\hfill
    \begin{subfigure}[t]{0.48\textwidth}
        \centering
        \includegraphics[width = \textwidth]{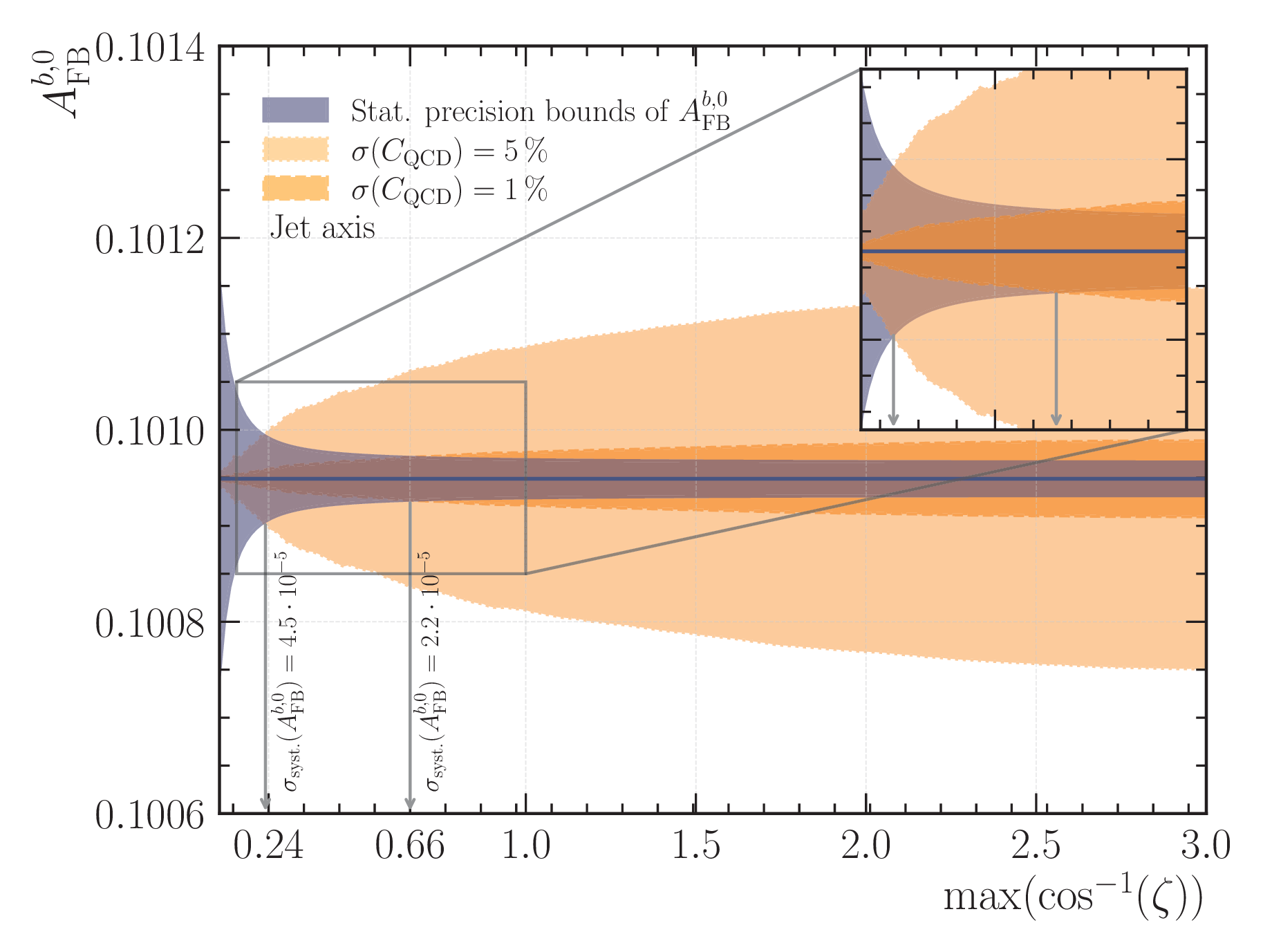}
        \caption{The QCD-corrected \AFBbeauty from the jet-axis direction. The decrease and increase of the systematic and statistical uncertainty are shown as orange and blue bands, respectively. The intersection defines the threshold value for two scenarios of the QCD corrections uncertainty.}
        \label{subfig:Zbb:b_quark_acollinearity}
    \end{subfigure}
    \caption{Jet-acollinearity cuts are valid estimators for the amount of direction distortion of the $b$ quark and the jet direction. For tighter cuts, both converge towards \AFBbeautyzero. By applying the QCD corrections, optimal thresholds at $\max(\cos^{-1}(\zeta))\leq 0.66$ and $\max(\cos^{-1}(\zeta))\leq 0.24$ have been found for two scenarios of the uncertainty on the QCD corrections, for which $\sigma_\text{stat.}(\AFBbeauty) = \sigma_\text{syst.}(\AFBbeauty)$.}
    \label{fig:Zbb:AFB_acollinearity_results}
\end{figure}
Fig.~\ref{fig:Zbb:AFB_acollinearity_results} illustrates the results as a function of jet acollinearity, employing a finer binning for the maximum allowed acollinearity cuts compared to previous analyses. The left panel displays \AFBbeauty without any QCD corrections at both the object and parton levels following gluon radiation in black and green, respectively. Both results converge towards the horizontal line, which indicates the parton-level value prior to gluon radiation and includes the expected statistical precision as an uncertainty band. It is evident that the statistical uncertainty grows considerably with very stringent acollinearity cuts. The discrepancy between the object-level and parton-level results stems from selection criteria and the impact of jet acollinearity, such as the requirement for at least two $b$-tagged jets. Further factors include the detector-acceptance effects of additional gluon jets. However, this effect reduces with tighter acollinearity cuts, thus reducing the discrepancy, since the general event topology of the jets becomes more \textit{back-to-back}. For acollinearity cuts below \num{0.8}, the difference becomes almost negligible.

In contrast, the QCD-corrected value \AFBbeautyzero from Eq.~\eqref{eqn:Zbb:QCD_correction_application} is illustrated on the right side of Fig.~\ref{fig:Zbb:AFB_acollinearity_results}. In this context, \AFBbeauty has been derived from the object-level distribution to overcome the discrepancies between the object- and parton-level quantities, as previously discussed, since the aim of this study is to determine an appropriate cut on the jet acollinearity, where the systematic precision matches the statistical precision. The systematic uncertainty is shown for the two scenarios ${\large\sfrac{\sigma(C_\text{QCD})}{C_\text{QCD}}} = [1,5]\,\si{\percent}$ in darker and lighter orange, respectively. The result shows that in both cases the systematic uncertainty reduces with tighter jet-acollinearity cuts. From the intersection $\sigma_\text{stat.}(\AFBbeautyzero) = \sigma_\text{syst.}(\AFBbeautyzero)$, the optimal thresholds have been determined to be $\max(\cos^{-1}(\zeta)) \leq 0.66$ and $\max(\cos^{-1}(\zeta)) \leq 0.24$ and the uncertainties on \AFBbeautyzero result to
\begin{align}
    \begin{split}
        \frac{\sigma(C_\text{QCD})}{C_\text{QCD}} &= \SI{5}{\percent}\;\Rightarrow\;\AFBbeautyzero = \mu(\AFBbeautyzero)\,\pm\,\num{4.5e-5}(\text{stat.})\,\pm\,\num{4.5e-5}(\text{syst.})\\
        \frac{\sigma(C_\text{QCD})}{C_\text{QCD}} &= \SI{1}{\percent}\;\Rightarrow\;\AFBbeautyzero = \mu(\AFBbeautyzero)\,\pm\,\num{2.2e-5}(\text{stat.})\,\pm\,\num{2.2e-5}(\text{syst.})\,.
    \end{split}
\end{align}
Here, it is assumed that in a more realistic scenario the statistical uncertainty behaves similarly when only one hemisphere has been exclusively reconstructed and the other one decays inclusively. 
With this result, the analysis of jet-acollinearity cuts as a means to reduce the QCD corrections is concluded. Nevertheless, the effects of the jet-clustering algorithm, $b$-jet tagging, and the transverse-momentum cut still require further investigation. To inherently address the aforementioned issues, the next section studies the potential of minimising the QCD correction through kinematic cuts on the $b$ hadron, eliminating the necessity of clustering and tagging $b$ jets in the event.

\begin{figure}[t]
  \begin{minipage}[b]{.48\textwidth}
    \centering
    \includegraphics[width = \textwidth]{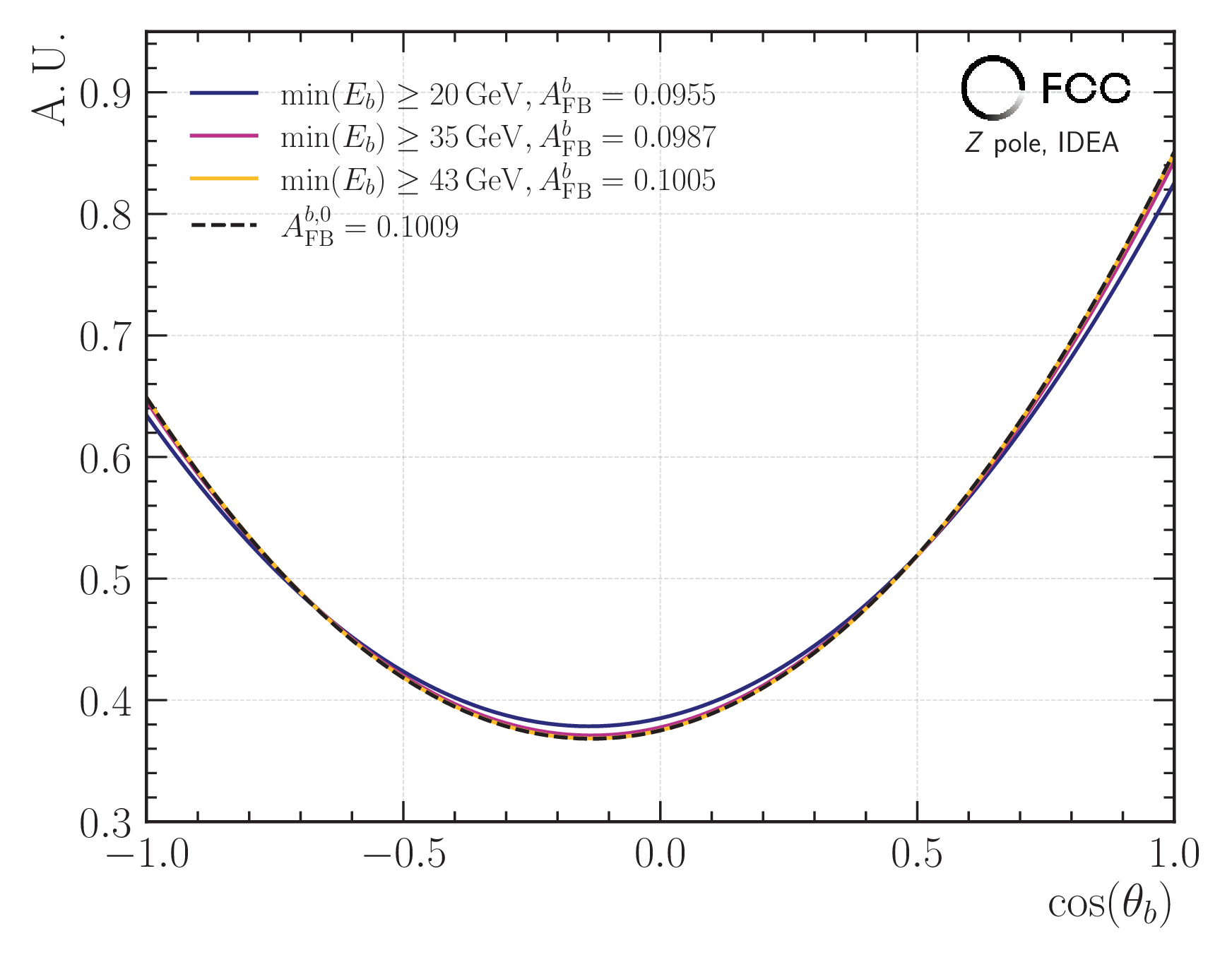}
    \captionsetup{width=0.9\textwidth}
    \caption{The fit result for different cuts on the $b$-quark energy. The $b$-quark forward-backward asymmetry without gluon radiation is shown as a dashed line. For tighter cuts, the results converge.}
    \label{fig:Zbb:bquark_energy}
  \end{minipage}\hfill
  \begin{minipage}[b]{.48\textwidth}
    \centering
    \begin{tabular}[t]{l|S[table-format=1.4(4)] S[table-format=1.4]} 
        \toprule
        Cut & {$A_\text{FB}^b$} & {$\sfrac{\alpha_\text{S}}{\pi}\cdot C_\text{QCD}$} \\ \midrule 
        10  & 0.0955(1) & 0.0535 \\ 
        20  & 0.0968(1) & 0.0406 \\ 
        30  & 0.0980(1) & 0.0287 \\ 
        35  & 0.0987(2) & 0.0218 \\ 
        40  & 0.0994(2) & 0.0156 \\ 
        41  & 0.0996(2) & 0.0129 \\ 
        42  & 0.0998(2) & 0.0109 \\ 
        43  & 0.1005(3) & 0.0040 \\ 
        44  & 0.1009(4) & 0.0000 \\ 
        \bottomrule
    \end{tabular}
    \captionsetup{width=0.9\textwidth}
    \captionof{table}{\AFBbeauty and $C_\text{QCD}$ computed from the $b$ quarks for different cuts on the $b$-quark energy. A convergence of \AFBbeauty towards \AFBbeautyzero can be observed due to a significant reduction of $C_\text{QCD}$.\\}
    \label{tab:Zbb:bquark_energy}
  \end{minipage}
\end{figure}

\subsection[$b$-hadron energy cuts]{\boldmath{$b$}-hadron energy cuts}\label{subsec:b_hadron_energy}

In order to minimise the QCD corrections, the kinematic characteristics of the reconstructed $b$-hadron in the event can be used. Analogously to the investigation of acollinearity cuts, the feasibility has initially been confirmed at the parton level with energy cuts on the $b$ quarks after gluon radiation. It is mentioned that, because both hemispheres are compelled to contain two $b$ hadrons, there are events with one or two reconstructed $b$ hadrons in the event. If both $b$ hadrons have been reconstructed, one has been chosen randomly to avoid biases in the determination of \AFBbeauty. To comply with this strategy at the parton level, only one of the $b$ quarks is randomly required to meet the energy criterion.

\paragraph{Parton level}
As first step, suitable energy cuts have been chosen to be
\begin{equation}
    \min(E) \geq \{10, 20, 30, 35, 40, 41, 42, 43, 44\}\,,
    \label{eqn:Zbb:energy_cut_range}
\end{equation}
with an emphasis on the higher energies, since it is expected that the reduction of the QCD corrections is driven by the highest-energetic $b$ quarks or hadrons. 

The parton-level result is illustrated in Fig.~\ref{fig:Zbb:bquark_energy} and is numerically detailed in Tab.~\ref{tab:Zbb:bquark_energy}. As expected, the magnitude of the QCD corrections is significantly reduced for higher $b$-quark energies following gluon radiation. Consequently, the QCD corrections are reduced by approximately an order of magnitude for $b$-quark energies above \SI{43}{\giga\electronvolt}.

\begin{figure}[t]
    \begin{subfigure}[t]{0.48\textwidth}
        \centering
        \includegraphics[width = \textwidth]{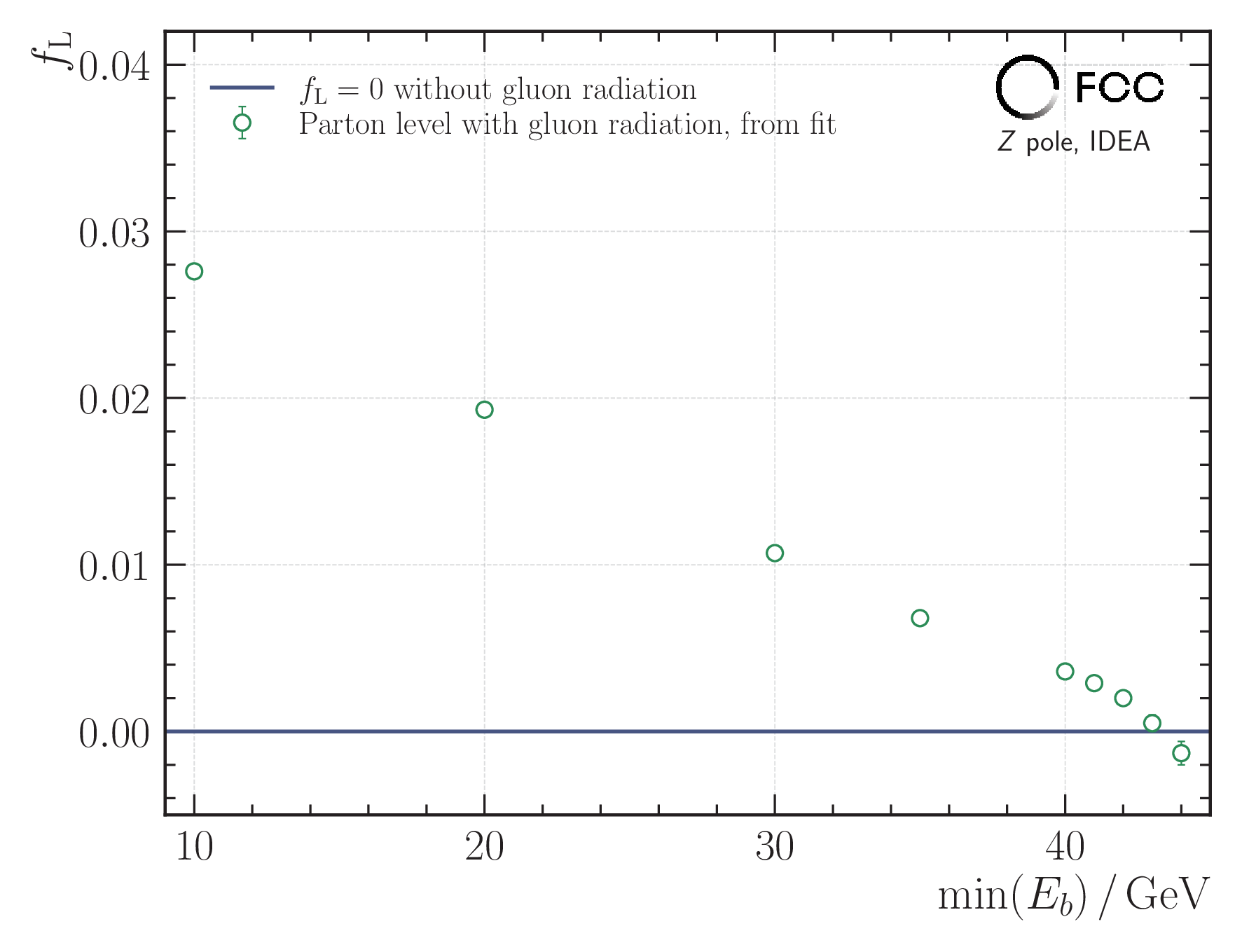}
        \caption{The $f_\text{L}$ parameter as function of the maximally allowed $b$-quark energy.}
        \label{subfig:Zbb:fL_acollinearity_bquark_energy}
    \end{subfigure}\hfill
    \begin{subfigure}[t]{0.48\textwidth}
        \centering
        \includegraphics[width = \textwidth]{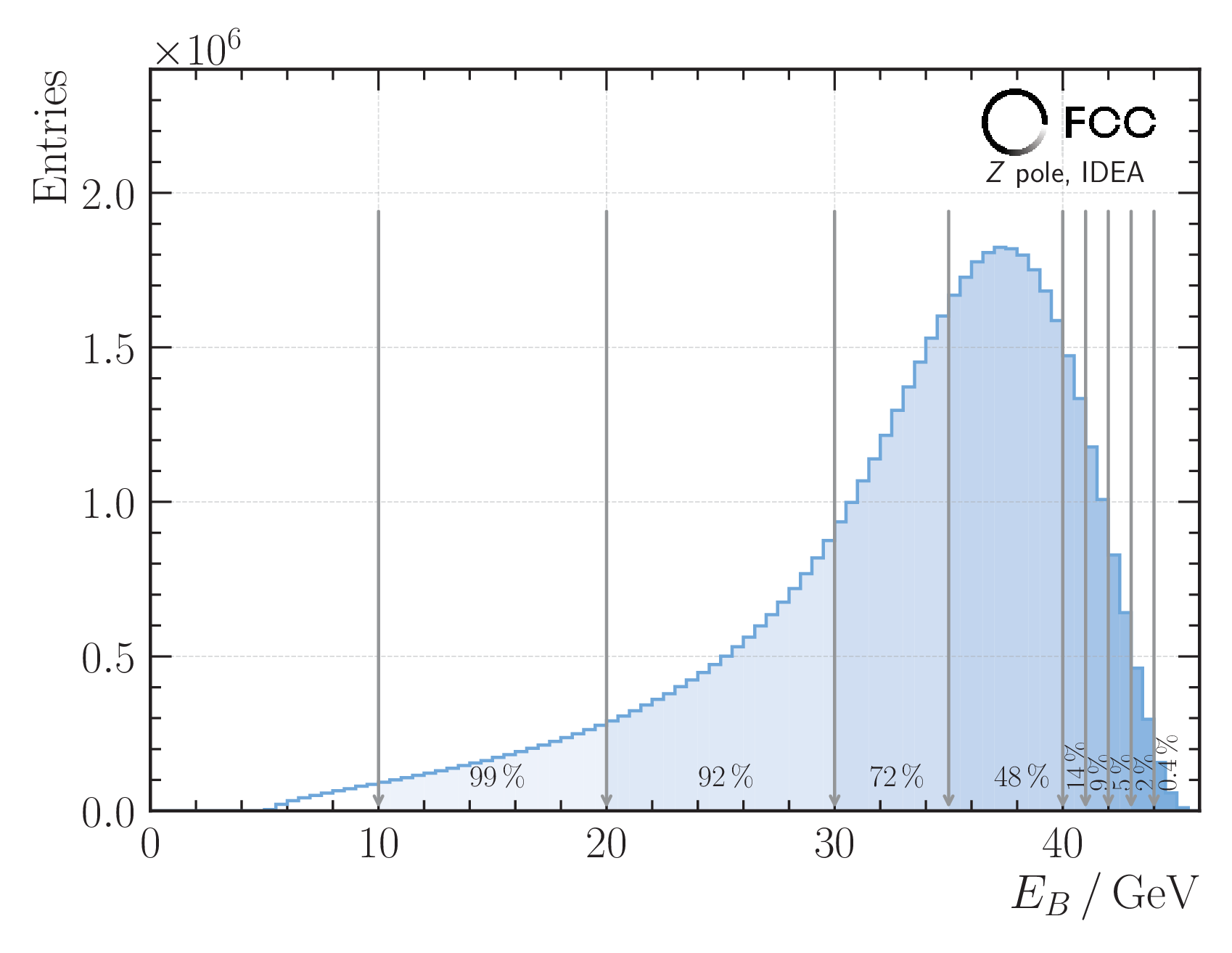}
        \caption{The energy distribution of the $b$ hadron, indicating also the fraction of events when applying the different energy cuts.}
        \label{subfig:Zbb:B_energy_cuts}
    \end{subfigure}
    \caption{The longitudinal component $f_\text{L}$ as function of the energy cuts at the parton level for the $b$ quarks after gluon radiation. Experimentally, cuts on the $b$-hadron energy reduce the number of events, where the remaining fraction of events is presented in the one-dimensional distribution in Fig.~\subref{subfig:Zbb:jet_acollinearity_cuts}.}
    \label{fig:Zbb:AFB_parton_level_result_and_B_energy}
\end{figure}

The impact on $f_\text{L}$ is presented on the left side of Fig.~\ref{fig:Zbb:AFB_parton_level_result_and_B_energy}, which shows a similar trend towards zero compared to the acollinearity cuts for higher-energy cuts on the $b$ quark. This result motivates the use of an effective reduction of the QCD corrections by applying cuts on the $b$-hadron energy in the paragraph below. 

\begin{figure}[t]
    \begin{subfigure}[t]{0.48\textwidth}
        \centering
        \includegraphics[width = \textwidth]{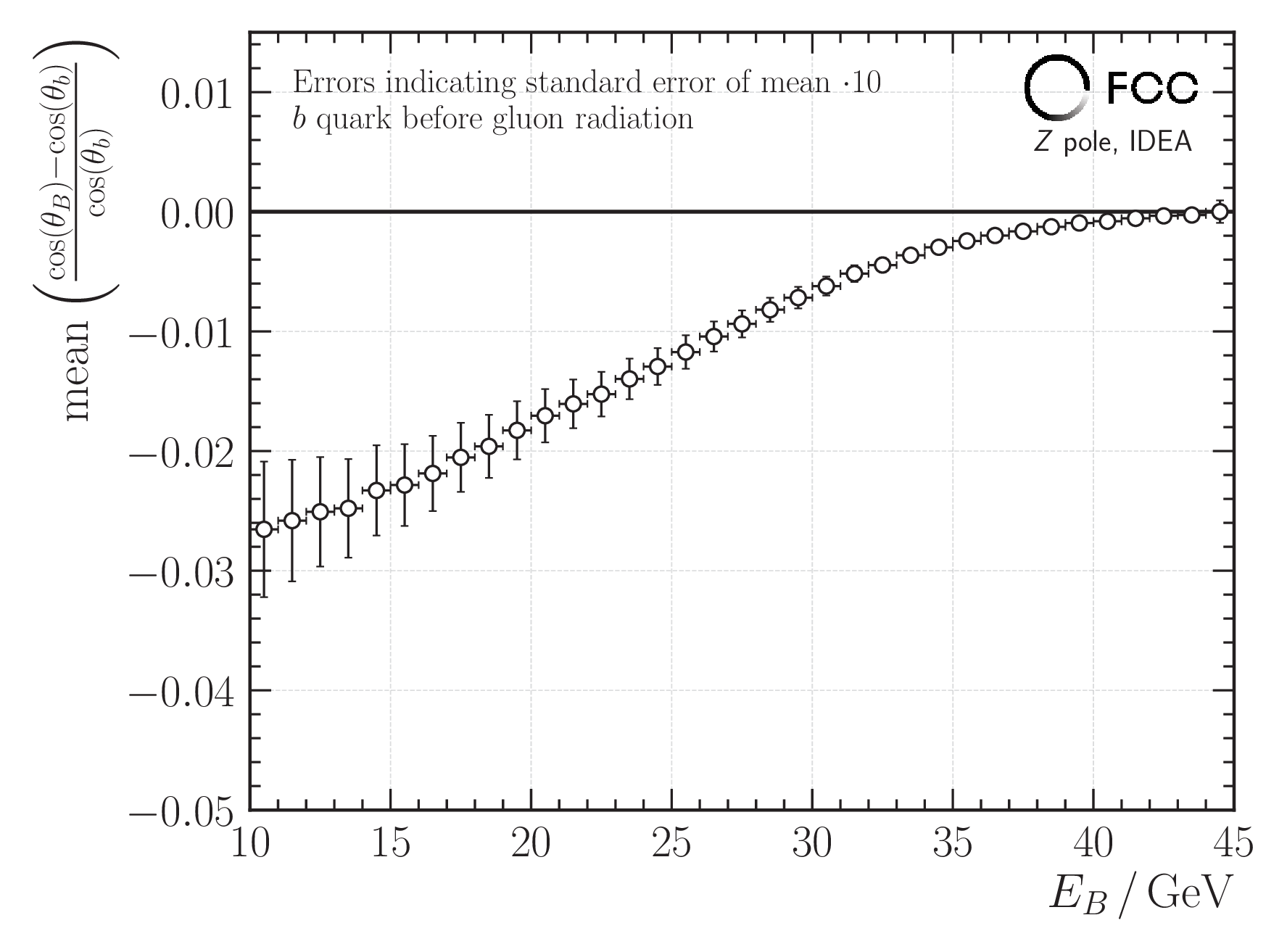}
        \caption{The mean of the relative difference of the $b$-quark polar-angle before gluon radiation and the reconstructed $B$ meson. The uncertainties shown refer to the standard error of the mean and have been scaled by a factor of ten for visibility.}
        \label{subfig:Zbb:b_B_energy_correlation}
    \end{subfigure}\hfill
    \begin{subfigure}[t]{0.48\textwidth}
        \centering
        \includegraphics[width = \textwidth]{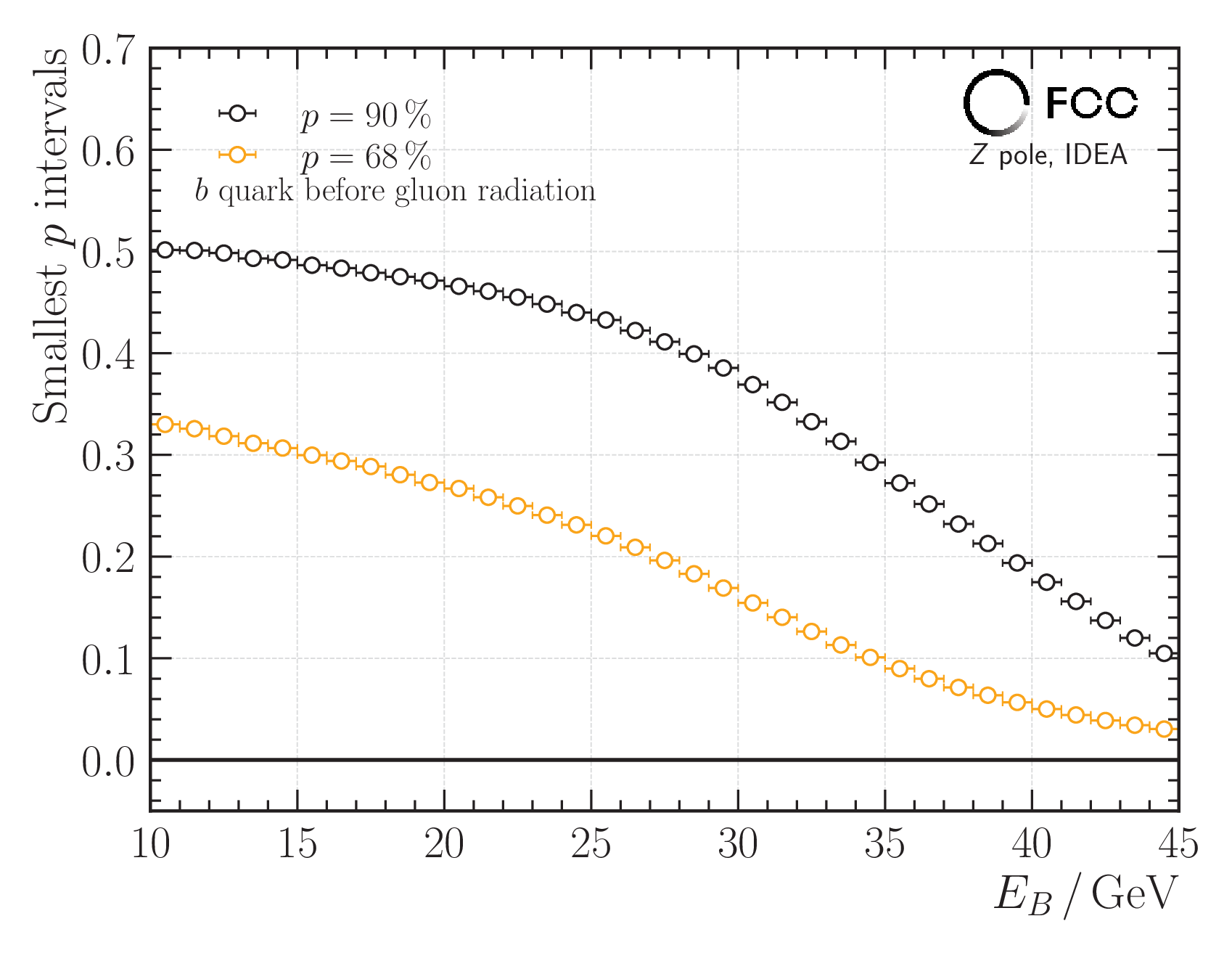}
        \caption{The \SI{68}{\percent} and \SI{90}{\percent} smallest intervals of the relative polar-angle resolution.}
        \label{subfig:Zbb:b_B_energy_correlation}
    \end{subfigure}
    \caption{The impact of energy cuts applied on the reconstructed $b$ hadrons on the accuracy of the $b$-quark direction estimation.}
    \label{fig:Zbb:B_energy_correlation_and_cuts}
\end{figure}

\paragraph{Object level}
Since the cuts on the $b$-hadron energy aim to improve the direction estimation of the initial $b$-quark prior to gluon radiation, a differential analysis of the relative polar-angle difference has been carried out, showing the mean of the relative difference in bins of the $b$-hadron energy in the left panel of Fig.~\ref{fig:Zbb:B_energy_correlation_and_cuts}. For clarity, the standard error of the mean has been increased by a factor of ten and is shown as an error bar on the points. The bias in the distribution, expressed by the difference of the points to the zero line, reduces to zero for the highest-energetic $b$ hadrons. Accordingly, the \SI{68}{\percent} and \SI{90}{\percent} smallest intervals of the relative polar-angle resolution are shown in the right panel of Fig.~\ref{fig:Zbb:B_energy_correlation_and_cuts}, which decrease significantly with increasing energy. The remaining ranges of the order \num{0.03} and \num{0.1} for the smallest \SI{68}{\percent} and \SI{90}{\percent} intervals at energies above \SI{44}{\giga\eV} are due to hadronisation effects.

Although a notable reduction in bias and a gain in direction-estimation precision can be achieved for the highest-energetic candidates, the loss in statistical precision is significant. This is evident when examining the $b$-hadron's energy distribution in the right panel of Fig.~\ref{fig:Zbb:AFB_parton_level_result_and_B_energy}, where only a small fraction of candidates of the order \SI{10}{\percent} remains after placing cuts on the $b$-hadrons energy above \SI{40}{\giga\eV}.

Nevertheless and similarly to the study of the jet-acollinearity cuts, the primary focus is on the threshold value that reduces the impact of $C_\text{QCD}$ on \AFBbeauty to a level such that the systematic uncertainty on \AFBbeauty becomes competitive with the statistical precision. Following this, \AFBbeauty has been derived from the polar-angle distribution of the negatively charged $b$-hadron.

\begin{figure}[t]
    \begin{subfigure}[t]{0.48\textwidth}
        \centering
        \includegraphics[width = 1\textwidth]{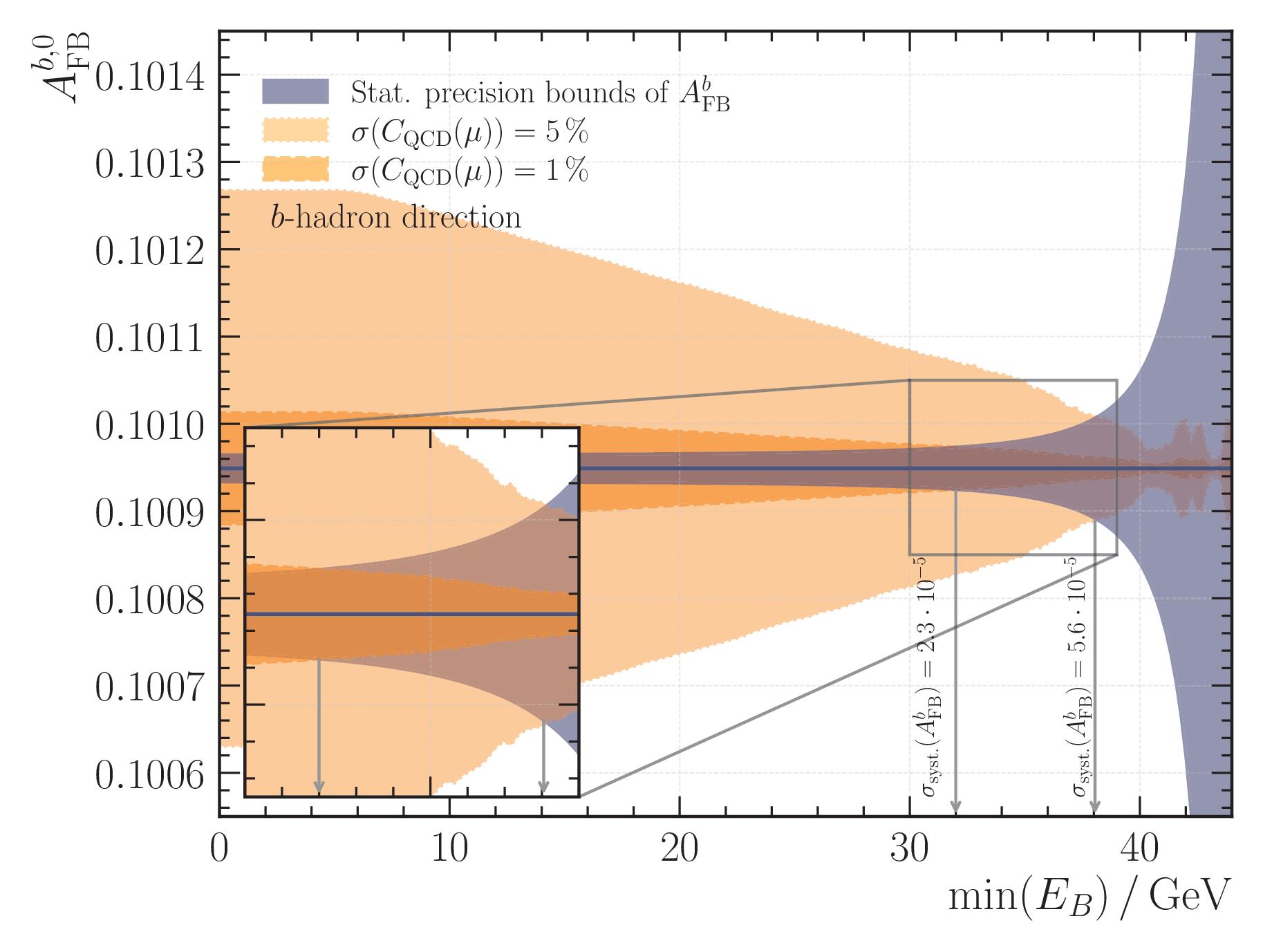}
        \caption{Similar to Fig.~\ref{subfig:Zbb:b_quark_acollinearity}, but for the $b$-hadron energy instead of the jet-acollinearity on the $x$ axis, since $E_B$ does not require to reconstruct and tag the flavour of the jets, which introduce another source of systematic uncertainty.}
        \label{subfig:Zbb:AFB_energy_results}
    \end{subfigure}\hfill
    \begin{subfigure}[t]{0.48\textwidth}
        \centering
        \includegraphics[width = 1\textwidth]{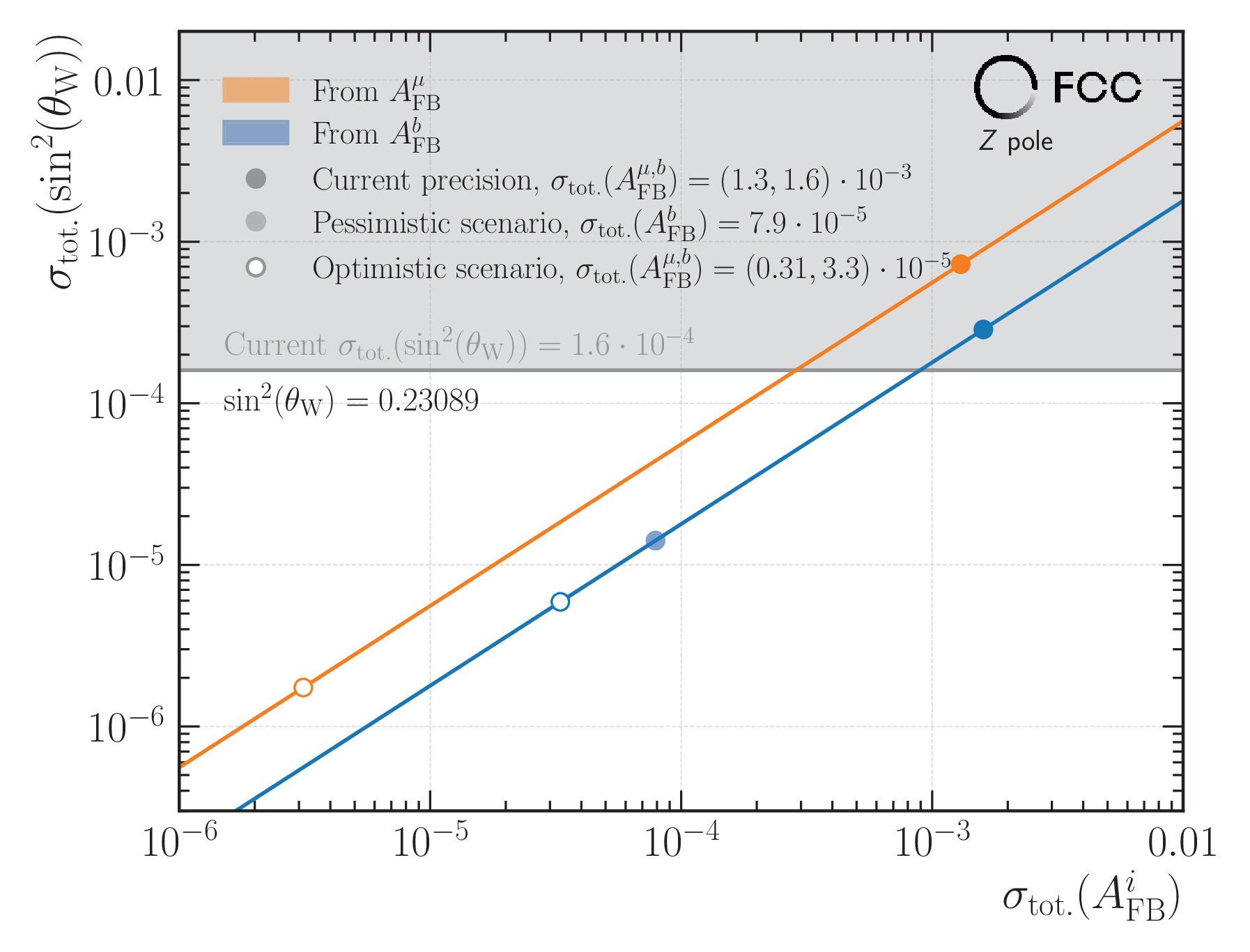}
        \caption{The total uncertainty on $\sin^2(\theta_\text{W})$ as function of the total uncertainty on $A_\text{FB}^i$ for $i\in [\mu, b]$, assuming that $\sigma_\text{tot.}(\sin^2(\theta_\text{W}))$ only arises from variations in $A_\text{FB}^i$. With the exclusively reconstructed $b$ hadrons as hemisphere tagger, a competitive method has been developed to challenge the precision on $\sin^2(\theta_\text{W})$ from $A_\text{FB}^\mu$.}
        \label{subfig:Zbb:sin2thetaW_uncertainties}
    \end{subfigure}
    \caption{The energy of the $b$ hadron directly serves as estimator for the amount of gluon radiation, which distort the initial $b$-quark direction. With a cut of $\min(E_B) \geq \SI{32}{\giga\eV}$ and a knowledge of the QCD corrections of \SI{1}{\percent}, $\sin^2(\theta_\text{W})$ can be derived with similar precision compared to $A_\text{FB}^\mu$.}
    \label{fig:Zbb:AFB_energy_results}
\end{figure}

In analogy to the studies in Sec.~\ref{subsec:Zbb:acollinearity}, a finer binning of the energy cuts has been chosen to properly identify the threshold, at which $\sigma_\text{stat.}(\AFBbeauty) = \sigma_\text{syst.}(\AFBbeauty)$ for the two scenarios of the relative QCD-correction uncertainty ${\large\sfrac{\sigma(C_\text{QCD})}{C_\text{QCD}}} = [1,5]\,\si{\percent}$. The left panel of Fig.~\ref{fig:Zbb:AFB_energy_results} presents the result normalised to the object level as a function of the reconstructed $b$-hadron energy, similar to Fig.~\ref{subfig:Zbb:b_quark_acollinearity}. The statistical precision is again shown in the grey uncertainty band, while the systematic uncertainty for \AFBbeautyzero for the more conservative and optimistic estimations for $\large\sfrac{\sigma(C_\text{QCD})}{C_\text{QCD}}$ are presented in lighter and darker orange, respectively. The threshold values have been found to be $\min(E_B) \geq \SI{32}{\giga\electronvolt}$ and $\min(E_B) \geq \SI{38.1}{\giga\electronvolt}$ for ${\large\sfrac{\sigma(C_\text{QCD})}{C_\text{QCD}}} = \SI{1}{\percent}$ and \SI{5}{\percent}, respectively. This results in the final uncertainties for \AFBbeautyzero
\begin{align}
    \begin{split}
        \frac{\sigma(C_\text{QCD})}{C_\text{QCD}} &= \SI{5}{\percent}\;\Rightarrow\;\AFBbeautyzero = \mu(\AFBbeautyzero)\,\pm\,5.6\cdot 10^{-5}\,(\text{stat.})\,\pm\,5.6\cdot 10^{-5}\,(\text{syst.})\,,\\
        \frac{\sigma(C_\text{QCD})}{C_\text{QCD}} &= \SI{1}{\percent}\;\Rightarrow\;\AFBbeautyzero = \mu(\AFBbeautyzero)\,\pm\,2.3\cdot 10^{-5}\,(\text{stat.})\,\pm\,2.3\cdot 10^{-5}\,(\text{syst.})\,.
    \end{split}
    \label{eqn:Zbb:final_AFB_uncertainties}
\end{align}
The total uncertainty on \AFBbeautyzero in the conservative and optimistic scenario results to \num{7.9e-5} and \num{3.3e-5}, respectively. Consequently, this would improve the measurement uncertainty from the LEP average in Eq.~\eqref{eqn:Zbb:AFB_LEP_world_average} by about a factor of 20 and 50.

\subsection{Constraints on the weak mixing angle}
The constraints on $\sin^2(\theta_\text{W})$ are expected to be competitive with those in reach from the muon forward-backward asymmetry, $A_\text{FB}^\mu$, with a notable improvement in the measurement of \AFBbeauty. The impact on $\sin^2(\theta_\text{W})$ is discussed in the following. 
From the $b$-quark forward-backward asymmetry, the weak mixing angle $\sin^2(\theta_\text{W})$ can be extracted via
\begin{equation}
    A_\text{FB}^b = \frac{3}{4}A_eA_b\,,\quad\text{with}\quad A_b = \frac{2v_ba_b}{v_b^2 + a_b^2}\,.
\end{equation}
using the explicit formulations of the vectorial ($v_b$) and axial ($a_b$) coupling of the $Z$ boson to the $b$ quark
\begin{align}
    \begin{split}
        a_b &= T_b\,,\\
        v_b &= T_b - 2Q_b \sin^2(\theta_\text{W})\,.
    \end{split}
    \label{eqn:intro:v_f}
\end{align}
In Eq.~\eqref{eqn:intro:v_f}, $T_b$ is the third component of the weak isospin and $Q_b$ refers to the charge of the $b$ quark.
Although the standard method for extracting $\sin^2(\theta_\text{W})$ is through $A_\text{FB}^\mu$, \AFBbeauty can provide a competitive method for comparing both methods to probe departures. This is because the sensitivity of \AFBbeauty is approximately three times higher than that of $A_\text{FB}^\mu$. Consequently, the uncertainty on $\sin^2(\theta_\text{W})$ derived from \AFBbeauty is about three times lower than that derived from $A_\text{FB}^\mu$, given the same total uncertainty. 

Therefore, it is worthwhile to calculate $\sigma_{\text{tot.}}(\sin^2(\theta_\text{W}))$ using the latest estimation~\cite{COM_ttbar} for the precision of $A_\text{FB}^\mu$
\begin{equation}
    \sigma_\text{tot.}(A_\text{FB}^\mu) = \sqrt{(\num{2e-6}(\text{stat.}))^2 + (\num{2.4e-6}(\text{syst.}))^2}
\end{equation}
and $\sigma_\text{tot.}(A_\text{FB}^b)$ (see Eq.~\eqref{eqn:Zbb:final_AFB_uncertainties}). In $\sigma_\text{tot.}(A_\text{FB}^\mu)$, the systematic uncertainty mainly arises from the knowledge of the COM energy. Numerical methods from the \texttt{scipy} library have been used to calculate
\begin{align}
    \begin{split}
        \sin^2(\theta_\text{W}) &= f(A_\text{FB}^i)\,,\\
        \sigma_\text{tot.}(\sin^2(\theta_\text{W})) &= \left(\frac{\partial f(A_\text{FB}^i)}{\partial A_\text{FB}^i}\right)\cdot\sigma_\text{tot.}(A_\text{FB}^i)\,,
    \end{split}
\end{align}
for $\sin^2(\theta_\text{W}) = 0.23089$.
The result is presented in the right panel of Fig.~\ref{fig:Zbb:AFB_energy_results} for the muon and $b$ quark in orange and blue, respectively. The grey-shaded area indicates the most precise uncertainty of $\sin^2(\theta_\text{W})$ to date. The filled points represent the currently most accurate measurements of $A_\text{FB}^{\mu,b}$~\cite{PDG}. The shaded and white blue dot show the pessimistic and optimistic scenario, where the pessimistic scenario assumes $\large\sfrac{\sigma(C_\text{QCD})}{C_\text{QCD}} = \SI{5}{\percent}$ and the optimistic scenario assumes $\large\sfrac{\sigma(C_\text{QCD})}{C_\text{QCD}} = \SI{1}{\percent}$. 

The figure reveals a bias between both representations, which originates from the higher sensitivity of \AFBbeauty to $\sin^2(\theta_\text{W})$. In the pessimistic case, an improvement in the precision of $\sin^2(\theta_\text{W})$ of about one order of magnitude is within reach. In contrast, the result becomes comparable to the one obtainable with $A_\text{FB}^\mu$ in the optimistic scenario, where the precision from \AFBbeauty compared to $A_\text{FB}^\mu$ is worse by about a factor of three. 

\section{Conclusions}\label{sec:Conclusions}

Although a powerful collider concept like FCC-ee unlocks unprecedented statistical precision with an enormous amount of $\mathcal{O}(10^{12})$ expected $Z$-boson decays, it requires careful consideration of controlling systematic uncertainties, which are not automatically reduced collecting more data. To substantially improve on fundamental SM parameters and to provide a competing method for their validation, new approaches to measuring the quantities of interest are needed. This has been demonstrated in the field of $b$-quark EWPOs by tagging the hemisphere flavour with exclusively reconstructed $b$-hadrons, eliminating over \SI{75}{\percent} of the systematic uncertainty in the measurement of \Rb and \AFBbeauty. Historically, both observables have suffered from contamination by lighter quarks. It could be shown through six representative decay modes that purities above \SI{99.8}{\percent} are achievable using exclusively reconstructed $b$-hadrons as hemisphere-flavour tagger. Further studies have addressed subleading systematic uncertainties, specifically the hemisphere correlation and the QCD correction for \Rb and \AFBbeauty, respectively. The sources of these uncertainties have been identified and mitigated by removing the PV dependence and using the energy of the $b$ hadron to estimate the angular distortion from radiated gluons prior to hadronisation. For \AFBbeauty, QCD corrections must be known with a relative precision of \SI{1}{\percent}, while it is sufficient to estimate the hemisphere correlation with \SI{10}{\percent} precision. With these assumptions, \Rb, \AFBbeauty and $\sin^2(\theta_\text{W})$ are expected to be measured with the following precisions at FCC-ee
\begin{alignat*}{4}
    \Rb &= \mu(\Rb) &\,&\pm\,&\,&2.22\cdot10^{-5}\,(\text{stat.}) &\,&\pm\,2.16\cdot 10^{-5}\,(\text{syst.})\,, \\
    \AFBbeauty &= \mu(\AFBbeauty) &\,&\pm\,&\,&2.30\cdot 10^{-5}\,(\text{stat.}) &\,&\pm\,2.30\cdot 10^{-5}\,(\text{syst.})\,, \\
    \Rightarrow \sin^2(\theta_\text{W}) &= \mu(\sin^2(\theta_\text{W})) &\,&\pm\,& &6 \cdot 10^{-6}\,.
\end{alignat*}
\section*{Acknowledgements}
This work has been funded by the Deutsche Forschungsgemeinschaft (DFG, German Research
Foundation) -- project number 465609373. Furthermore, LR would like to acknowledge the financial support of the Heinrich Hertz foundation and the Franco-German University (FGU).
\clearpage
\appendix
\section{Appendices}

\subsection[$R_b$ analysis]{\boldmath{$R_b$} analysis}\label{sec:app:R_b}

The following sections provide additional material for the $R_b$ analysis.

\subsubsection[List of \boldmath{$b$}-hadron decay modes]{List of \boldmath{$b$}-hadron decay modes}\label{sec:app:decay_modes}

In this section of the appendix, the list of all $b$-hadron decay modes to be included to reach a tagging efficiency of $\approx\!\SI{1}{\percent}$ is first presented before the results of the remaining representative decay modes are presented.

The following tables present the $b$-hadron decay modes, separately for the different $b$ hadrons: $B^\pm$ in Tab.~\ref{app:tab:decay_modes_Bplus}, $B^0$ in Tab.~\ref{app:tab:decay_modes_B0}, $B_s^0$ in Tab.~\ref{app:tab:decay_modes_Bs0} and for the $\Lambda_b^0$ baryon in Tab.~\ref{app:tab:decay_modes_LambdaB}. If available, the subsequent decay of, for example, heavy $c$-hadrons and baryons is indicated in the third column. The sum of Brs in percentage values, which quantifies the overall tagging efficiency, is given in the last column. Heavy $c$-hadron decays in the $B^0_{(s)}$ and $\Lambda_b^0$ decays are expected to be the ones of Tab.~\ref{app:tab:decay_modes_Bplus}.
\begin{table}[ht!]
    \caption{List of possible $B^+$ decay-modes. The decay modes in bold indicate the first decay stage of the $B^+$ meson followed by the subsequent decays in the third column. The hadronisation fraction of a $b$ quark to a $B^+$ is \SI{40.7}{\percent} and is not included in the branching fraction calculations.}
    \centerline{
    \rotatebox{90}{
        \begin{tabular}{ll|ll|c} 
        \toprule
        Mode & {$\text{Br}(B^+ \to XY)\,/\,\%$} & \multicolumn{2}{c|}{$\text{Br}(X \to \text{final state})\,/\,\%$} & {$\sum\text{Br}\,/\,\%$} \\ \midrule
        \multirow{2}{*}{\boldmath{$J/\psi\,K^+$}} & \multirow{2}{*}{\num{0.102(002)}} & \textcolor{orange}{$J/\psi \to e^+e^-$}     & \num{5.971(032)} & \multirow{2}{*}{\textbf{0.012}} \\ 
                                                                &                                   & \textcolor{orange}{$J/\psi \to \mu^+\mu^-$} & \num{5.961(033)} & \\ \midrule
        \multirow{5}{4.8cm}[0ex]{\boldmath{$\bar{D}^0\,\rho^+$} \\[.1ex]\boldmath{$\bar{D}^0\,\pi^+\pi^-\pi^+$} \\[.1ex] \boldmath{$\bar{D}^0\,\pi^+$} \\[.1ex] \boldmath{$[\bar{D}^0 \pi^+]_{D^*(2010)^+}\,\pi^-\pi^-\pi^0$}} & \multirow{5}{2.5cm}[0ex]{\num{1.340(180)} \\[.1ex] \num{0.560(210)} \\[.1ex] \num{0.468(13)} \\[.1ex] \num{10.160(4740)}} & \textcolor{orange}{$\bar{D}^0 \to K^+\pi^-\pi^0$}    & \num{14.400(500)}   & \multirow{5}{1cm}[0ex]{\textbf{0.545} \\[.1ex] \textbf{0.723} \\[.1ex] \textbf{0.909} \\[.1ex] \textbf{0.950}}\\
        && \textcolor{orange}{$\bar{D}^0 \to K^+ \pi^- 2\pi^0$} & \num{8.860(230)} & \\
        && \textcolor{orange}{$\bar{D}^0 \to K^+ 2\pi^- \pi^+$} & \num{8.220(140)} & \\
        && $\bar{D}^0 \to K^+ 2\pi^- \pi^+\pi^0$                & \num{4.300(400)} & \\
        && \textcolor{orange}{$\bar{D}^0 \to K^+\pi^-$} & \num{3.947(030)} & \\ \midrule
        \multirow{2}{*}{\boldmath{$D^-\,\pi^+\pi^-$}}  & \multirow{2}{*}{\num{0.107(5)}} & $D^+ \to K^- 2\pi^+$ & \num{9.380(160)} & \multirow{2}{*}{\textbf{0.966}}\\
        && $D^+ \to K^- 2\pi^+\pi^0$ & \num{6.250(180)} & \\ \midrule
        \multirow{7}{*}{\boldmath{$D_s^+\,\bar{D}^0$}} & \multirow{7}{*}{\num{0.900(090)}} & $D_s^+ \to [\pi^+\pi^-\pi^0]_\eta\,\pi^+\pi^0$ & \num{9.500(500)} & \multirow{7}{*}{\textbf{1.081}} \\
        && $D_s^+ \to [\pi^+\pi^-\pi^0]_\eta [\pi^+\pi^0]_{\rho^+}$ & \num{8.900(800)} & \\
        && $D_s^+ \to K^+K^-\pi^+\pi^0$                   & \num{5.500(240)}           & \\
        && \textcolor{orange}{$D_s^+ \to K^+K^-\pi^+$}    & \num{5.380(100)}           & \\
        && $D_s^+ \to 2\pi^+\pi^-$                        & \num{1.080(040)}           & \\
        && $D_s^+ \to K^+K^- 2\pi^+\pi^-$                 & \num{0.860(150)}           & \\
        && $D_s^+ \to 3\pi^+2\pi^-$                       & \num{0.790(080)}           & \\
        \bottomrule
    \end{tabular}
    }
    }
    \label{app:tab:decay_modes_Bplus}
\end{table}

\begin{table}[ht!]
    \centering
    \caption{List of possible $B^0$ decay-modes. The subsequent decays of the $J/\psi$ and $c$ mesons are not shown as they can be found in Table~\ref{app:tab:decay_modes_Bplus}. The hadronisation fraction of a $b$ quark to a $B^0$ is \SI{40.7}{\percent} and is not included in the branching fraction calculations.}
    \begin{tabular}{lc|c}
        \toprule
        Mode & {$\text{Br}(B^0 \to\text{final state})\,/\,\%$} & {$\sum_{}^{} \text{Br}\,/\,\%$}\\ \midrule
        \boldmath{$J/\psi\,K^+\pi^-$}                   & 0.014 & \textbf{0.014}\\ \midrule
        \boldmath{$D^*(2010)^-\,\pi^+\pi^+\pi^-\pi^0$}  & 0.473 & \textbf{0.487}\\
        \boldmath{$D^*(2010)^-\,\pi^+\pi^0$}            & 0.403 & \textbf{0.891}\\
        \boldmath{$D^*(2010)^-\,\pi^+\pi^+\pi^-$}       & 0.194 & \textbf{1.084}\\
        \boldmath{$D^-\,\pi^+\pi^+\pi^-$}               & 0.094 & \textbf{1.178}\\
        \boldmath{$D^*(2010)^-\,\pi^+$}                 & 0.074 & \textbf{1.252}\\
        \boldmath{$D^*(2010)^-\,D_s^+$}                 & 0.069 & \textbf{1.321}\\
        \boldmath{$D^-\,\pi^+$}                         & 0.039 & \textbf{1.360}\\
        \boldmath{$D^-\,D_s^+$}                         & 0.036 & \textbf{1.396}\\
        \boldmath{$D^*(2010)^-\,D^0\,K^+$}              & 0.026 & \textbf{1.422}\\
        \boldmath{$D^-\,D^0\,K^+$}                      & 0.007 & \textbf{1.429}\\
        \bottomrule
    \end{tabular}
    \label{app:tab:decay_modes_B0}
\end{table}

\begin{table}[ht!]
    \centering
    \caption{List of possible $B_s^0$ decay-modes. The subsequent decays of the $c$ mesons are not shown as they can be found in Tab.~\ref{app:tab:decay_modes_Bplus}. The hadronisation fraction of a $b$ quark to a $B_s^0$ is \SI{10.1}{\percent} and is not included in the branching fraction calculations.}
    \begin{tabular}{lc|c}
        \toprule
        Mode & {$\text{Br}(B_s^0 \to\text{final state})\,/\,\%$} & {$\sum_{}^{} \text{Br}\,/\,\%$}\\ \midrule
        \boldmath{$D_s^-\,[\pi^+\pi^0]_{\rho^+}$} & 0.218 & \textbf{0.218} \\
        \boldmath{$D_s^-\,\pi^+\pi^+\pi^-$}       & 0.195 & \textbf{0.413} \\
        \boldmath{$D^*(2010)^-\,\pi^+\pi^+\pi^-$} & 0.194 & \textbf{0.607} \\
        \boldmath{$D_s^-\,\pi^+$}                 & 0.095 & \textbf{0.702} \\
        \boldmath{$D_s^+\,D_s^-$}                 & 0.045 & \textbf{0.747} \\
        \boldmath{$D^0\,K^-\pi^+$}                & 0.041 & \textbf{0.789} \\
        \bottomrule
    \end{tabular}
    \label{app:tab:decay_modes_Bs0}
\end{table}

\begin{table}[ht!]
    \centering
    \caption{List of possible $\Lambda_b^0$ decay-modes. The hadronisation fraction of a $b$ quark to a $\Lambda_b^0$ is \SI{8.4}{\percent} and is not included in the branching fraction calculations.}
    \centerline{
    \begin{tabular}{ll|ll|c}
        \toprule
        Mode & {$\text{Br}(\Lambda_b^0 \to XY)\,/\,\%$} & \multicolumn{2}{c|}{$\text{Br}(X \to \text{final state})\,/\,\%$} & {$\sum\text{Br}\,/\,\%$} \\ \midrule
        \multirow{2}{*}{\boldmath{$\Lambda_b^0 \to \Lambda_c^+ \pi^+\pi^-\pi^-$}} & \multirow{2}{*}{\num{0.760(110)}} & $\Lambda_c^+ \to pK^-\pi^+$ & \num{6.280(320)} & \multirow{2}{*}{\textbf{0.082}} \\ 
        && $\Lambda_c^+ \to pK^-\pi^+\pi^0$ & \num{4.460(300)} & \\ 
        \bottomrule
    \end{tabular}
    }
    \label{app:tab:decay_modes_LambdaB}
\end{table}

\clearpage
\subsubsection{Reconstruction of the remaining \boldmath{$b$}-hadron decay modes}\label{app:subsec:Zbb:rest_of_the_modes}

The characteristics of the decay modes in the aforementioned tables are represented in the listing of six decay modes in Sec.~\ref{subsec:Zbb:exclusive_reconstruction_mode}. As an example, the decay $B^+ \to [K^+\pi^-\pi^0]_{\bar{D}^0}\pi^+$ has been reconstructed. In the following, the results of the remaining five decay modes are shown. In general, the assumptions made for the vertex-resolution emulation and for kinematic cuts on intermediate particles have been applied similarly.

\begin{figure}[ht]
    \begin{subfigure}[t]{0.48\textwidth}
        \centering
        \includegraphics[width=1\linewidth]{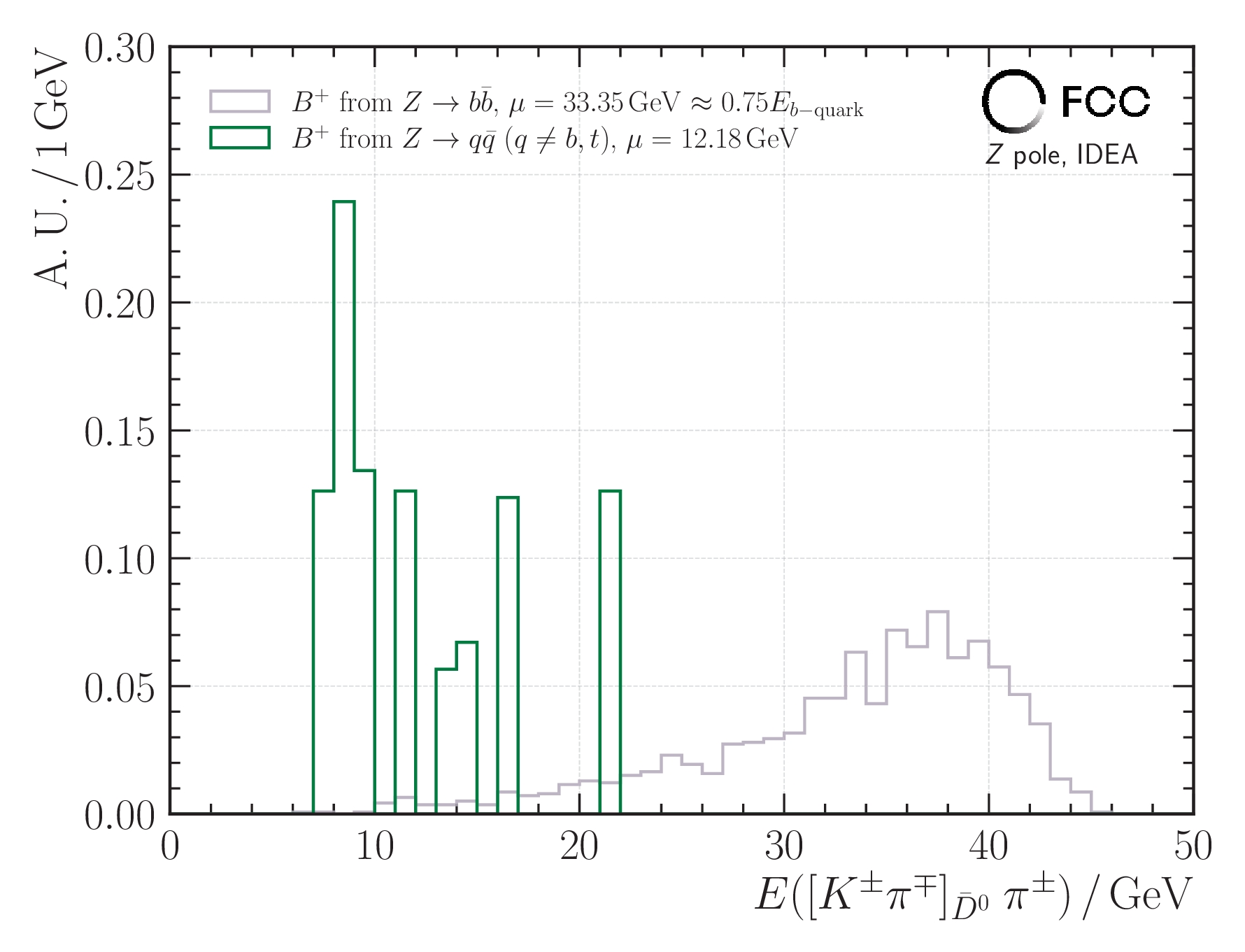}
        \caption{}
        \label{app:subfig:Bplus_energy}
    \end{subfigure}\hfill
    \begin{subfigure}[t]{0.48\textwidth}
        \centering
        \includegraphics[width=1\linewidth]{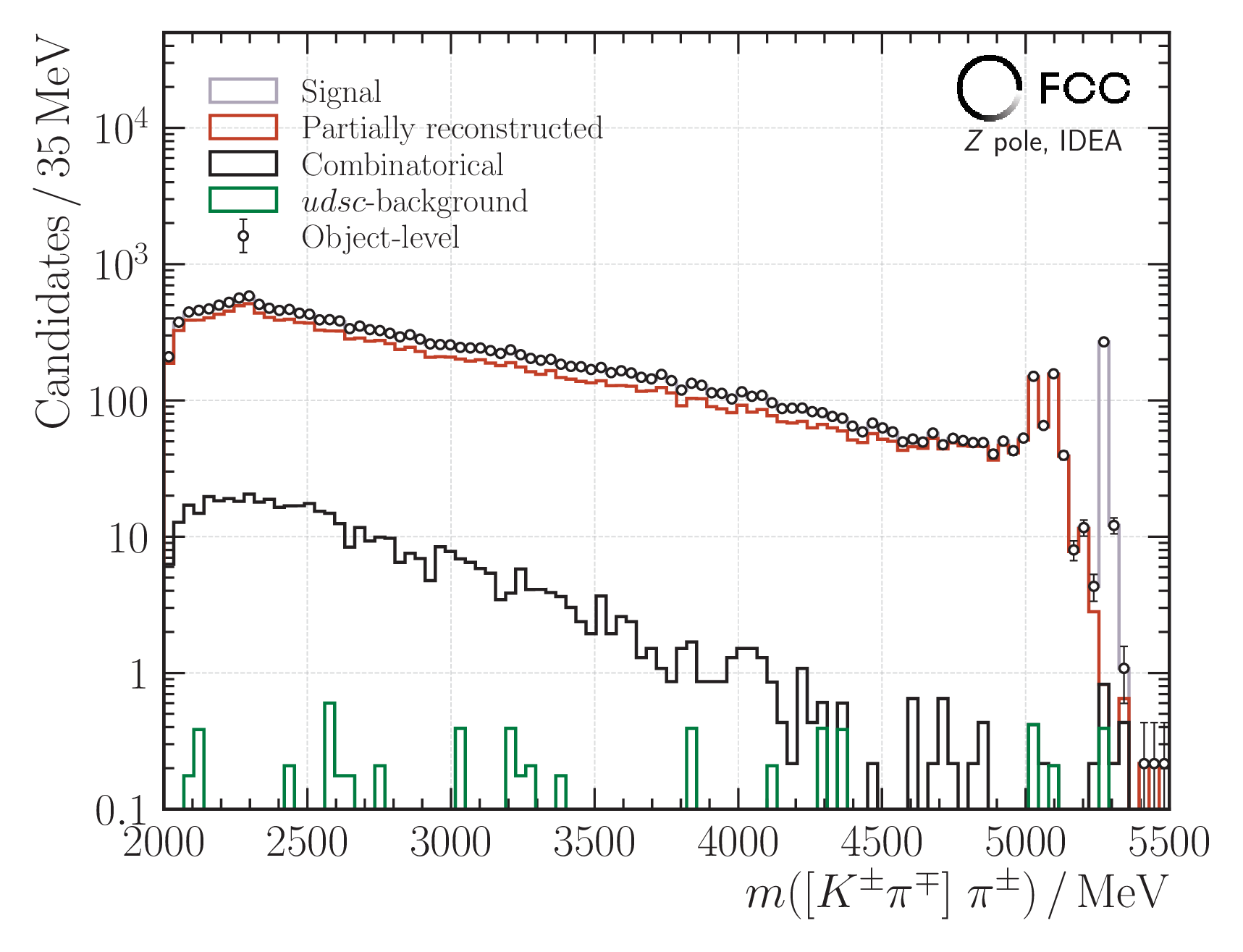}
        \caption{}
        \label{app:subfig:Bplus_invariant_mass}
    \end{subfigure}
    \caption{Energy and invariant-mass distribution of the signal and background candidates in Figs.~\subref{app:subfig:Bplus_energy} and~\subref{app:subfig:Bplus_invariant_mass}, respectively. The energy cut has been set to \SI{20}{\giga\eV}.}
    \label{app:fig:Bplus_fully_charged}
\end{figure}

\paragraph{Fully charged \boldmath{$D^0$} decay} 
The invariant-mass distribution of the $B^+$ meson from $B^+ \to [K^+\pi^-]_{\bar{D}^0} \pi^+$ is shown on the right side of Fig.~\ref{app:fig:Bplus_fully_charged} after an energy cut on the $B^+$ candidates of \SI{20}{\giga\eV} has been applied. The distribution of the energy is presented on the left side of Fig.~\ref{app:fig:Bplus_fully_charged}. With an energy cut of $E_{B^+} \geq \SI{20}{\giga\eV}$, a purity of \SI{99.93(11)}{\percent} has been achieved, where the uncertainty originates only from the size of the available sample.

\begin{figure}[ht]
    \begin{subfigure}[t]{0.48\textwidth}
        \centering
        \includegraphics[width=1\linewidth]{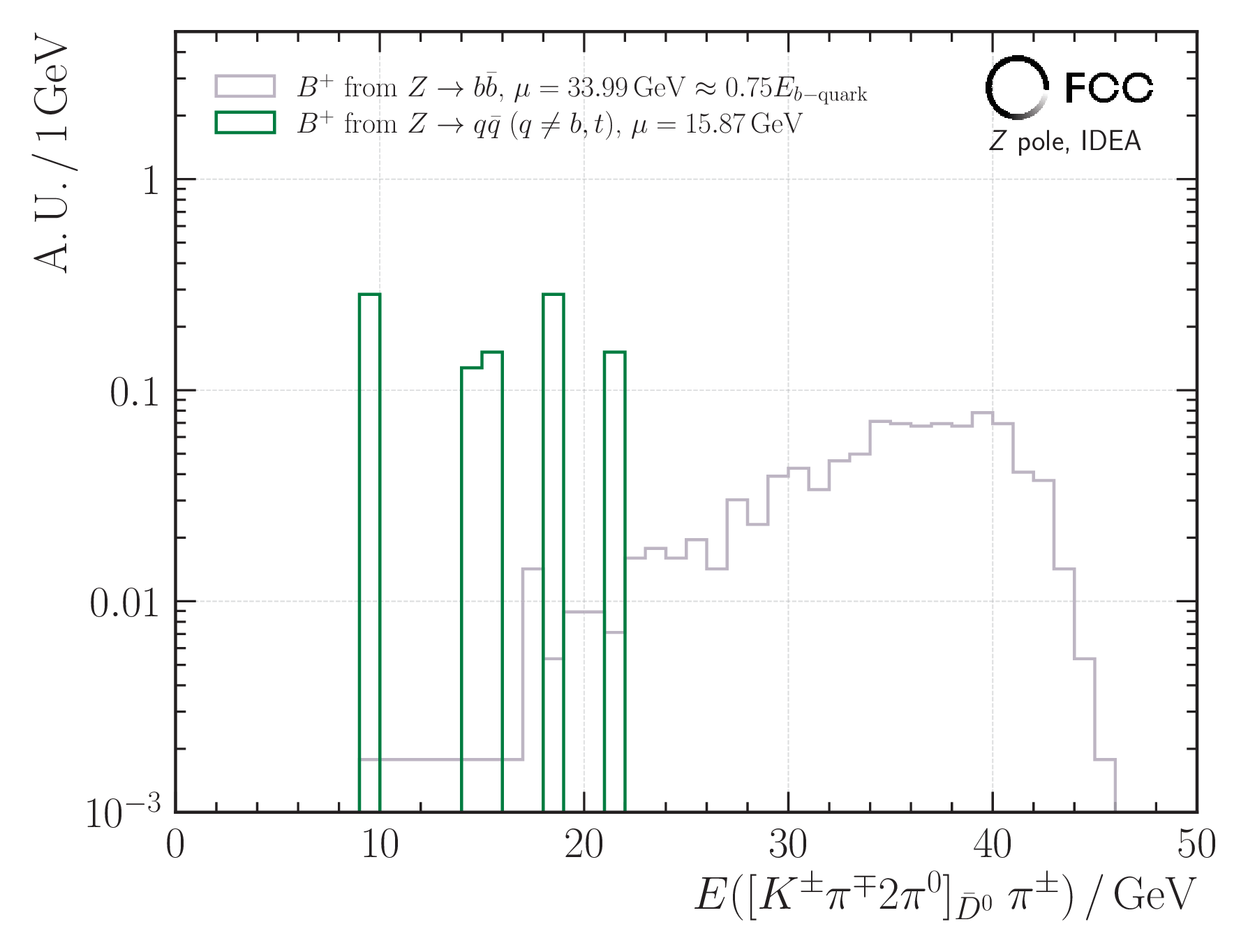}
        \caption{}
        \label{app:subfig:Bplus_D000_energy}
    \end{subfigure}\hfill
    \begin{subfigure}[t]{0.48\textwidth}
        \centering
        \includegraphics[width=1\linewidth]{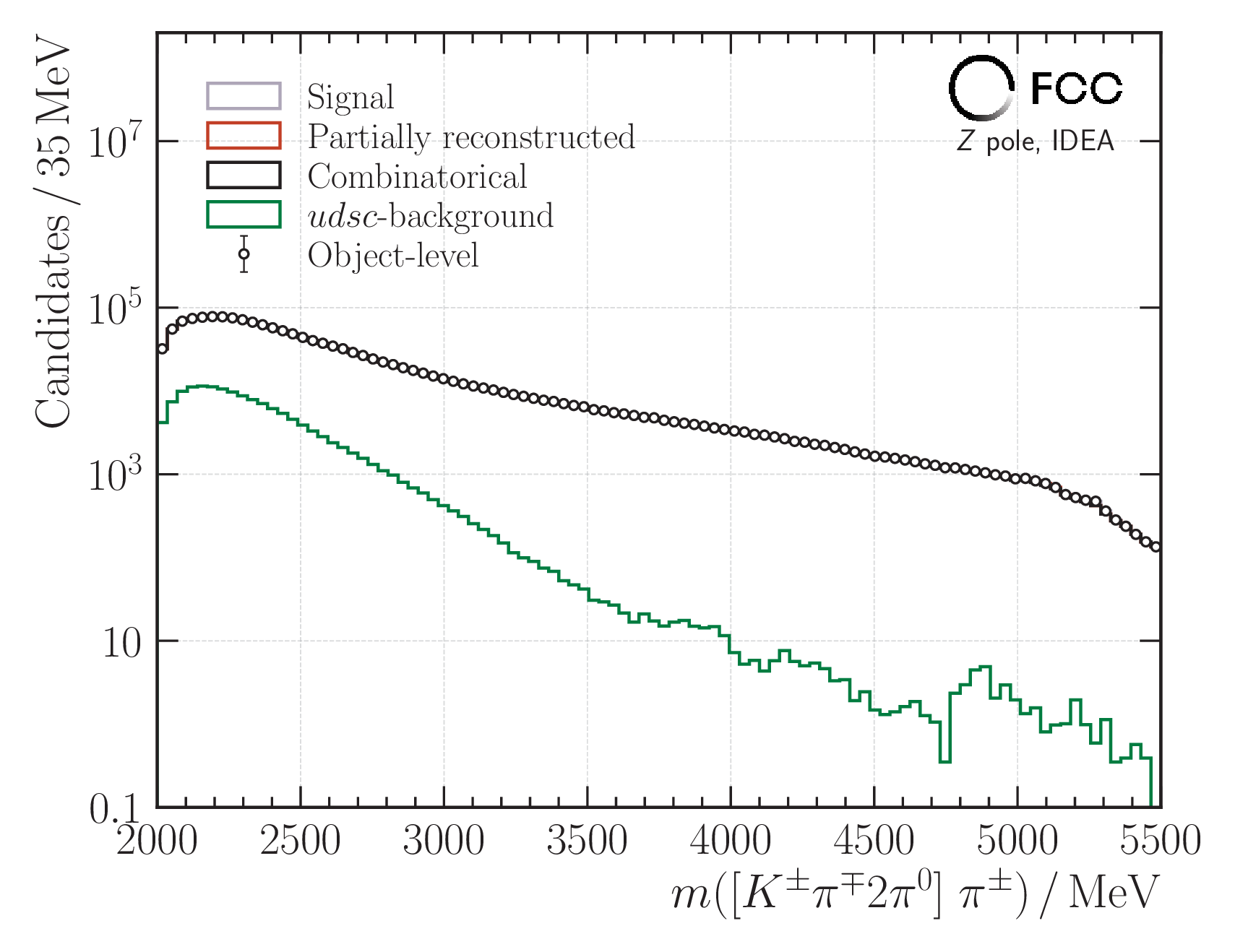}
        \caption{}
        \label{app:subfig:Bplus_D000_invariant_mass}
    \end{subfigure}
    \caption{Energy and invariant-mass distribution of the signal and background candidates in Figs.~\subref{app:subfig:Bplus_D000_energy} and~\subref{app:subfig:Bplus_D000_invariant_mass}, respectively. The energy cut has been set to \SI{20}{\giga\eV}.}
    \label{app:fig:Bplus_2pi0}
\end{figure}

\clearpage
\paragraph{\boldmath{$D^0$} decay with two \boldmath{$\pi^0$}} 
The invariant-mass distribution of the $B^+$ meson from $B^+ \to [K^+\pi^-\pi^0\pi^0]_{\bar{D}^0} \pi^+$ is shown on the right side of Fig.~\ref{app:fig:Bplus_2pi0} after an energy cut on the $B^+$ candidates of \SI{20}{\giga\eV} has been applied. The distribution of the energy is presented on the left side of Fig.~\ref{app:fig:Bplus_2pi0}. With an energy cut of $E_{B^+} \geq \SI{20}{\giga\eV}$, a purity of \SI{99.81(07)}{\percent} has been achieved, where the uncertainty originates only from the size of the available sample.

\begin{figure}[ht]
    \begin{subfigure}[t]{0.48\textwidth}
        \centering
        \includegraphics[width=1\linewidth]{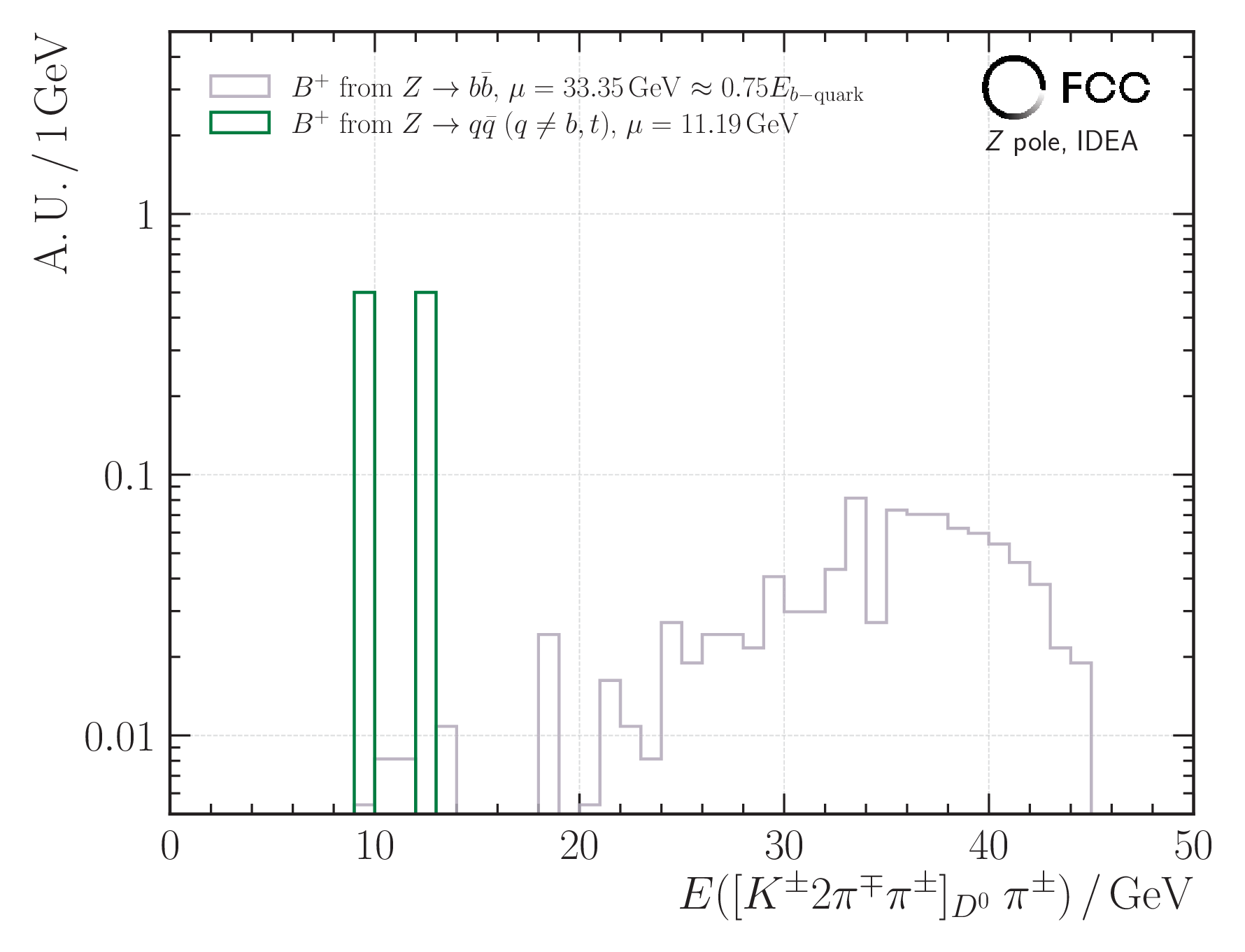}
        \caption{}
        \label{app:subfig:Bplus_4ct_energy}
    \end{subfigure}\hfill
    \begin{subfigure}[t]{0.48\textwidth}
        \centering
        \includegraphics[width=1\linewidth]{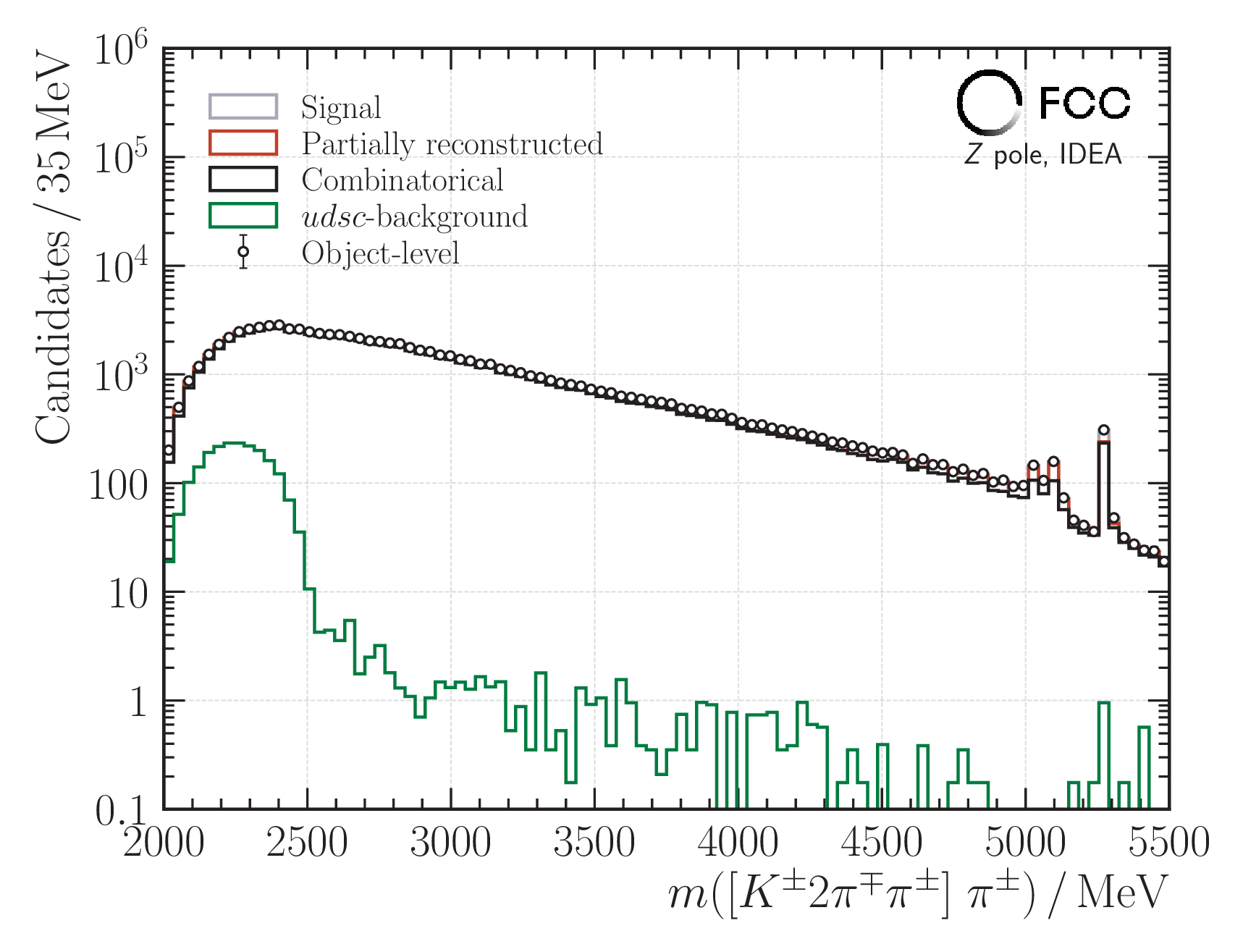}
        \caption{}
        \label{app:subfig:Bplus_4ct_invariant_mass}
    \end{subfigure}
    \caption{Energy and invariant-mass distribution of the signal and background candidates in Figs.~\subref{app:subfig:Bplus_4ct_energy} and~\subref{app:subfig:Bplus_4ct_invariant_mass}, respectively. The energy cut has been set to \SI{20}{\giga\eV}.}
    \label{app:fig:Bplus_4ct}
\end{figure}

\clearpage
\paragraph{Four charged tracks at the \boldmath{$D^0$} decay-vertex} 
The invariant-mass distribution of the $B^+$ meson from $B^+ \to [K^+\pi^-\pi^-\pi^+]_{\bar{D}^0} \pi^+$ is shown on the right side of Fig.~\ref{app:fig:Bplus_4ct} after an energy cut on the $B^+$ candidates of \SI{20}{\giga\eV} has been applied. The distribution of the energy is presented on the left side of Fig.~\ref{app:fig:Bplus_4ct}. With an energy cut of $E_{B^+} \geq \SI{20}{\giga\eV}$, a purity of \SI{99.73(27)}{\percent} has been achieved, where the uncertainty originates only from the size of the available sample.

\begin{figure}[ht]
    \centering
    \includegraphics[width=0.7\linewidth]{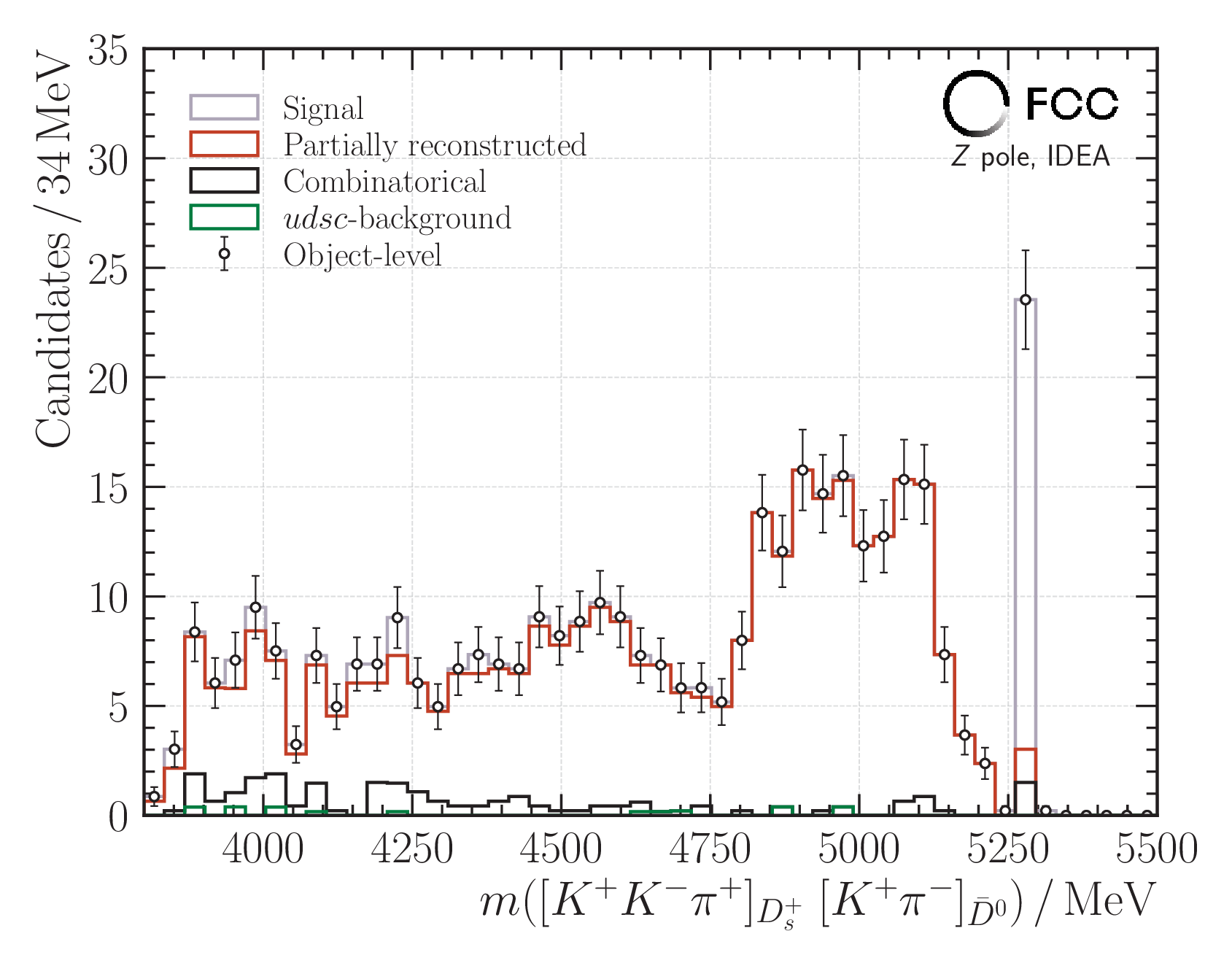}
    \caption{The invariant-mass distribution without any energy cut. However, it is expected to have some light-quark contamination when considering the full event statistics.}
    \label{app:fig:Bplus_Dspm_energy}
\end{figure}

\clearpage
\paragraph{Including two \boldmath{$c$} mesons} 
The invariant-mass distribution of the $B^+$ meson from $B^+ \to \bar{D}^0D_s^+$ with the subsequent decays $\bar{D}^0 \to K^+\pi^-$ and $D_s^+ \to K^+K^-\pi^+$ is shown in Fig.~\ref{app:fig:Bplus_Dspm_energy} without any energy cut on the $B^+$ candidates, since with the limited amount of simulated events, no $udsc$ events have been found in the signal mass window. Therefore, a purity of \SI{100.00(0)}{\percent} has been achieved, which is expected to be slightly lowered with more events available.

\begin{figure}[ht]
    \begin{subfigure}[t]{0.48\textwidth}
        \centering
        \includegraphics[width=1\linewidth]{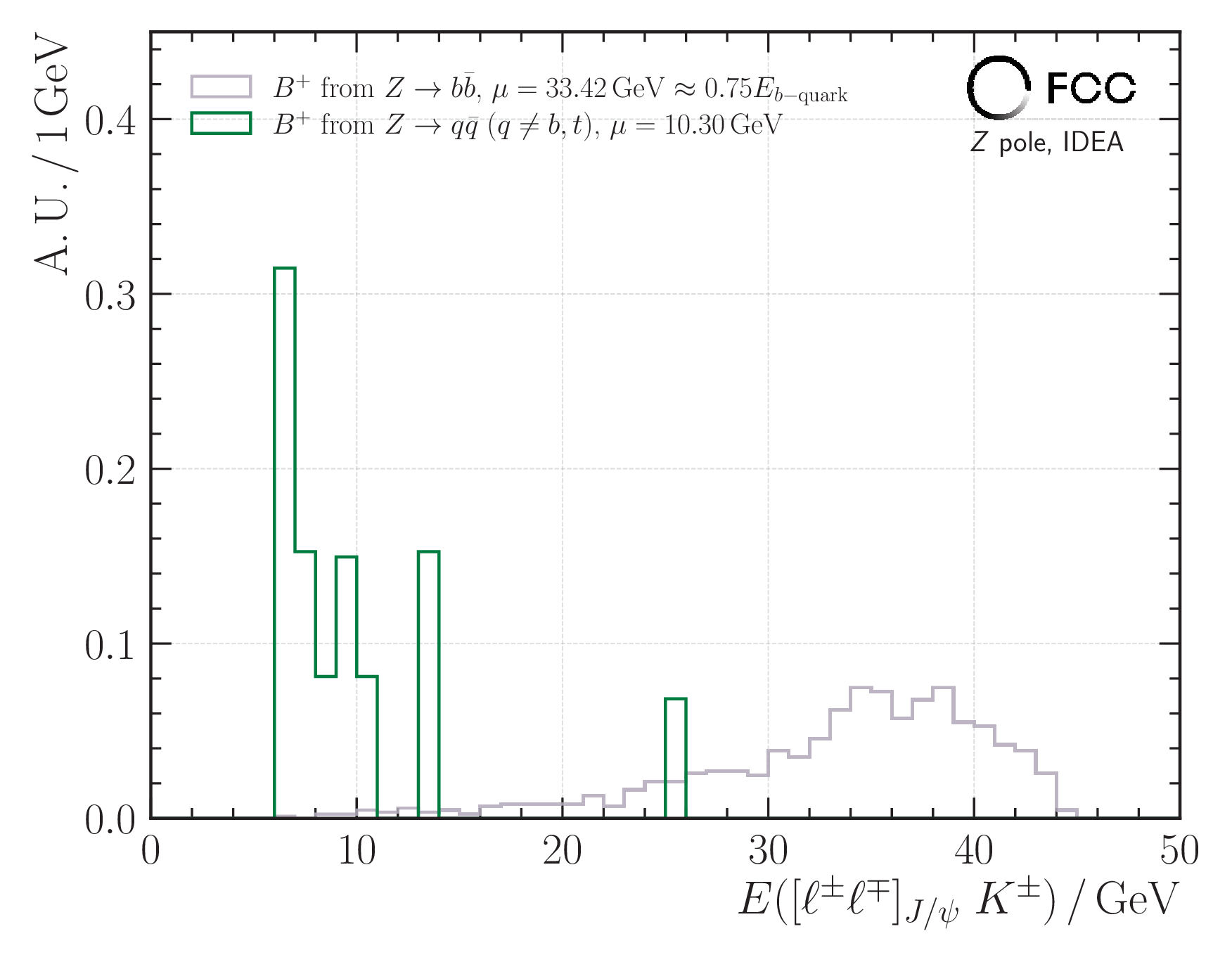}
        \caption{}
        \label{app:subfig:Bplus_JPsi_energy}
    \end{subfigure}\hfill
    \begin{subfigure}[t]{0.48\textwidth}
        \centering
        \includegraphics[width=1\linewidth]{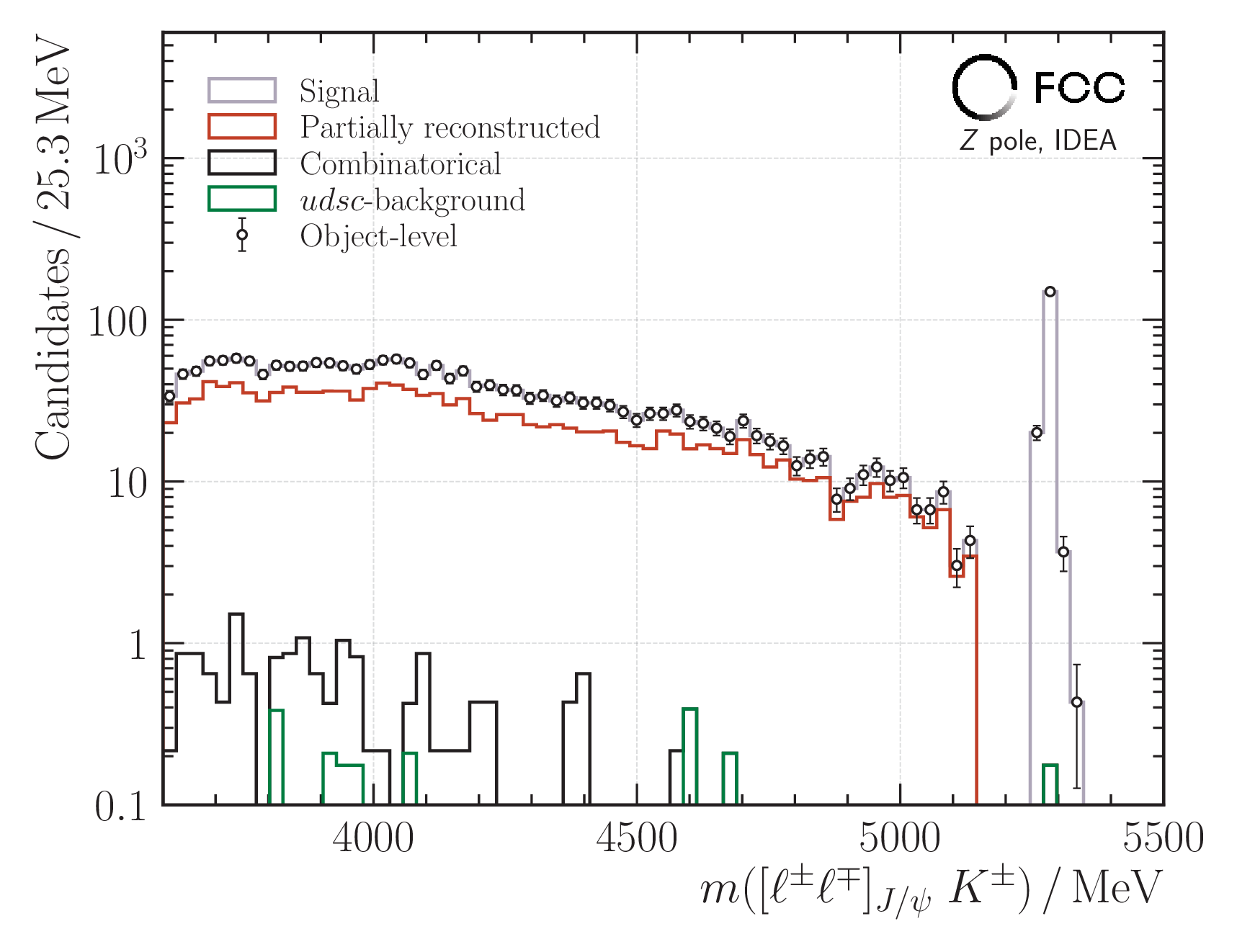}
        \caption{}
        \label{app:subfig:Bplus_JPsi_invariant_mass}
    \end{subfigure}
    \caption{Energy and invariant-mass distribution of the signal and background candidates in Figs.~\subref{app:subfig:Bplus_JPsi_energy} and~\subref{app:subfig:Bplus_JPsi_invariant_mass}, respectively. The energy cut has been set to \SI{20}{\giga\eV}.}
    \label{app:fig:Bplus_JPsi}
\end{figure}

\clearpage
\paragraph{Including a \boldmath{$c\bar{c}$} meson} 
The invariant-mass distribution of the $B^+$ meson from $B^+\to [\ell^+\ell^-]_{J/\psi}\,K^+$ with $\ell \in [e, \mu]$ is shown on the right side of Fig.~\ref{app:fig:Bplus_JPsi} after an energy cut on the $B^+$ candidates of \SI{20}{\giga\eV} has been applied. The distribution of the energy is presented on the left side of Fig.~\ref{app:fig:Bplus_JPsi}. With an energy cut of $E_{B^+} \geq \SI{20}{\giga\eV}$, a purity of \SI{99.90(24)}{\percent} has been achieved, where the uncertainty originates only from the size of the available sample.

\clearpage
\subsection[$A_\text{FB}^b$ analysis]{\boldmath{$A_\text{FB}^b$} analysis}\label{sec:app:AFB_b}
In the following additional information for the analysis of \AFBbeauty is provided.

\subsubsection{Theory of QCD correction and longitudinal boson-polarisation}\label{subsec:app:C_QCD_and_fL}

In the following, the theoretical basis is provided, mainly motivated and adapted by the studies in Ref.~\cite{Revisiting_QCD_correction_really_a_problem}. The analytical expressions for $C_\text{QCD}$ are provided as follows
\begin{equation}
    C_\text{QCD}(\mu) \approx \int_{x_\text{min}}^{x_\text{max}}\int_{\bar{x}_\text{min}(x)}^{\bar{x}_\text{max}(x)} \frac{2\bar{x}^2 (1 - \cos(\zeta(x, \bar{x}, \mu)))}{3 (1 - x) (1 - \bar{x})}\,\text{d}\bar{x}\,\text{d}x\,,
    \label{eqn:app:C_mu_integral}
\end{equation}
with the energy fractions of the $b$ and $\bar{b}$ quark $x = \large\sfrac{2E_{b}}{\sqrt{s}}$ and $\bar{x} = \sfrac{2E_{\bar{b}}}{\sqrt{s}}$, respectively. The acollinearity here is written explicitly as a function of $(x, \bar{x}, \mu)$ and the definition is given according to Ref.~\cite{Note_on_QCD_corretions} as
\begin{equation}
    \cos(\zeta(x,\bar{x},\mu)) = \frac{x\bar{x} + \mu^2 + 2(1 - x - \bar{x})}{\sqrt{x^2 - \mu^2}\sqrt{\bar{x}^2 - \mu^2}}\,.
\end{equation}
The integral limits in Eq.~\eqref{eqn:app:C_mu_integral} are derived from the possible configurations for the $b$ and $\bar{b}$ quarks. This means $x_\text{min} = \mu$, $x_\text{max} = 1$ (either carrying no momentum or the full momentum of $\large\sfrac{\sqrt{s}}{2}$), such that 
\begin{align}
    \bar{x}_\text{min}(x) &= 1 - \frac{x + \sqrt{x^2 - \mu^2}}{2} + \frac{\mu^2}{2 - x - \sqrt{x^2 - \mu^2}}\,,\\
    \bar{x}_\text{max}(x) &= 1 - \frac{x - \sqrt{x^2 - \mu^2}}{2} + \frac{\mu^2}{2 - x + \sqrt{x^2 - \mu^2}}\,.
\end{align}
The analytical expression for $f_\text{L}$ is given as
\begin{equation}
    f_\text{L}(\mu) \approx \int_{x_\text{min}}^{x_\text{max}}\int_{\bar{x}_\text{min}(x)}^{\bar{x}_\text{max}(x)} \frac{4\alpha_\text{S}\sqrt{\bar{x}^2 - \mu^2} (1 - \cos^2(\zeta(x,\bar{x},\mu)))}{3\pi (1 - x)(1 - \bar{x})}\,\text{d}x\,\text{d}\bar{x}\,.
\end{equation}

\printbibliography
\end{document}